\documentclass[a4paper,UKenglish,cleveref,autoref]{lipics-v2019}

\newif\ifextended
\extendedtrue

\usepackage[utf8]{inputenc}
\usepackage{microtype}

\usepackage{graphicx}
\usepackage{booktabs}
\usepackage{subcaption}
\usepackage{multirow}
\usepackage{numprint}
\usepackage[compress,sort]{cite}

\ifextended
\usepackage[bibliography=separate]{apxproof}
\else
\usepackage[bibliography=separate,appendix=strip]{apxproof}
\fi

\usepackage{etoolbox}

\usepackage{pgfplots}
\pgfplotsset{compat=1.14}

\pgfplotsset{every axis/.append style={font=\normalsize, line width=0.5pt, tick
 style={line width=0.5pt}}}

\tikzset{every mark/.append style={scale=2}}

\renewcommand{\phi}{\varphi}
\renewcommand{\epsilon}{\varepsilon}
\renewcommand{\leq}{\leqslant}
\renewcommand{\geq}{\geqslant}

\ifextended
\else
\EventEditors{Pablo Barcelo and Marco Calautti}
\EventNoEds{2}
\EventLongTitle{22nd International Conference on Database Theory (ICDT 2019)}
\EventShortTitle{ICDT 2019}
\EventAcronym{ICDT}
\EventYear{2019}
\EventDate{March 26--28, 2019}
\EventLocation{Lisbon, Portugal}
\EventLogo{}
\SeriesVolume{127}
\ArticleNo{9}
\fi

\ifextended
\relatedversion{This is an extended version of an article published in
the proceedings of ICDT 2019.}
\else
\relatedversion{An extended version of this paper is available at
\url{TODO}.}
\fi

\title{An Experimental Study of the Treewidth of Real-World Graph Data
\ifextended (Extended Version)\fi}

\author{Silviu Maniu}{LRI, CNRS, Universit\'e Paris-Sud, Universit\'e
Paris-Saclay, [Orsay], France}{silviu.maniu@lri.fr}{}%{https://orcid.org/0000-0002-8623-1533}
{}

\author{Pierre Senellart}{DI ENS, ENS, CNRS, PSL University \& Inria \&
LTCI, T\'el\'ecom ParisTech, [Paris], France}{pierre@senellart.com}{}%{https://orcid.org/0000-0002-7909-5369}
{}

\author{Suraj Jog}{University of Illinois at Urbana–Champaign, [Urbana-Champaign], USA}{sjog2@illinois.edu}{}{}

\authorrunning{S. Maniu, P. Senellart, and S. Jog}

\Copyright{Silviu Maniu, Pierre Senellart, and Suraj Jog}%mandatory, please use full first names. LIPIcs license is "CC-BY";  http://creativecommons.org/licenses/by/3.0/


\acknowledgements{We are grateful to Antoine Amarilli and Mikaël Monet for feedback on
  parts of this paper.}

\ifextended
\hideLIPIcs
\fi

\nolinenumbers

\begin{document}
  
\maketitle

\begin{abstract}
  Treewidth is a parameter that measures how tree-like a
  relational instance is, and whether it can reasonably be decomposed into a tree. Many computation tasks are known to be
  tractable on databases of small treewidth, but computing the treewidth
  of a given instance is intractable.
  This article is the first large-scale experimental study of treewidth and tree
  decompositions of real-world database instances (25~datasets
  from 8 different domains, 
  with sizes ranging from a few thousand to a
  few million vertices). The goal is to determine which data, if any, can
  benefit of the wealth of algorithms for databases of small treewidth.
  For each dataset, 
  we obtain upper and lower bound estimations of their
  treewidth, and study the properties of their tree decompositions.
  We show in particular that, even when treewidth is high,
  using partial tree decompositions can result in data structures that can
  assist algorithms.
\end{abstract}

\maketitle

\section{Introduction and Related Work}\label{sec:intro}
A number of data management tasks related to query evaluation are
computationally intractable when rich query languages or complex tasks
are involved, even when the query is assumed to be fixed (that is, when
we consider \emph{data complexity}~\cite{Vardi82}). For example:
\begin{itemize}
  \item query evaluation of Boolean
    monadic second-order (MSO) queries is hard for every level of the
    polynomial hierarchy~\cite{AjtaiFS00}; 
  \item unless $\mathrm{P}=\mathrm{NP}$, there is no polynomial-time
    enumeration or counting algorithm for first-order (FO)
    queries with free second-order variables~\cite{SALUJA1995493,DBLP:conf/csl/DurandS11};
  \item computing the probability of conjunctive queries (CQs) over 
    tuple-independent databases, a very simple model
    of
    probabilistic databases, is
    $\mathrm{\#P}$-hard~\cite{DBLP:conf/pods/DalviS07a};
  \item unless $\mathrm{P}=\mathrm{NP}$, there is no polynomial-time
    algorithm to construct a
    deterministic decomposable negation normal form (\mbox{d-DNNF})
    representation of the Boolean provenance of some CQ
    \cite{DBLP:conf/pods/DalviS07a,DBLP:journals/mst/JhaS13};
  furthermore, there is no polynomial bound on the size of a \emph{structured} d-DNNF
    representation of the Boolean provenance of unions of conjunctive
    queries with disequalities \cite[Theorem~33]{DBLP:conf/icdt/AmarilliMS18}.
\end{itemize}

Other problems yield complexity classes usually considered
tractable, such as $\textrm{AC}_0$ for Boolean FO query evaluation
\cite{abiteboul1995foundations}, but may still result in
impractical running times on large database instances. 

To face this intractability and practical inefficiency, one possible
approach has been to determine conditions on the structure of databases
that ensure tractability, often through a series of \emph{algorithmic
meta-theorems} \cite{kreutzer2009algorithmic}. This has led, for
instance, to the introduction of the notions of \emph{locally
tree-decomposable structures} for near-linear-time evaluation of Boolean
FO queries~\cite{DBLP:journals/jacm/FrickG01}, or to that of
\emph{structures of bounded expansion} for constant-delay enumeration of
FO queries \cite{DBLP:conf/pods/KazanaS13}.

\subparagraph*{Treewidth} 
A particularly simple and widely used way to restrict database instances
that
ensures a wide class of tractability results is to bound the
\emph{treewidth} of the instance (this is actually a special case of both
notions of locally tree-decomposable and bounded expansion). \emph{Treewidth}~\cite{robertson1984graph} is a graph-theoretic parameter that
characterizes how tree-like a graph, or more generally a relational
instance, is, and hence whether it can be reasonably transformed into a
tree structure (a \emph{tree decomposition}). Indeed:
\begin{itemize}
  \item query evaluation of MSO queries is linear-time
    over bounded-treewidth structures \cite{courcelle1990monadic,flum2002query};
  \item counting \cite{arnborg1987complexity} and enumeration
    \cite{bagan2006mso,DBLP:conf/icalp/AmarilliBJM17} of MSO queries on bounded-treewidth structures is
    linear-time;
  \item computing the probability of MSO queries over a bounded-treewidth
    tuple-independent database is linear-time assuming constant-time
    rational arithmetic \cite{amarilli2015provenance};
  \item a linear-sized structured d-DNNF representation of the provenance of any MSO
    query over bounded-treewidth databases can be computed in linear-time
    \cite{amarilli2016tractable,DBLP:conf/icdt/AmarilliMS18}.
\end{itemize}
These results mostly stem from the fact that, on trees, MSO queries
can be rewritten to tree automata~\cite{thatcher1968generalized}, though
this process is non-elementary in general (which impacts the
\emph{combined complexity}~\cite{Vardi82}, but not the data complexity).
We can see these results as \emph{fixed-parameter tractability},
with a complexity in $\mathcal O(f(|Q|,k)\times |D|)$ where $|D|$ is the size of the
database, $k$ its treewidth, $|Q|$ the size of the query, and $f$ some
computable function.
Note that another approach for tractability, out of the scope of this
paper, is to restrict the queries instead of the
instances, e.g., by enforcing low treewidth on the queries
\cite{grohe2001when} or on the provenance of the queries
\cite{jha2012tractability}.

Such results have been, so far, of mostly theoretical interest -- mainly
due to the high complexity of the function~$f$ of $|Q|$ and~$k$.
However, algorithms that exploit the low treewidth of instances have been
proposed and successfully applied to real-world and synthetic data:
for shortest path queries in graphs \cite{wei2010tedi,
planken2012computing}, distance queries in probabilistic graphs
\cite{maniu2017indexing}, or ad-hoc queries compiled to tree
automata \cite{DBLP:conf/sigmod/Monet16}.
In other domains, low treewidth is an indicator for efficient evaluation of quantified Boolean formulas~\cite{pulina2010empirical}.

Sometimes, treewidth even seems to be the \emph{sole} criterion that may
render an intractable problem tractable, under some technical
assumptions: \cite{kreutzer2010lower} shows that, unless the
exponential-time hypothesis is false, MSO$_2$ query evaluation
is intractable over \emph{subinstance-closed} families of instances of
treewidth \emph{strongly unbounded poly-logarithmically};
\cite{ganian2014lower,amarilli2016tractable} show that MSO query
evaluation is intractable over \emph{subinstance-closed} families of
instances of treewidth that are \emph{densely unbounded poly-logarithmically} (a
weaker notion); \cite{amarilli2016tractable} shows that counting MSO
query results is intractable over \emph{subinstance-closed} families of
instances \emph{of unbounded treewidth that are treewidth-constructible} (an
even weaker notion, simply requiring large-treewidth instances to be
efficiently constructible, see \cite[Definition~4.1]{amarilli2016tractable}); finally,
\cite{amarilli2015provenance,amarilli2016tractable} shows that one can
exhibit FO queries whose probability evaluation is polynomial-time on
structures of bounded treewidth, but \#P-hard on \emph{any}
treewidth-constructible family of instances of unbounded treewidth.

For
this reason, and because of the wide variety of problems that become
tractable on bounded-treewidth instances, treewidth is an especially
important object of study.

\subparagraph*{Treewidth of real-world databases}

If there is hope for practical applicability of treewidth-based
approaches, one needs to answer the following two questions: \emph{Can one
efficiently compute the treewidth of real-world databases?}
and \emph{What is the treewidth of real-world data?} The latter
question is the central problem addressed in this paper.

The answer to the former is that, unfortunately, treewidth cannot
reliably be computed efficiently in practice. Indeed, computing the
treewidth of a graph is an NP-hard problem~\cite{arnborg1987complexity} and, in practice, exact computation of treewidth is possible only for very
small instances, with no more than dozens of vertices~\cite{bodlaender2012exact}.
An additional theoretical result is that, given a width~$w$, it is possible
to check whether a graph has treewidth~$w$ and produce a tree
decomposition of the graph in linear time \cite{bodlaender1996linear}; however, the large constant
terms make this procedure impractical. Known exact treewidth computation
algorithms~\cite{bodlaender2012exact} may be usable on small graphs, but
they are impossible to apply for our purposes. Indeed,
in~\cite{bodlaender2012exact}, the largest graph for which algorithms
finished running had a mere 40 vertices.

A more realistic approach is to compute \emph{estimations} of 
treewidth, i.e., an interval formed of a \emph{lower bound} and an
\emph{upper bound} on
the treewidth.
Upper bound
algorithms (surveyed in \cite{bodlaender2010treewidth}) use multiple
approaches for estimation, which all output a tree decomposition. One
particularly important class of
methods for generating tree decompositions relies on \emph{elimination
orderings},
that also appear
in junction tree algorithms used in belief propagation~\cite{lauritzen1990local}. For
lower bounds (surveyed in \cite{bodlaender2011treewidth}), where
no decomposition can be obtained, one can use degree-based or minor-based
measures on graphs, which themselves act as proxies for treewidth.

Some upper bound and lower bound algorithms have been
implemented and experimented with in \cite{vandijk2006computing,
bodlaender2010treewidth, amir2010approximation, bodlaender2011treewidth}. However, in all cases
these algorithms were evaluated on graphs that are either very small (of
the order of dozens of vertices), as in \cite{vandijk2006computing}, or
on slightly larger synthetic graphs (with up to $1\,000$ vertices)
generated with exact treewidth values in mind, as in
\cite{bodlaender2010treewidth, bodlaender2011treewidth}. The main purpose
of these experiments was to evaluate the estimators' performance.
Recently, the PACE challenge has had a track dedicated to the estimation
of treewidth~\cite{dell2017pace}: exact treewidth on relatively small
graphs, upper bounds on treewidth on larger graphs. Local improvements of
upper bounds have been also evaluated on small graphs
in~\cite{fichte2017sat}. Since all these works aim at comparing estimation
algorithms, they do not investigate the actual treewidth of real-world data.

Another relevant work is \cite{adcock2016tree}, which studied the
\emph{core-periphery structure} of social networks, by building tree
decompositions via node elimination ordering heuristics, but without
establishing any treewidth bounds. In this work, we use the same
heuristics to compute bounds on treewidth.

Finally, there have been some work on analyzing properties of
real-world \emph{queries}. Queries are usually
much smaller than database instances, but it turns out that they are also
much simpler in structure: \cite{DBLP:conf/sigmod/PicalausaV11} shows
that an astounding 99.99\% of conjunctive patterns present in a SPARQL
query log are acyclic, i.e., of treewidth~1.
\cite{DBLP:journals/pvldb/BonifatiMT17}~similarly showed that the
overwhelming majority of graph pattern queries in SPARQL query logs had
treewidth 1, less than 0.003\% had treewidth 2, and a single one (out of
more than 15 million) had treewidth~3. We shall see that the situation is
much different with the treewidth of database instances. Note that, in many settings,
low-treewidth of queries does not suffice for tractability: in
probabilistic databases, for instance, \#P-hardness holds even for
acyclic queries~\cite{DBLP:conf/pods/DalviS07a}.

\subparagraph*{Contributions}
In this experimental study, our contributions are twofold. 

First, using previously studied
algorithms for treewidth estimation, we set out to find classes of
real-world data that may exhibit relatively low values of treewidth,
thus identifying potential cases in which treewidth-based approaches are
of practical interest. For this, after formally defining tree
decompositions and treewidth
(Section~\ref{sec:prelim}), we  select the algorithms that are
able to deal with large-scale data instances, for both lower- and
upper-bound estimations (Section~\ref{sec:algorithms}). Our aim here is
not to propose new algorithms for treewidth estimation, and not to
exhaustively evaluate existing treewidth estimation algorithms, but rather
to identify algorithms that can give acceptable treewidth estimation values
in reasonable time, in order to apply them to real-world data.
Then, we use these algorithms to obtain lower and upper bound
intervals on treewidth for 25 databases from 8 different domains
(Section~\ref{sec:results}). We mostly consider graph data, for which
the notion of treewidth was initially designed (the treewidth of an arbitrary
relational instance is simply defined as that of its Gaifman graph).
The graphs we consider, all obtained from
real-world applications, have between several thousands and several millions
of vertices.
To the best of our knowledge, this is the first comprehensive study of the treewidth of 
real-world data of large scale
from a variety of application domains.

Our finding is that, \emph{generally}, the treewidth is too large to be able to use treewidth-based algorithms directly with any hope of efficiency.

Second, from this finding, we investigate how a \emph{relaxed (or partial) decomposition}
can be used on real-world graphs. In short, we no longer look for complete tree
decompositions; instead, we allow the graph to be only partially decomposed. In
complex networks, there often exists a dense core together with a
tree-like fringe structure \cite{newman2001random}; it is hence possible to
decompose the fringe into a tree, and to place the rest of
the graph in a dense ``root''. It has been shown that this approach can improve the efficiency
of some graph algorithms \cite{wei2010tedi, akiba2012shortest,
maniu2017indexing}. In Section~\ref{sec:partial}, we
analyze its behavior on real-world graphs.
We conclude the paper in Section~\ref{sec:discussion} with a discussion
of lessons learned, as to
which real-world data admit (full or partial) low-treewidth tree
decompositions, and how this impacts query evaluation tasks.

\ifextended\else
Due to lack of space, some details and additional experiments can be
found in an extended version of this article \cite{extended}.
\fi

\section{Preliminaries on Treewidth}\label{sec:prelim}
\begin{figure*}[t]
  \centering
  \begin{subfigure}{0.4\linewidth}
    \includegraphics[width=.8\textwidth]{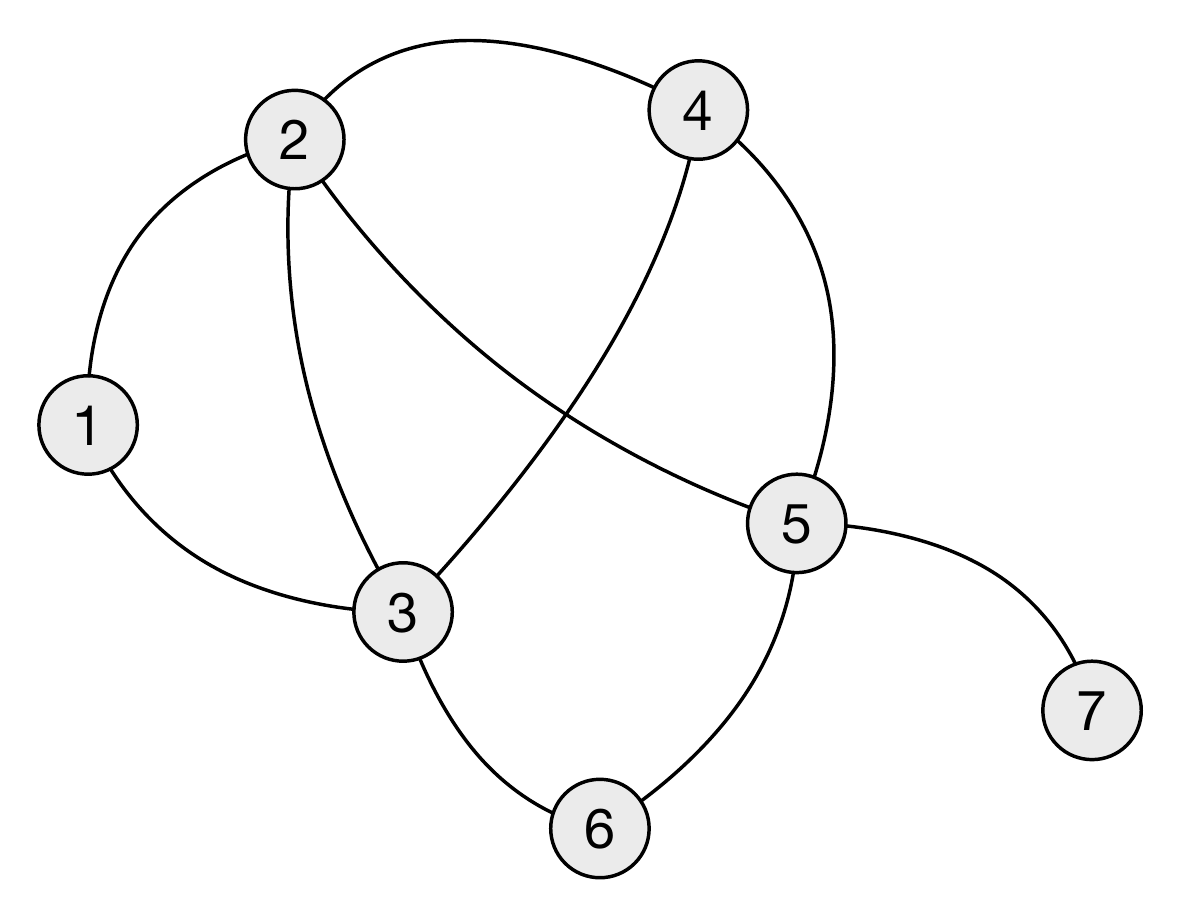}
  \end{subfigure}
  \qquad
  \begin{subfigure}{0.4\linewidth}
    \includegraphics[width=.8\textwidth]{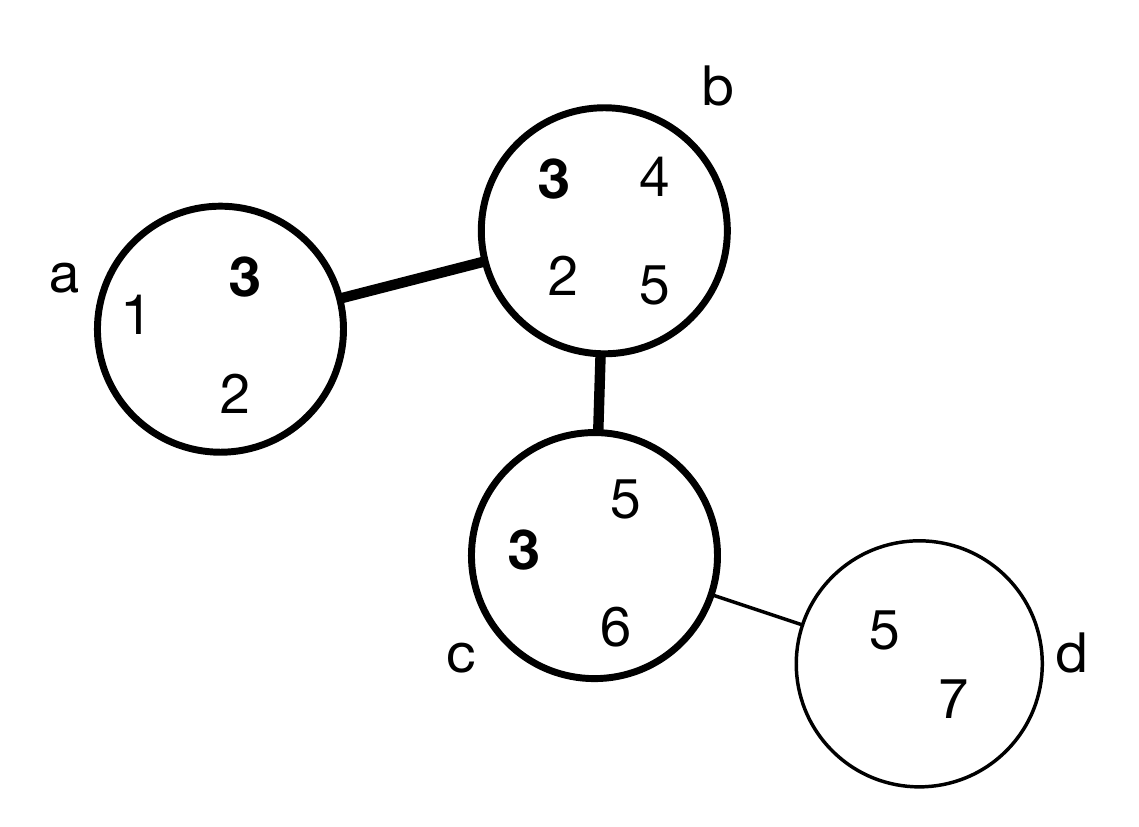}
  \end{subfigure}

  \vspace{-1em}
  
  \caption{Example undirected, unlabeled, graph (left) and decomposition
    of width 3 (right)}\label{fig:decomposition}
  
  \vspace{-1em}
\end{figure*}

To make the concepts in the following clear, we start by formally introducing the concept
of \emph{treewidth}.
Following the original definitions
in~\cite{robertson1984graph}, we first define a tree decomposition:

\begin{definition}[(Tree Decomposition)]\label{def:tree_decomposition}
    Given an \emph{undirected graph} $G=(V,E)$, where $V$ represents the set of
    vertices (or nodes) and $E\subseteq V\times V$ the set of edges, a \emph{tree
    decomposition} is a pair $(T,B)$ where $T=(I,F)$ is a tree and $B:I\to 2^V$
    is a labeling of the nodes of $T$ by subsets of $V$ (called
    \emph{bags}), with the following
    properties:
    \begin{enumerate}
    \item $\bigcup_{i\in I}B(i)=V$;
    \item $\forall(u,v)\in E$, $\exists i\in I$ s.t.\ $\{u,v\}\subseteq B(i)$; and
    \item $\forall v\in V$, $\{i\in I\mid v\in B(i)\}$ induces a subtree of $T$.
    \end{enumerate}
\end{definition}

Intuitively, a tree decomposition groups the vertices of a graph into
bags so that they form a tree-like structure, where a link between bags
is established when there exists common vertices in both bags. 

\begin{example}
Figure~\ref{fig:decomposition} illustrates such a decomposition. The resulting
decomposition is formed of 4 bags, each containing a subset of the nodes in the
  graph. The bags containing node~$3$ (in bold) form a connected subtree
  of the tree decomposition.
\end{example}

Based on the
number of vertices in a bag, we can define the concept of \emph{treewidth}:

\begin{definition}[(Treewidth)]\label{def:treewidth}
  Given a graph $G=(V,E)$ the \emph{width} of a tree decomposition $(T,B)$
  is equal to $\max_{i\in I} (|B(i)|-1)$. The \emph{treewidth} of $G$,
  $w(G)$, is equal to the minimal width of all tree decompositions of~$G$.
\end{definition}

It is easy to see that an isolated point has treewidth 0, a tree
treewidth 1, a cycle treewidth~2, and a $(k+1)$-clique (a complete graph
of $k$ nodes) treewidth $k$.

\begin{example}
  The width of the decomposition in Figure~\ref{fig:decomposition} is~3.
  This tells us the graph has a treewidth of at most~3. The treewidth of
  this graph is actually exactly 3: indeed, the 4-clique, which has treewidth~3, is a \emph{minor} of the graph in
  Figure~\ref{fig:decomposition} (it is obtained by removing nodes $1$ and
  $7$, and by contracting the edges between $3$ and $6$ and $5$ and $6$), and
  treewidth never increases when taking a minor (see, for instance,
  \cite{harvey2014treewidth}).
\end{example}

As previously mentioned, the treewidth of an arbitrary relational
instance is defined as that of its \emph{Gaifman graph}, the graph whose
vertices are constants of the instances and where there is an edge
between two vertices if they co-occur in the same fact. We will therefore
implicitly represent relational database instances by their Gaifman
graphs in what follows.

We are now ready to present algorithms for lower and upper bounds on
treewidth.

\section{Treewidth Estimation}\label{sec:algorithms}
The objective of our experimental evaluation is to obtain 
reasonable estimations of treewidth, using algorithms
with reasonable execution time on real-world graphs.

Once we know we do not have the luxury of an exact computation of the treewidth, we
are left with estimations of the range of possible treewidths, between a
\emph{lower bound} and an \emph{upper bound}. For the purposes of this
experimental survey, we restrict ourselves to the most efficient
estimation algorithms from the literature. We refer the reader to
\cite{bodlaender2010treewidth} and \cite{bodlaender2011treewidth},
respectively, for a more complete survey of treewidth upper and lower
bound estimation algorithms on synthetic data.

\subparagraph*{Treewidth Upper Bounds}

As we have defined, the treewidth is the smallest width among all
possible tree decompositions. In other words, the width of any
decomposition of a graph is an
upper bound of the actual treewidth of that graph. A treewidth upper
bound estimation algorithm can thus be seen as an algorithm to find
a decomposition whose width is as close as possible to the treewidth
of the graph. To understand how one can do that, we need to introduce the
classical concept
of \emph{elimination ordering} and to explain its connection to treewidth.

We start by introducing \emph{triangulations} of graphs, which transform a
graph $G$ into a graph~$G^{\Delta}$ that is \emph{chordal}:

\begin{definition}
  A \emph{chordal} graph is a graph~$G$
such that every cycle in $G$ of at least four vertices has a
\emph{chord} -- an edge between two non-successive vertices in the cycle.

  A \emph{triangulation} (or \emph{chordal completion}) of a graph~$G$ is a minimal chordal supergraph $G^{\Delta}$
  of~$G$: a graph
  obtained from~$G$ by adding a minimal set of edges
to obtain a chordal graph.
\end{definition}

\begin{figure*}
  \centering
  \begin{subfigure}{0.4\linewidth}
    \includegraphics[width=.8\textwidth]{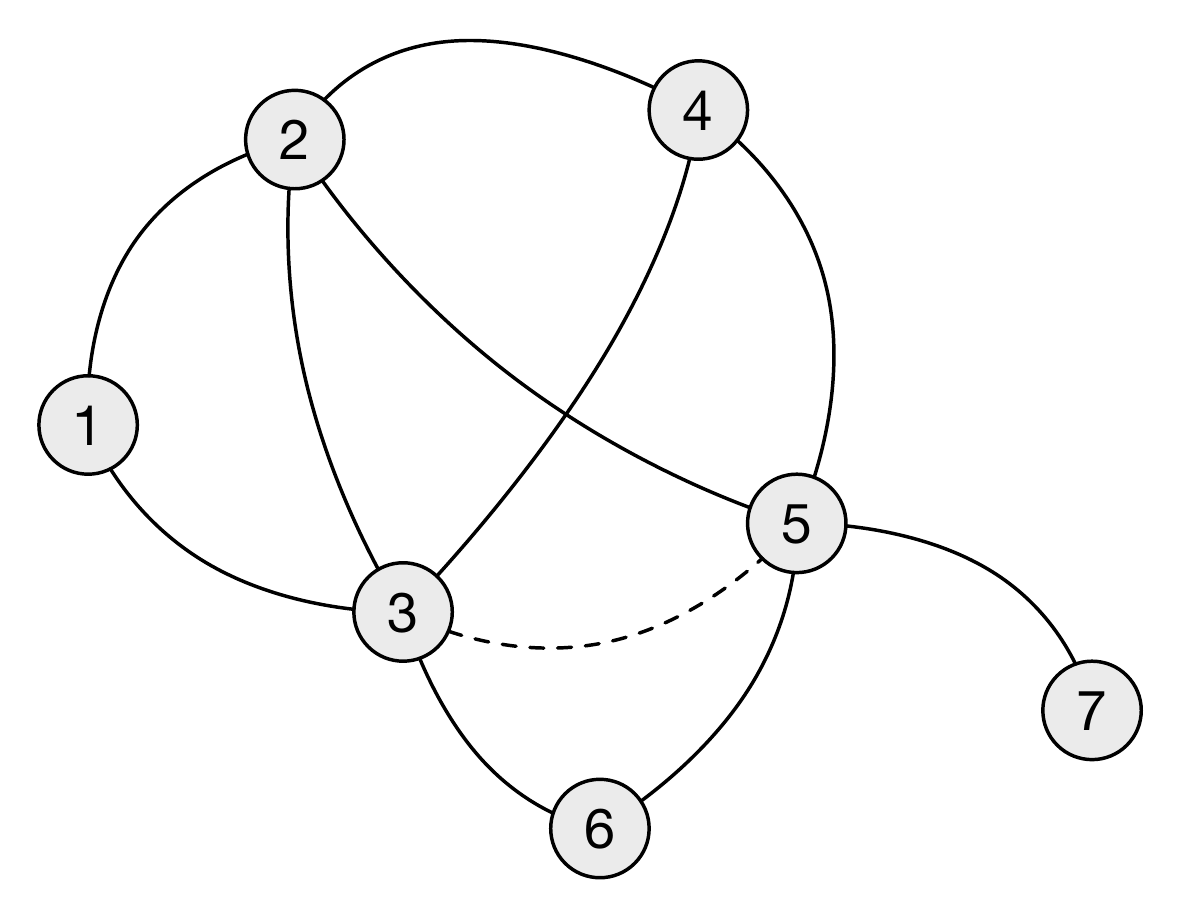}
  \end{subfigure}
  \qquad
  \begin{subfigure}{0.4\linewidth}
    \includegraphics[width=\textwidth]{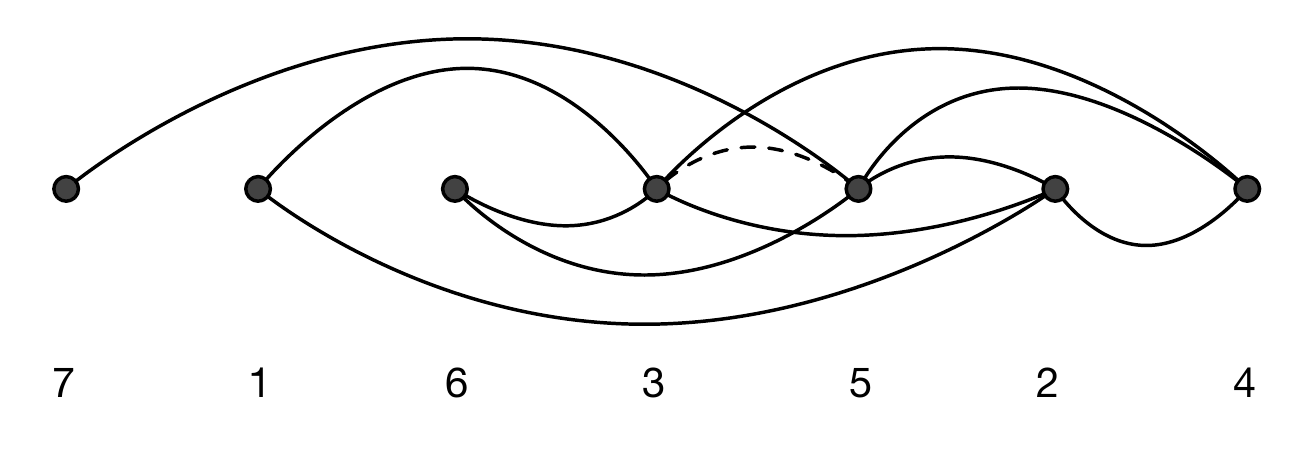}
  \end{subfigure}

  \vspace{-1em}
  \caption{Graph triangulation for the graph of
    Figure~\ref{fig:decomposition} (left) and its elimination ordering
  (right)}\label{fig:ordering}
  \vspace{-1em}
\end{figure*}

\begin{example}
  The graph in Figure~\ref{fig:decomposition} is not chordal, since, for
  example, the
  cycle $3$--$4$--$5$--$6$--$3$ does not have a chord. If one adds an edge between
  $3$ and~$5$, as in Figure~\ref{fig:ordering} (left), one can verify that
  the resulting graph is chordal, and thus a triangulation of the graph
  of Figure~\ref{fig:decomposition}.
\end{example}

One way to obtain triangulations of graphs is \emph{elimination orderings}. An
elimination ordering $\omega$ of a graph $G=(V,E)$ of
$n$ nodes is an ordering of the vertices of $G$, i.e., it can be seen as
a bijection from $V$ onto $\{1,\dots,n\}$.
From this ordering, one obtains a triangulation by applying sequentially
the following \emph{elimination procedure} for each vertex $v$: first, edges are added between 
remaining neighbors of $v$ as needed so that they 
form a clique, then $v$ is \emph{eliminated} (removed) from the graph.
For every elimination ordering~$\omega$, $G$ along with all
edges added to~$G$ in the elimination procedure forms a graph, denoted
$G^{\Delta}_{\omega}$. This graph is chordal (indeed, we know that the
two neighbors of the first node of any cycle we encounter in the
elimination ordering have been connected by a chord by the elimination
procedure). It is also a supergraph of~$G$, and it can be shown it is a
minimal chordal supergraph, i.e., a
triangulation of~$G$.

\begin{example}
  Figure~\ref{fig:ordering} (right) shows a possible elimination ordering
  $(7,1,6,3,5,2,4)$ of the graph
of Figure~\ref{fig:decomposition}. The elimination procedure adds a
  single edge, when processing node~$6$, between nodes $3$ and~$5$. The
  resulting triangulation is the graph on the left of
  Figure~\ref{fig:ordering}.
\end{example}
  
Elimination orderings are connected to treewidth by the following result:
\begin{theorem}~{\upshape\cite{bodlaender2010treewidth}} Let $G=(V,E)$ a graph, and
  $k\leq n$. The following are equivalent:
  \begin{enumerate}
    \item $G$ has treewidth $k$.
    \item $G$ has a triangulation $G^{\Delta}$, such that the maximum clique in
      $G^{\Delta}$ has size $k+1$.
    \item There exists an elimination ordering $\omega$ such that the maximum
      clique size in $G^{\Delta}_{\omega}$ is~$k+1$.
    \end{enumerate}
\end{theorem}

Obtaining the treewidth of the graph is thus equivalent to finding an optimal
elimination ordering. Moreover, constructing a tree decomposition from an
elimination ordering is a natural process: each time a vertex is processed, a
new bag is created containing the vertex and its neighbors. Note that, in practice, we do not need to compute the full elimination ordering: we can simply stop when we know that the number of remaining vertices is lower that the largest clique found thus far. 

\begin{example}
  In the triangulation of Figure~\ref{fig:ordering} (left), corresponding to
  the elimination ordering on the right, the maximum clique has size 4:
  it is induced by the vertices ${2,3,4,5}$. This proves the existence of
  a tree decomposition of width~$3$. Indeed, it is exactly the tree decomposition in
  Figure~\ref{fig:decomposition} (right): bag $\mathsf{d}$ is constructed
  when $7$ is eliminated, bag $\mathsf{a}$ when $1$ is eliminated, bag~$\mathsf{c}$ when $6$ is eliminated, and finally bag $\mathsf{b}$ when $3$
  is eliminated.
\end{example}

Finding a ``good'' upper bound on the treewidth can thus be done by finding a
``good'' elimination ordering. This is still an intractable problem, of
course, but there are various heuristics for generating elimination
orderings leading to good treewidth upper bounds. One important class of such
elimination ordering heuristics are the greedy heuristics. Intuitively, the
elimination ordering is generated in an incremental manner: each time a new node has to
be chosen in the elimination procedure, it is chosen using a criterion based on its neighborhood. In our
study, we have implemented the following greedy criteria (with ties 
broken arbitrarily):
\begin{itemize}
  \item\textsc{Degree}. The node with the minimum degree is chosen.~\cite{markovitz1957elimination, berry2003minimum}
  \item\textsc{FillIn}. The node with the minimum needed ``fill-in''
    (i.e., the minimum number of missing edges for its neighbors to form
    a clique) is chosen.~\cite{bodlaender2010treewidth}
  \item\textsc{Degree+FillIn}. The node with the minimum sum of degree
  and fill-in is chosen.
\end{itemize}

\begin{example}
  The elimination ordering of Figure~\ref{fig:ordering} (right) is an
  example of the use of \textsc{Degree+FillIn}. Indeed,
  $7$ is first chosen, with value 1, then $1$ with value~2, then $6$ with value
  $\text{2\/}+\text{1}=\text{3}$. After that, the order is arbitrary since
  $2$, $3$, $4$, and $5$ form a
  clique (and thus have initial value 3).
\end{example}

Previous studies~\cite{vandijk2006computing, koller2009probabilistic,
bodlaender2010treewidth} have found these greedy criteria give the
closest estimations of the real treewidth. An alternative way of generating an
elimination ordering is based on maximum cardinality
search \cite{koller2009probabilistic, bodlaender2010treewidth}; however, it is
both less precise than the greedy algorithms -- due to its reliance on how the
first node in the ordering is chosen -- and slower to run.

\subparagraph*{Treewidth Lower Bounds}

In contrast to upper bounds, obtaining treewidth lower bounds is not
constructive. In other words, lower bounds do not generate decompositions;
instead, the estimation of a lower bound is made by computing other measures on
a graph, that are a proxy for treewidth. In this study, we implement algorithms
using three approaches: subgraph-based bounds, minor-based bounds, and
bounds obtained by constructing improved graphs.

Given a graph $G=(V,E)$, let $\delta(G)$ be its lowest degree, and 
$\delta_2(G)\geq\delta(G)$
its second lowest degree (i.e., the degree of the second vertex when
ordered by degree). It is known that $\delta_2(G)$ is
itself a lower bound on the treewidth~\cite{koster2005degree}. This, however, is too coarse an
estimation, and we need better bounds. We shall use two degeneracy measures of
the graphs. The first, the \emph{degeneracy} of $G$, $\delta D(G)$, is the
maximum value of $\delta(H)$ over all subgraphs $H$ of~$G$. Similarly, the
$\delta_2$\emph{-degeneracy}, $\delta_2D(G)$, of a graph is the maximum value of
$\delta_2(H)$ over all subgraphs $H$. 

We have the following lemma:
\begin{lemma}~{\upshape\cite{bodlaender2011treewidth}}
  Let $G=(V,E)$ be a graph, and $W\subseteq V$ be a set of vertices. The
  treewidth of the subgraph of $G$ induced by~$W$ is at most the treewidth of $G$. 
  \label{lem:treewidth-subgraph}
\end{lemma}

A corollary of the above lemma is that the values $\delta D(G)$ and
$\delta_2D(G)$ are themselves lower bounds of treewidth:

\begin{corollary}~{\upshape\cite{bodlaender2011treewidth}}
  For every graph $G$, the treewidth of $G$ is at least $\delta_2D(G)\geq
  \delta D(G)$.
\end{corollary}

To compute $\delta D(G)$ and $\delta_2 D(G)$ exactly, the following
natural algorithms can be used \cite{koster2005degree,
bodlaender2011treewidth}: repeatedly remove a vertex of smallest degree
-- or smallest except for some fixed node~$v$, respectively -- from the
graph, and keep the maximum value thus encountered. As
in~\cite{bodlaender2011treewidth}, we refer to these algorithms as
\textsc{Mmd} (Maximum Minimum Degree) and \textsc{Delta2D}, respectively.
Ties are broken arbitrarily.

\begin{example}
  Let us apply the \textsc{Mmd} algorithm to the graph of
  Figure~\ref{fig:decomposition} (left). The algorithm may remove, in order,
  $7$ (degree 1), $1$ (degree 2), $6$ (degree~2), $3$ (degree~2), $5$
  (degree 2), $4$
  (degree 1), $2$ (degree~0). This gives a lower bound of~2 on the
  treewidth, which is not tight as we saw.
  \label{exa:mmd}
\end{example}

An equivalent of Lemma~\ref{lem:treewidth-subgraph}  on treewidth also holds for \emph{minors} of a
graph $G$: if $H$ is a minor of $G$, then the treewidth of $H$ is at most the
treewidth of $G$~\cite{harvey2014treewidth}. A minor $H$ of a graph $G$ is a graph obtained by
allowing, in addition to edge and node deletion as when taking subgraphs,
\emph{edge contractions}.
%-- an operation where an edge is removed and its endpoints merged. 
Then the concepts of \emph{contraction degeneracy}, $\delta
C(G)$, and $\delta_2$\emph{-contraction degeneracy}, $\delta_2C(G)$, are defined
analogously to $\delta D(G)$ and $\delta_2 D(G)$ by considering all minors
instead of all subgraphs:
\begin{lemma}~{\upshape\cite{bodlaender2011treewidth}}
  For every graph $G$, the treewidth of $G$ is at least $\delta_2C(G)\geq
  \delta C(G)$.
\end{lemma}

Unfortunately, computing $\delta C(G)$ or $\delta_2 C(G)$ is NP-hard~\cite{bodlaender2004contraction}; hence,
only heuristics can be used. One such heuristic for $\delta C(G)$ is a simple
change to the \textsc{Mmd} algorithm, called
\textsc{Mmd+} \cite{bodlaender2004contraction, gogate2004complete}: at each
step, instead of removing a vertex, a neighbor is chosen and the corresponding
edge is contracted. Choosing a neighbor node to contract requires some heuristic
also; in line with previous studies, in this study we choose the neighbor node
that has the least overlap in neighbors -- this is called the \emph{least-c}
heuristic~\cite{wolle2005computational}.

Finally, another approach to treewidth lower bounds that we consider are
\emph{improved graphs}, an approach that can be used in combination with any of
the lower bound estimation algorithms presented so far. Consider a graph~$G$ and an integer
$k$ and the following operation: while there are non-adjacent vertices $v$ and
$w$ that have at least $k+1$ common neighbors, add the edge $(v,w)$ to the
improved graph $G'$. The resulting graph $G'$ is the $(k+1)$-neighbor improved
graph of $G$. Using these improved graphs can lead to a lower bound on
treewidth: the $(k+1)$-neighbor improved graph $G'$ of a graph $G$ having at most
treewidth $k$ also has treewidth at most~$k$.

To use this property, one can start from an already computed estimation $k$ of a
lower bound (by using \textsc{Mmd}, \textsc{Mmd+}, or \textsc{Delta2D} for
example) and then repeatedly generate a $(k+1)$-neighbor improved graph,
estimate a new lower bound on treewidth, and repeat the process until the graph
cannot be improved. This algorithm is known as \textsc{Lbn} in the
literature~\cite{clautiaux2003lower}, and can be combined with any other lower
bound estimation algorithm. A refinement of \textsc{Lbn}, that alternates
improvement and contraction steps, \textsc{Lbn+}, has also been
proposed~\cite{bodlaender2004contraction}.

\begin{example}
  Let us illustrate the use of \textsc{Lbn} together with \textsc{Mmd} on
  the graph of Figure~\ref{fig:decomposition} (left). As shown in
  Example~\ref{exa:mmd}, a first run of \textsc{Mmd} yields $k=\text{2}$. We
  compute a 3-neighbor improved graph for $G$ by adding an edge between
  nodes $3$ and $5$ (that share neighbors $2$, $4$,~$6$).
  Now, running \textsc{Mmd} one more time yields the possible sequence
  $7$
  (degree~1), $1$ (degree~2), $6$ (degree~2), $3$ (degree~3), $5$
  (degree~2), $4$
  (degree~1), $2$ (degree~0). We thus obtain a lower bound of $3$ on the
  treewidth, which is this time tight.
\end{example}

\section{Estimation Results}\label{sec:results}
We now present our main experimental study, first introducing the 25
datasets we are considering, then upper and lower bound results, running
time and estimators, and an aside discussion of the treewidth of
synthetic networks. All experiments were made on a server using an 8-core
Intel Xeon 1.70GHz CPU, having 32GB of RAM, and using 64bit Debian Linux.
All datasets were given at least two weeks to finish, after which the algorithms were stopped and the best lower and upper bounds were recorded.

\subparagraph*{Datasets} For our study, we have evaluated the treewidth estimation
algorithms on 25 datasets from 8 different domains (see
Appendix~A
\ifextended\else of \cite{extended} \fi
for descriptions of how they were obtained):
infrastructure networks (road networks, public transportation, power
grid), social networks (explicit as in social networking sites, or
derived from interaction patterns), web-like networks, a communication
network, data with a hierarchical structure (genealogy trees), knowledge
bases, traditional OLTP data, as well as a biological interaction
network.

\begin{table*}[h]\small
  \setlength{\tabcolsep}{6pt}
  \centering
    \vspace{-1em}
  \caption{Datasets and summary of lower and upper
    bounds. Bounds with a~${}^*$ are partial (the decomposition process was interrupted before it finished).\label{tab:summary}}
    \vspace{-1em}
    \begin{tabular}{rrn{9}{0}n{9}{0}n{9}{0}n{9}{0}}
    \toprule
      \multicolumn{4}{c}{{\bfseries Dataset}}& \multicolumn{1}{c}{\bfseries Lower}& \multicolumn{1}{c}{\bfseries Upper}\\ 
    {\bfseries type} & {\bfseries name} & \multicolumn{1}{r}{\bfseries nodes} & \multicolumn{1}{r}{\bfseries edges} & \multicolumn{1}{c}{\bfseries width}& \multicolumn{1}{c}{\bfseries width}\\
    \midrule
    \bfseries infrastructure & \textsc{Ca} & 1965206 & 2766607 & 5 & 252\\
     & \textsc{Pa}  & 1088092 & 1541898 & 5 & 333 \\
     & \textsc{Tx} & 1379917 & 1921660 & 5 & 189 \\ 
     & \textsc{Bucharest}  & 189732 & 223143 & 21 &  139 \\
     & \textsc{HongKong}  & 321210 & 409038 & 32 & 145 \\
     & \textsc{Paris}   & 4325486 & 5395531 & 55 & 521 \\
     & \textsc{London}   & 2099114 & 2588544 & 57 & 507 \\
     & \textsc{Stif} & 17720 & 31799 & 28 & 86 \\
     & \textsc{USPowerGrid} & 4941 & 6594 & 10 & 18 \\
    \midrule
    \bfseries social & \textsc{Facebook} & 4039 & 88234 & 142 & 247 \\
     & \textsc{Enron} & 36692 & 183831 & 257 & 1989 \\
     & \textsc{WikiTalk} & 2394385 & 4659565 & 1113 & 12843 \\
     & \textsc{CitHeph} & 34546 & 420877 & 469 & 9498 \\
     & \textsc{Stack-TCS} & 25232 & 69026 & 143 & 717 \\
     & \textsc{Stack-Math} & 1132468 & 2853815 & 850 & 11100 \\
     & \textsc{LiveJournal} & 3997962 & 34681189 & 360 & 919532$^*$\\
    \midrule
    \bfseries web & \textsc{Wikipedia} & 252335 & 2427434 & 1007 & 19876 \\
     & \textsc{Google} & 875713 & 4322051 & 621 & 17571 \\
    \midrule
    \bfseries communication & \textsc{Gnutella} & 65586 & 147892 & 244 & 9374 \\
    \midrule
    \bfseries hierarchy & \textsc{Royal} & 3007 & 4862 & 11 & 24 \\
    & \textsc{Math}  & 101898 & 105131 & 56 & 515 \\
    \midrule
    \bfseries ontology & \textsc{Yago} & 2635315 & 5216293 & 836 & 79059$^*$ \\
    & \textsc{DbPedia} & 7697211 & 30622392 & 28 & 538805$^*$ \\
    \midrule
    \bfseries database & \textsc{Tpch} & 1381291 & 79352127 & 699 & 124316$^*$ \\
    \midrule
    \bfseries biology & \textsc{Yeast} & 2284 & 6646 & 54 & 255 \\ 
    \bottomrule
  \end{tabular}
\end{table*}

\begin{figure*}
  \centering
  \begin{subfigure}{0.8\linewidth}
    \includegraphics[width=\textwidth]{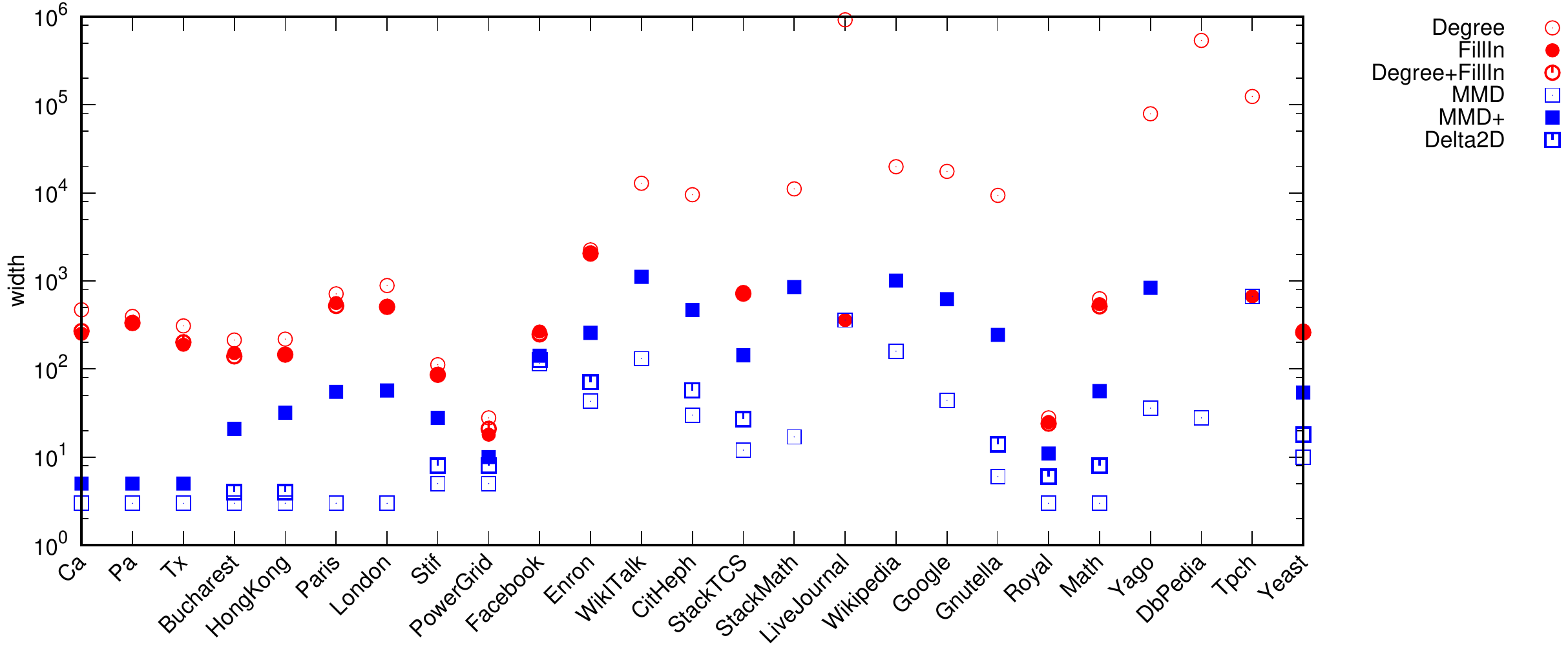}
    \caption{absolute values}
  \end{subfigure}

  \begin{subfigure}{0.8\linewidth}
    \includegraphics[width=\textwidth]{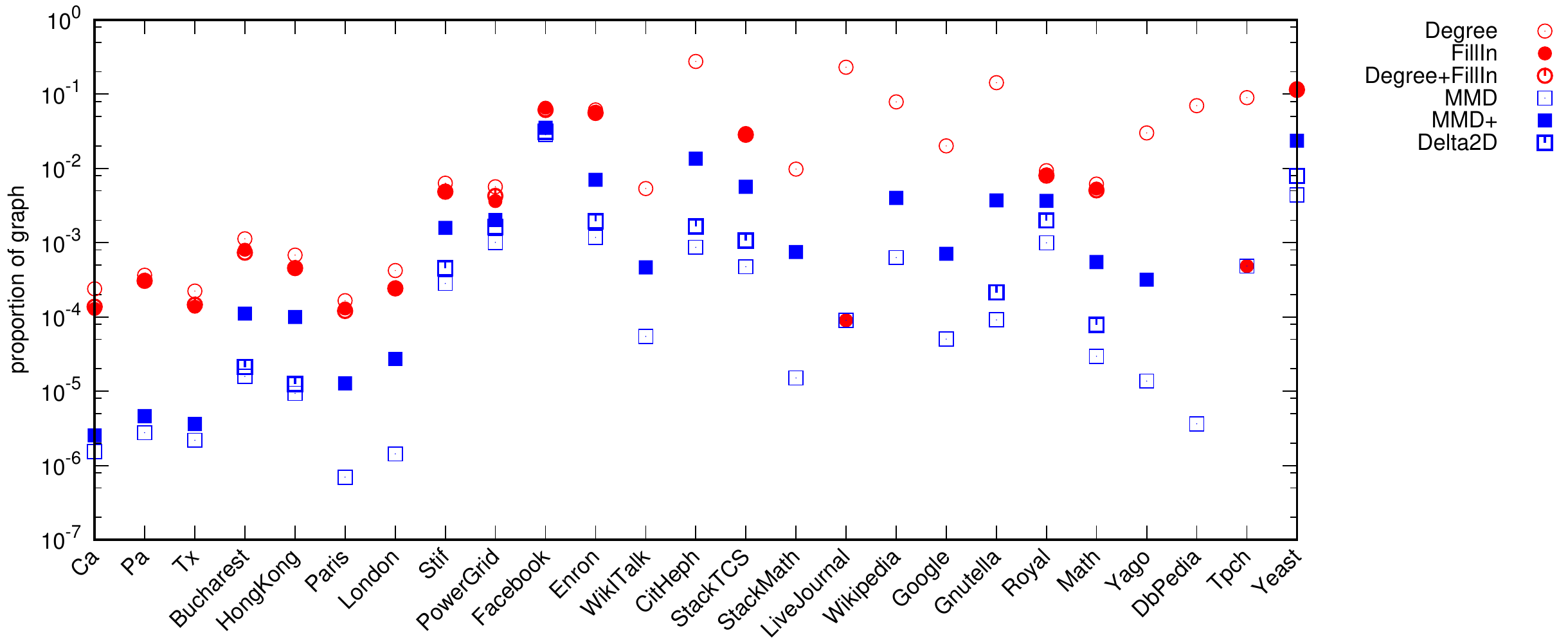}
    \caption{relative values}
  \end{subfigure}
  
  \vspace{-1em}
  \caption{{Treewidth estimation} of different algorithms
    (logarithmic scale)\label{fig:bounds}}
  \vspace{-1em}
\end{figure*}

\begin{toappendix}
  \section{Datasets}\label{app:datasets}
We describe below the source and pre-processing of the datasets used in our experimental study:

\begin{description}
  \item[Infrastructure.] For this domain, we have collected
    four types of datasets. The first are road networks of three US states,
    \textsc{Ca}, \textsc{Pa}, and \textsc{Tx}, downloaded from the site of the
    9th DIMACS Challenge on shortest
    paths\footnote{\url{http://www.dis.uniroma1.it/challenge9/}}. Second, we
    have extracted, from the XML export of
    OpenStreetMap\footnote{\url{https://www.openstreetmap.org/}}, the city maps
    of \textsc{Bucharest}, \textsc{HongKong}, \textsc{Paris}, and
    \textsc{London}.  The OpenStreetMap format consists of \emph{ways} which
    represent routes between the nodes in the graph. We have extracted all the
    ways appearing in each map, and eliminated the nodes which appeared in more
    than one way.  For evaluating public transport graphs, we used the
    \textsc{Stif} dataset from the open public transit data of the Paris
    metropolitan area\footnote{\url{https://opendata.stif.info/page/home/}},
    where the nodes are the stations of the system (bus, train, metro) and an
    edge appears if at least one transport line exists between the two stations. Finally, we used the Western US power grid network, \textsc{USPowerGrid}, first used in~\cite{watts1998collective}, in which the edges are the major power lines and the nodes represent electrical equipment (transformers, generators, etc.).
  \item[Social Networks.] The networks used in this domain are mainly taken
    from the Stanford SNAP
    repository\footnote{\url{https://snap.stanford.edu/}}; this is the case for
    the \textsc{Facebook} (ego networks), \textsc{Enron} (Enron email
    conversations), \textsc{WikiTalk} (convesations between Wikipedia
    contributors), \textsc{CitHeph} (citations between authors in high-energy
    physics), and \textsc{LiveJournal} (social network of the Livejournal
    site). Other social datasets were extracted from the StackOverflow Q\&A
    site\footnote{\url{http://stackoverflow.com/}}, for the mathematics
    sub-domain
    (\textsc{Stack-Math}) and the theoretical computer science sub-domain
    (\textsc{Stack-TCS}). The nodes represent users of the site, and an edge
    exists when a reply, vote, or comment occurs between the edge endpoints.
  \item[Web Networks.] For Web-like graphs, we evaluated treewidth on the \textsc{Wikipedia}
    network of articles, and the \textsc{Google} Web graph (provided by Google
    during its 2002 programming contest); the versions we used were downloaded from the
    Stanford SNAP website.
  \item[Hierarchical.] The next category is of data that
    has by nature a hierarchical structure.
    \textsc{Royal} is extracted from a genealogy of royal
    families in Europe, originally published in GEDCOM format by Brian Tompsett and that
    has informally circulated on the Web since
    1992\footnote{\url{https://www.hull.ac.uk/php/cssbct/genealogy/royal/nogedcom.html}};
    edges are created between spouses and between a parent and its child.
    \textsc{Math} is a graph of academic advisor--advisee data in mathematics and
    related areas, crawled from \url{http://www.genealogy.ams.org/};
    edges are created between an advisor and each of their students.
   \item[Ontologies.] We used subsets of two of the most popular knowledge bases
    available on the Web:
    \textsc{Yago}\footnote{\url{https://www.yago-knowledge.org/}} and
    \textsc{DbPedia}\footnote{\url{https://www.dbpedia.org}}. For \textsc{Yago},
    we downloaded its core facts subset, and removed the semantics on the
    links; hence, an edge exists if there exists at least a fact between two
    concepts (nodes). For \textsc{DbPedia}, we downloaded the ontology subset
    containing its RDF type statements, and we generated the edges using the
    same approach as for \textsc{Yago}. 
  \item[Others.] In addition to the above, we have evaluated
    treewidth on other types of networks: a communication network
    (\textsc{Gnutella}) \cite{ripeanu2002mapping}, a protein--protein interaction network
    (\textsc{Yeast})~\cite{bu2003topological}, and the Gaifman graph of the TPCH relational database benchmark (\textsc{Tpch})\footnote{\url{http://www.tpc.org/tpch/}}, generated using official tools and default parameters.
\end{description}
\medskip

\end{toappendix}

Table~\ref{tab:summary} summarizes the datasets, their size, and the best
treewidth estimations we have been able to compute. For reproducibility
purposes, all datasets,
along with the code that has been used to compute the treewidth
estimations, can
be freely downloaded from \url{https://github.com/smaniu/treewidth/}.

\subparagraph*{Upper Bounds}

We show in Figure~\ref{fig:bounds} the results of our estimation algorithms.
Lower values mean better treewidth estimations.  Focusing on the upper bounds
only (red circular points), we notice that, in general, \textsc{FillIn} does give the
smallest upper bound of treewidth, in line with previous
findings~\cite{bodlaender2011treewidth}.
Interestingly, the \textsc{Degree} heuristic is quite competitive with the other
heuristics. This fact, coupled with its lower running time, means that it can be
used more reliably in large graphs. Indeed, as can be seen in the figure, on
some large graphs only the \textsc{Degree} heuristic actually finished at all;
this means that, as a general rule, \textsc{Degree} seems the best fit for a
quick and relatively reliable estimation of treewidth.

We plot both the absolute values of the estimations
in Figure~\ref{fig:bounds}a, but also their relative values
(in Figure~\ref{fig:bounds}b, representing the ratio of the estimation over
the number of nodes in the graph), to allow for an easier comparison between networks. The
absolute value, while interesting, does not yield an intuition on how
the bounds can differ between network types. If we look at the relative values
of treewidth, it becomes clear that infrastructure networks have a treewidth that is much
lower than other networks; in general they seem to be consistently under one
thousandth of the original size of the graph. This suggests that, indeed,
this type of network may have properties that make them have a lower treewidth. For
the other types of networks, the estimations can vary considerably: they can go
from one hundredth (e.g., \textsc{Math}) to one tenth (e.g., \textsc{WikiTalk})
of the size of the graph.

As further explained in Appendix~B%
\ifextended\else\ of \cite{extended}\fi,
the bounds obtained here on 
infrastructure networks are consistent with a conjectured $\mathcal{O}(\sqrt[3]{n})$ 
bound on the treewidth of road
networks~\cite{dibbelt2016customizable}. One relevant property is
their \emph{low highway dimension}~\cite{abraham2010highway}, which helps with
routing queries and decomposition into contraction hierarchies. Even more
relevant to our results is the fact that they tend to be ``almost planar''. More
specifically, they are $k$-planar: each edge can allow up to $k$ crossing in
their plane embedding. It has been shown in~\cite{eppstein2015structure} that
$k$-planar graphs have treewidth $\mathcal{O}(\sqrt{(k+1)n})$, a relatively low
treewidth that is consistent with our results.

\begin{toappendix}
  \section{Predicting Treewidth of Transport Networks}\label{sec:app_regression}

The difference observed in Section~\ref{sec:results} between road networks and 
other 
networks raises the following
question: since road networks are so well-behaved, can we predict the
relation between their original size and the treewidth upper bound? To test
this, we need to predict a treewidth value $t$ as a function of the nodes of
the original graph, $n$:
\begin{align*}
  t &\simeq \alpha~n^{\beta},
\end{align*}
or, by taking the logarithm, solve the following linear regression problem:
\begin{align*}
  \log t &\simeq \beta~\log n + \log \alpha.
\end{align*}

\begin{table}[hb]
  \setlength{\tabcolsep}{6pt}
  \centering
  \caption{Regression of treewidth as function of number of 
  nodes\label{tab:estimation}}
  \begin{tabular}{crrrr}
    \toprule
    \bfseries{type} & \multicolumn{1}{c}{$\log \alpha$} &
    \multicolumn{1}{c}{$\beta$} & \multicolumn{1}{c}{$R^2$} &
    \multicolumn{1}{c}{$p$-val}\\
    \midrule
    \bfseries road & 1.1874 & 0.3180 & 0.7867 & 0.003 \\
    \bfseries social & 0.6853 & 0.5607 & 0.6976 & 0.038 \\ 
    \bottomrule
  \end{tabular}
\end{table}

\begin{figure}
  \centering
  \begin{subfigure}{0.48\linewidth}
    \includegraphics[width=\textwidth]{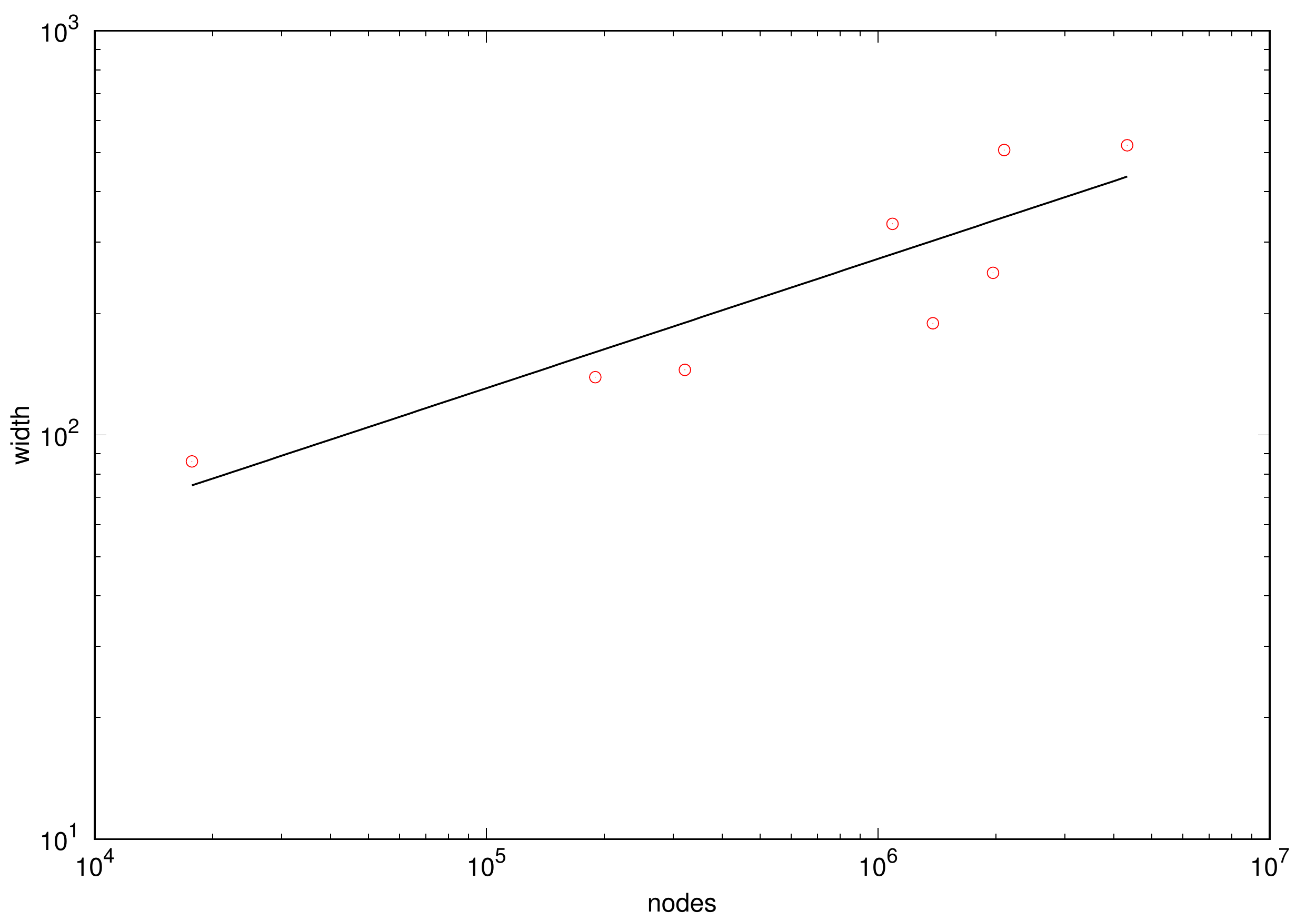}
    \caption{road/transport networks}
  \end{subfigure}
  \begin{subfigure}{0.48\linewidth}
    \includegraphics[width=\textwidth]{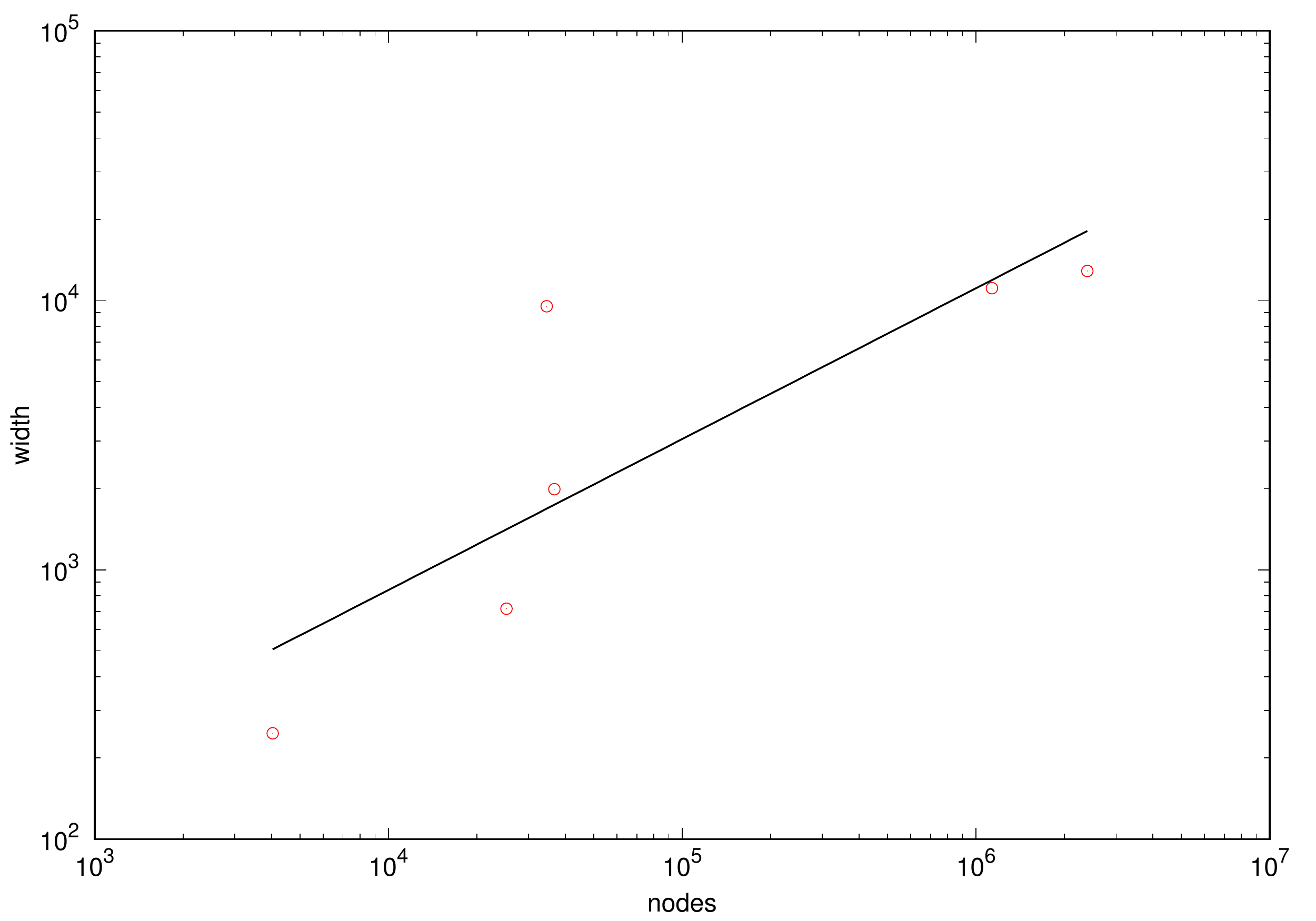}
    \caption{social networks}
  \end{subfigure}
  
  \caption{{Best treewidth estimations} and line of best 
  fit.\label{fig:fit}}
\end{figure}

We train this model on the road and social networks, and report in
Table~\ref{tab:estimation} the results in terms of the coefficients $\alpha$,
$\beta$, the goodness-of-fit $R^2$, and the resulting $p$-value. The results
give an indication that it is indeed easier to predict the treewidth of road
networks, as opposed to that of other networks. A further visual indication on
the better fit for road networks is given in Fig.~\ref{fig:fit}. Hence a rough
estimation of road network treewidth is given by the formula:
\begin{align*}
  t_{\text{road}} \simeq 3.28~n^{0.32} 
\end{align*}

The above result is consistent with previous findings that road networks 
perform well when
tree-like decompositions are used for query answering~\cite{maniu2017indexing},
and a conjectured $\mathcal{O}(\sqrt[3]{n})$ bound on the treewidth of road
networks~\cite{dibbelt2016customizable}.
\end{toappendix}

The treewidth of hierarchical networks is surprisingly high, but not
for trivial reasons: in both \textsc{Royal} and \textsc{Math}, 
largest cliques have size 3. More complex structures (cousin marriages,
scientific communities) impact the treewidth, along with the fact that
treewidth cannot exploit the property that both networks are actually
DAGs.

In upper bound algorithms, ties are broken arbitrarily which causes a
non-deterministic behavior. We show in Appendix~C
\ifextended\else of \cite{extended} \fi
that 
variations due to this non-determinism is of little
significance.

\begin{toappendix}
\section{Variations in Treewidth Upper Bounds}\label{sec:app_ties}

We plot, in Figure~\ref{fig:ties}, the variation in treewidth estimations
due to the way ties are broken in upper bound algorithms. To test this,
we have replaced the original tie breaking condition (based on node id)
with a random tie break, and we ran the upper bound algorithm
\textsc{Degree} 100 times. The results show that, even when ties are
broken randomly, the treewidth generally stays in a range of 10 between
the minimal and maximal values. The only exception is \textsc{Bucharest}
which registers a range of 100, but for a graph containing more than
100\,000 nodes.

\begin{figure}[h]
  \centering
  
  \includegraphics[width=0.9\textwidth]{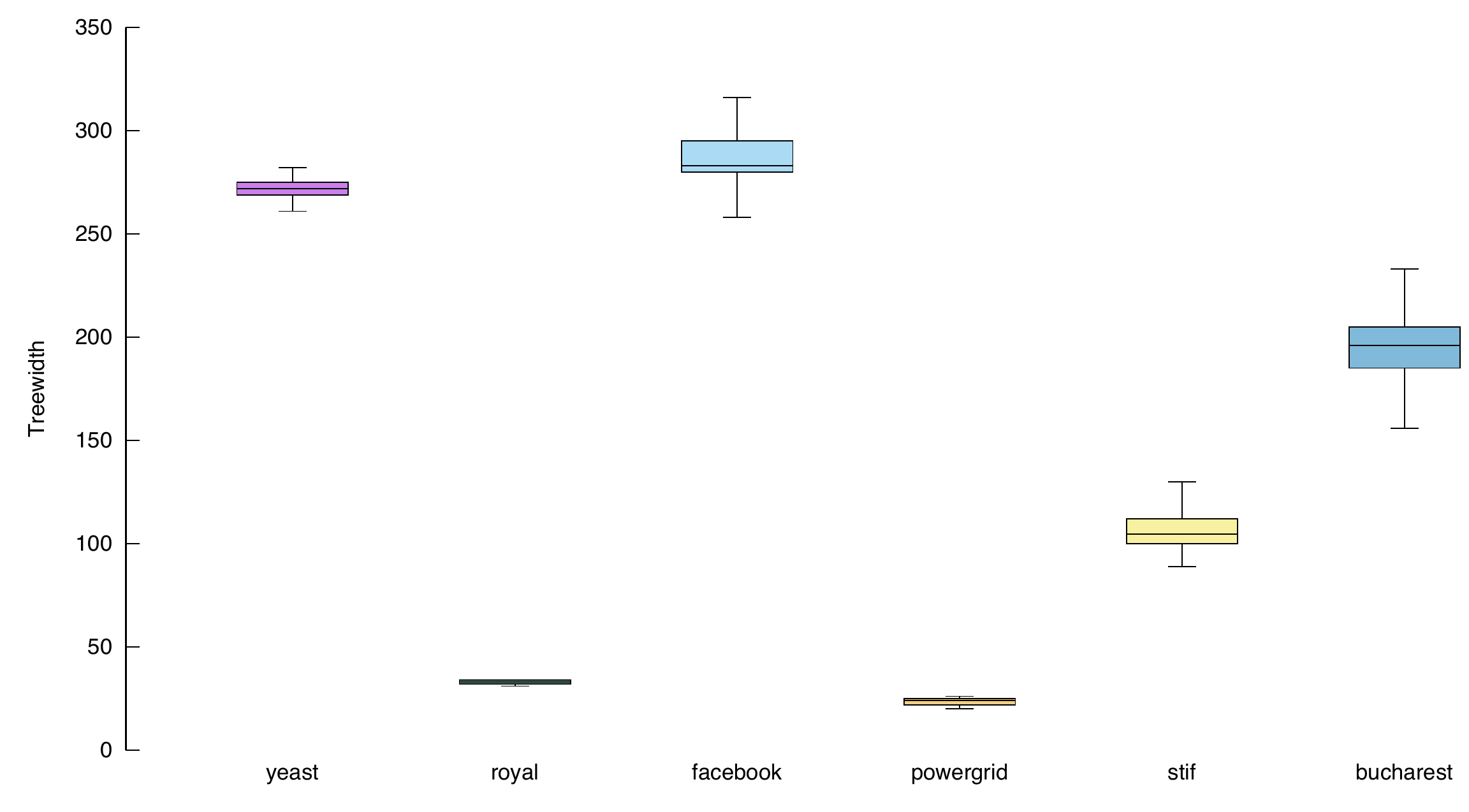}

  \caption{{Variation in treewidth.} Sizes of treewidth due to ties being broken randomly.\label{fig:ties}}

\end{figure}
\end{toappendix}

\subparagraph*{Lower Bounds}

Figure~\ref{fig:bounds} also reports on the lower bound estimations (blue
rectangular points). Now, higher values represent a better estimation. The same
differentiation between infrastructure networks and the other networks holds in the case of
lower bounds -- treewidth lower bounds are much lower in comparison with other
networks. We observe that the variation between upper and lower bound
estimations can be quite large.  Generally, we find that degree-based
estimators, \textsc{Mmd} and \textsc{Delta2D}, give bounds that are very weak.
The contraction-based estimator, \textsc{Mmd+}, however, consistently seems to give
the best bounds of treewidth, and returns lower bounds that are much larger than
the degree-based estimations.

Interestingly, in the case of the networks \textsc{Ca}, \textsc{Pa}, and
\textsc{Tx}, the values returned for \textsc{Mmd+} and \textsc{Mmd} are always 5
and 3, respectively. This has been remarked on
in~\cite{bodlaender2004contraction} -- there exist instances where heuristics of
lower bounds perform poorly, giving very low lower bounds. In our case, this only
occurs for some road networks, all coming from the same data source, i.e., the
DIMACS challenge on shortest paths. 

In Figure~\ref{fig:bounds}, we have not plotted the estimations resulting from
\textsc{Lbn} and \textsc{Lbn+}. The reason is that we have found that these
algorithms do not generally give any improvement in the estimation of the lower
bounds.  Specifically, we have found an improvement only in two datasets for the
\textsc{Delta2D} heuristic: for \textsc{Facebook} from 126 originally to 157
for \textsc{Lbn(Delta2D)} and 159 for \textsc{Lbn+(Delta2D)}; and for
\textsc{Stack-TCS} from 27 originally to 30 for \textsc{Lbn+(Delta2D)}. In all
cases, however, \textsc{Mmd+} is always competitive by itself.

\subparagraph*{Running Time}

\begin{toappendix}
\section{Computational Analysis of Estimation Algorithms}\label{sec:complexity}

An important aspect of estimations of treewidth are how much computational power
they use, as a function of the size of the original graph.

In the case of \emph{upper bounds}, the cost depends quadratically on the upper
bound $w$ found by the decomposition. This is due to the fact that, at each
step, when a node is chosen, a fill-in of $\mathcal{O}(w^2)$ edges is computed
between the neighbors; the bound is also tight, as there can be a case in which the neighbors of the node are not previously connected. Depending on the greedy algorithm used and the criteria
for generating the ordering, the complexity of updating the queue for extracting
the new node is $\mathcal{O}(w\log w)$ for \textsc{Degree}, and $\mathcal{O}(w^2
\log w)$ for the others (the neighbors of neighbor have to be taken into account
as possible updates). Hence, in the worst case, the complexities
of the greedy heuristics are the following: for \textsc{Degree}
$\mathcal{O}(|V|^2\log |V|)$ and for \textsc{FillIn}, \textsc{DegreeFillIn} the cost
increases to $\mathcal{O}(|V|^3\log |V|)$. 

For \emph{lower bounds}, the cost greatly depends on the algorithm chosen. In
the case of \textsc{Mmd} and \textsc{Mmd+}, the costliest operation is to sort
and update a queue of degrees, and the whole computation can be done in
$\mathcal{O}(|E|+|V|\log |V|)$. For \textsc{Delta2D} the cost is known to be
$\mathcal{O}(|V|\times|E|)$~\cite{koster2005degree}. \textsc{Lbn} and \textsc{Lbn+} add a
$\mathcal{O}(|V|)$ factor.

Surprisingly, the best estimations do not
necessarily come from the costliest algorithms. Indeed, in our
experiments, the simplest algorithms tend to also give the best estimations.
%\clearpage

\begin{figure}[t]
  \centering
  \begin{subfigure}{.99\linewidth}
    \includegraphics[width=\textwidth]{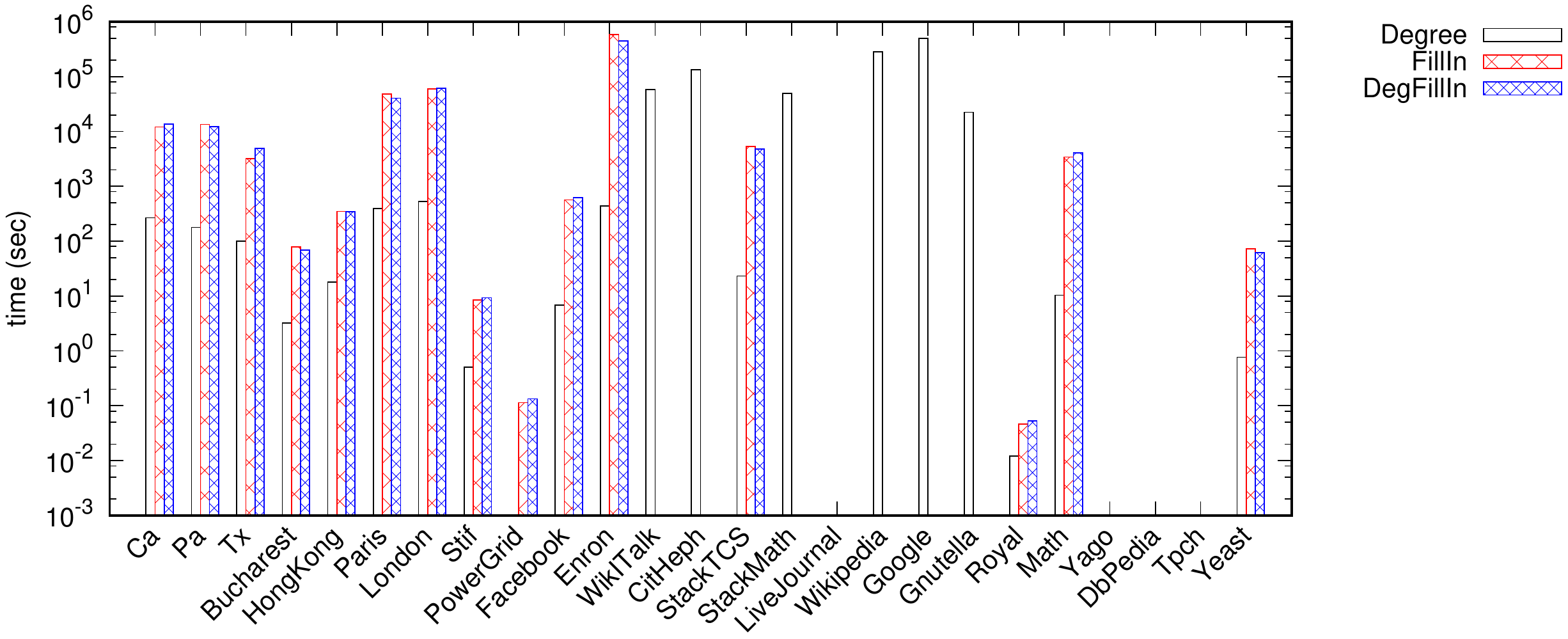}
    \caption{upper bounds}
  \end{subfigure}
  
  \begin{subfigure}{.99\linewidth}
    \includegraphics[width=\textwidth]{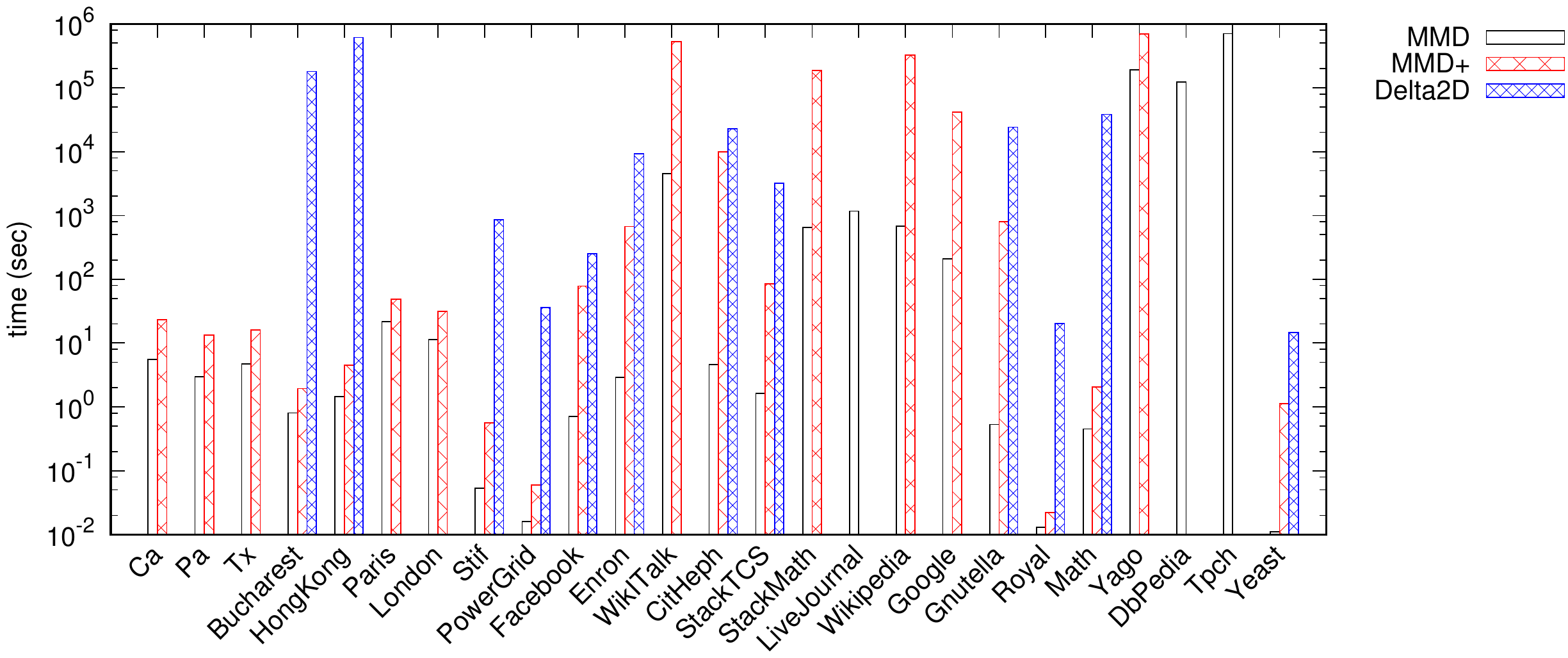}
    \caption{lower bounds}
  \end{subfigure}
  
  \begin{subfigure}{.99\linewidth}
    \includegraphics[width=\textwidth]{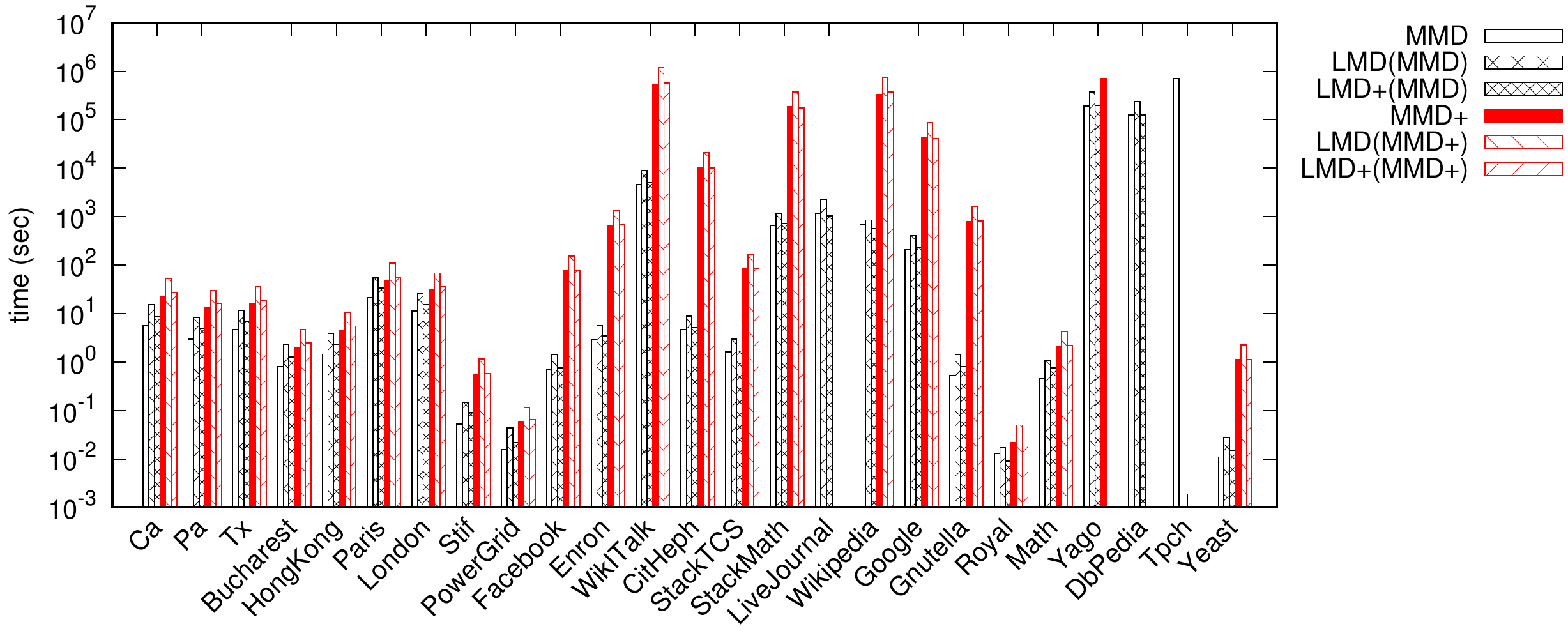}
    \caption{improved lower bounds}
  \end{subfigure}
  
  \vspace{-1em}
  \caption{{Running time} of algorithms (logarithmic
    scale)\label{fig:time}}
  \vspace{-1em}
\end{figure}

We detail the full running time results in Figure~\ref{fig:time}, on a logarithmic scale. For the upper bounds (Figure~\ref{fig:time}a), there is a clear distinction
between \textsc{Degree} on the one hand and \textsc{FillIn} or
\textsc{Degree+FillIn} on the other:
\textsc{Degree} is always at least one order of magnitude faster than the
others. Moreover, for larger networks, only \textsc{Degree} has a
chance to finish. This also occurs for lower bounds (Figure~\ref{fig:time}b):
\textsc{Mmd} and \textsc{Mmd+} have reasonable running times, while
\textsc{Delta2D} has a running time that is much larger -- by several orders of
magnitude. On the other hand, the \textsc{Lbn} and \textsc{Lbn+} methods
(Figure~\ref{fig:time}c) do not add much to the running time of \textsc{Mmd}
and \textsc{Mmd+}; we conjecture that this occurs due to the fact that not many
iterations are performed, which also explains the fact that they generally do
not give improvements in the estimated bounds.

\end{toappendix}

The complexity of different
estimation algorithms
(see Appendix~D%
\ifextended\else\ of \cite{extended}\fi), 
varies a lot, ranging from quasi-linear for low-treewidth to
cubic time. Even if all of the algorithms exhibit polynomial complexity, the
cost can become quickly prohibitive, even for graphs of relatively small sizes; indeed, not all algorithms finished on all datasets
within the time bound of two weeks of computation time: indeed,
only the fastest algorithms for each bound -- \textsc{Degree} and \textsc{Mmd}
respectively -- finish processing all datasets. In some cases, for upper
bounds, even \textsc{Degree} timed out -- in this case, we took the value found
at the moment of the time-out; this still represents an upper bound. The
datasets for which this occurred are \textsc{LiveJournal}, \textsc{Yago},
\textsc{DbPedia}, and \textsc{Tpch}.\footnote{We show in
Figure~9 of Appendix~D
\ifextended\else of \cite{extended} \fi
the full running
time results for the upper and lower bound
algorithms.}

\subparagraph*{Phase Transitions in Synthetic Networks}

We have seen that graphs other than the infrastructure networks have treewidth
values that are relatively high. An interesting question that these results lead
to is: \emph{is it the case for all relatively complex networks, or is it a
particularity of the chosen datasets?}

To give a possible answer to this, we have evaluated the treewidth of a range
of synthetic graph models and their parameters:

\begin{description}
  \item[Random.] This synthetic network
    model due to Erdős and Rényi~\cite{erdos1959random} is an entirely random one: given $n$ nodes and a parameter
    $p\in(0,1]$, each possible edge between the $n$ nodes is added with
    probability $p$. We have generated several network of $n=10\,000$ nodes,
    and with values of $p$ ranging from $10^{-5}$ to~$10^{-3}$.
  \item[Preferential Attachment.]
    This network model~\cite{albert2002statistical} is a generative model, and aims to simulate link
    formation in social networks: each time a new node is added to the network,
    it attaches itself to $m$ other nodes, each link being formed with a
    probability proportional to the number of neighbors the existing node
    already has. We have generated graphs of $n=10\,000$ nodes, with the
    parameter $m$ varying between $1$ and~$20$.
  \item[Small-World.] The model~\cite{watts1998collective}
    generates a small-world graph by the following process: first, the $n$ nodes
    are organized in a ring graph where each node is connected with $m$ other
    nodes on each side; finally, each edge is rewired (its endpoints are
    changed) with probability $p$. In the experiments, we have generated graphs
    of $n=10\,000$, with $p=\{0.1,0.2,0.5\}$ and $m$ ranging between 1 and~$20$.
\end{description}
\medskip

Our objective with evaluating treewidth on these synthetic instances is to evaluate
whether some parameters of these models lead to lower treewidth values, and if
so, in which model the ``phase transition'' occurs, i.e., when a
low-treewidth regime gives way to a high-treewidth one. For our purpose, and for
reasons we explain in Section~\ref{sec:partial}, we consider a treewidth under
$\sqrt{n}$ to be low.

\begin{figure*}[ht]
  \centering
  \begin{subfigure}{0.32\linewidth}
    \includegraphics[width=\textwidth]{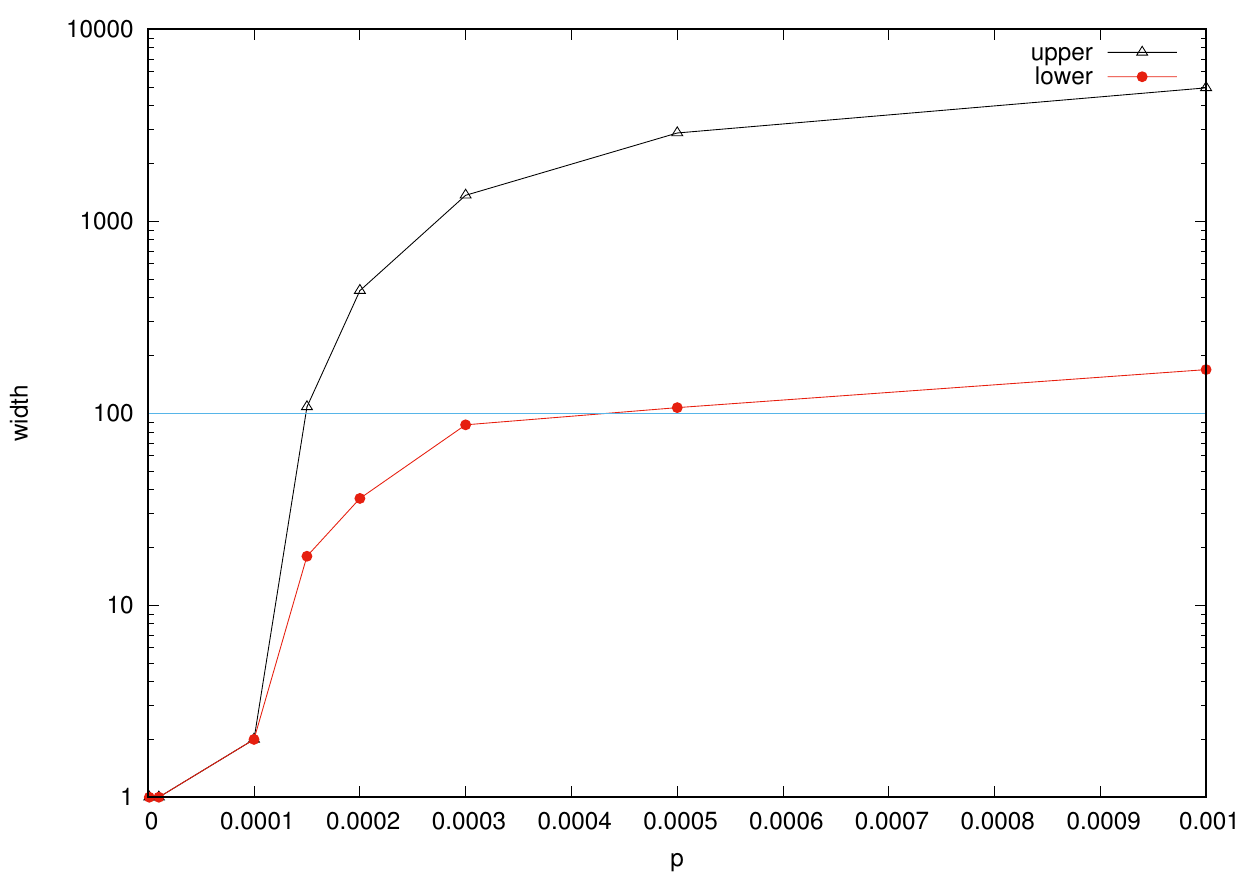}
    \caption{random}
  \end{subfigure}
  \begin{subfigure}{0.32\linewidth}
    \includegraphics[width=\textwidth]{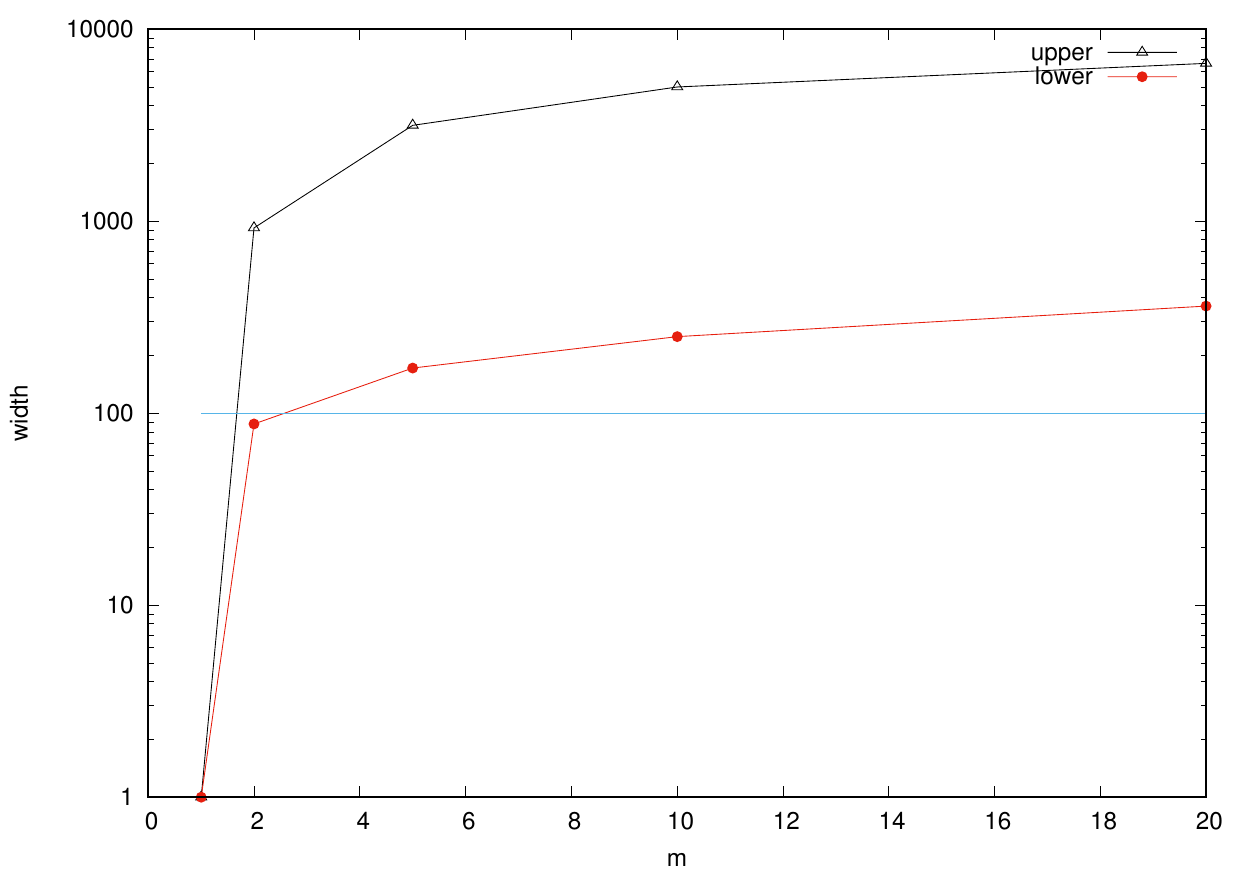}
    \caption{preferential attachment}
  \end{subfigure}
  \begin{subfigure}{0.32\linewidth}
    \includegraphics[width=\textwidth]{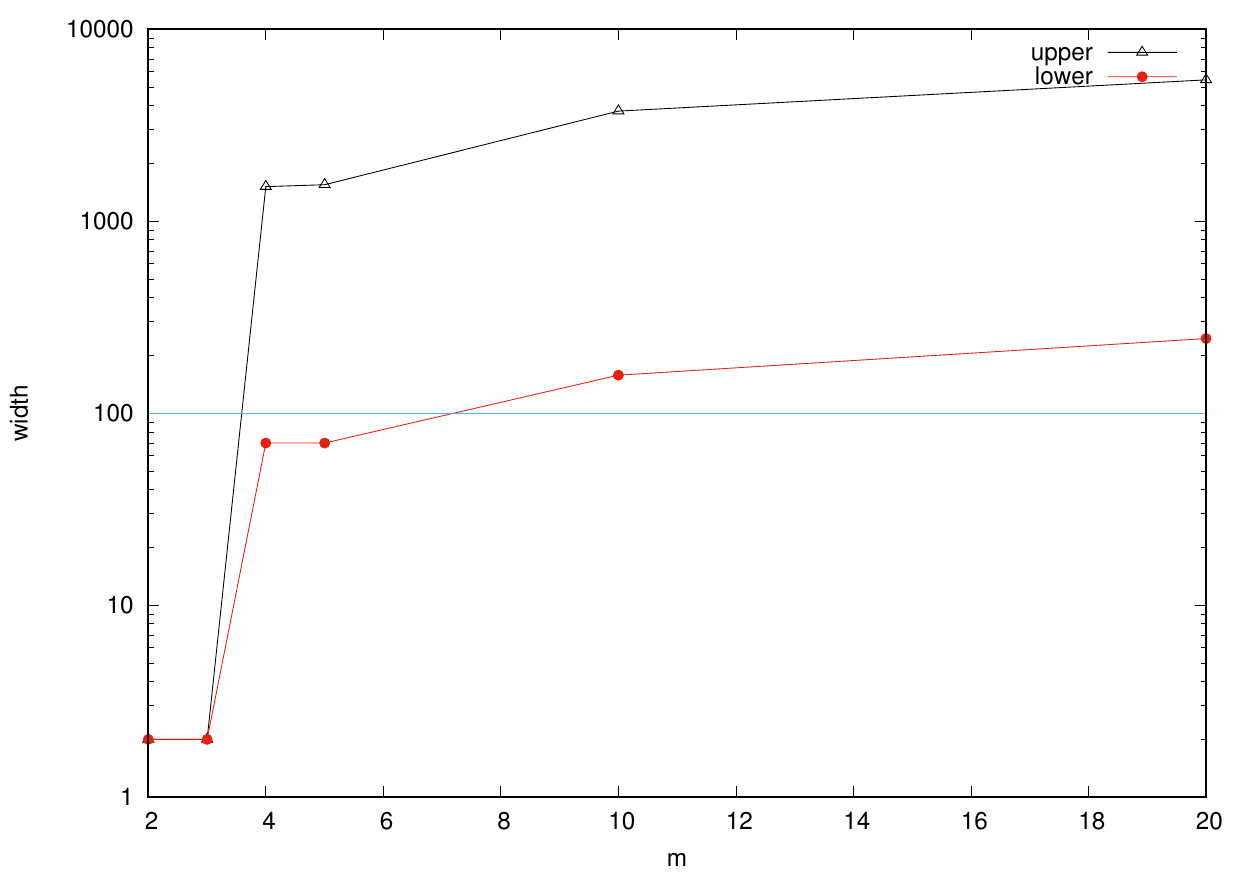}
    \caption{small world}
  \end{subfigure}

  \vspace{-1em}
  \caption{Lower and upper bounds of treewidth in synthetic networks, as a
    function of generator parameters. The blue line represent the low
    width regime, where treewidth is less than~$100$.\label{fig:synthetic}}
  \vspace{-1em}
\end{figure*}

Our first finding -- see Figure~\ref{fig:synthetic} -- is that the
high-treewidth regime arrives very early, relative to the parameter values. For
random graphs, only relatively small value of $p$ allow for a low treewidth;
for small-world and preferential attachment only trivial values of $m$ allow
low-treewidth -- and, even so, it is usually 1 or 2, possibly due to the fact
that the graph, in those cases, is composed of several small connected
components, i.e., the well-known subcritical regime of random 
graphs~\cite{barabasi2016network}.
The low-treewidth regime for random networks seems to be limited to values immediately
after $p=\frac{1}{n}$; after this point, the treewidth is high, in line with
findings of a linear dependency between graph size and treewidth in random
graphs~\cite{gao2012treewidth}.  Moreover, we notice that there is no smooth
transition for preferential attachment and small world networks; the treewidth jumps from
very low values to high treewidth values. This is understandable in scale-free
networks -- resulting from the preferential attachment model -- where a few hubs
may exist with degrees that are comparable to the number of nodes in the
graph. Comparatively, random networks exhibit a smoother progression -- one
can clearly see the shift from trivial values of treewidth, to relatively low
values, and to high treewidth values. Finally, the gap between lower
bound and upper bound 
tends to increase with the parameter; that is, however, not surprising since all
three model graphs tend to have more edges with larger parameter values.

\section{Partial Decompositions}\label{sec:partial}
Our results show that, in practice, the treewidths of real networks are quite
high. Even in the case of road networks, having relatively low treewidths,
their value can go in the hundreds, rendering most algorithms whose time is exponential
time in the treewidth (or worse) unusable.
In practical applications, however, we can still adapt treewidth-based approaches
for obtaining data structures -- not unlike indexes -- which can help with some
important graph queries like shortest distances and paths~\cite{wei2010tedi,
akiba2012shortest} or probability estimations~\cite{maniu2017indexing,
amarilli2016challenges}.

\begin{figure}[!h]
  \centering
  \includegraphics[width=0.6\linewidth]{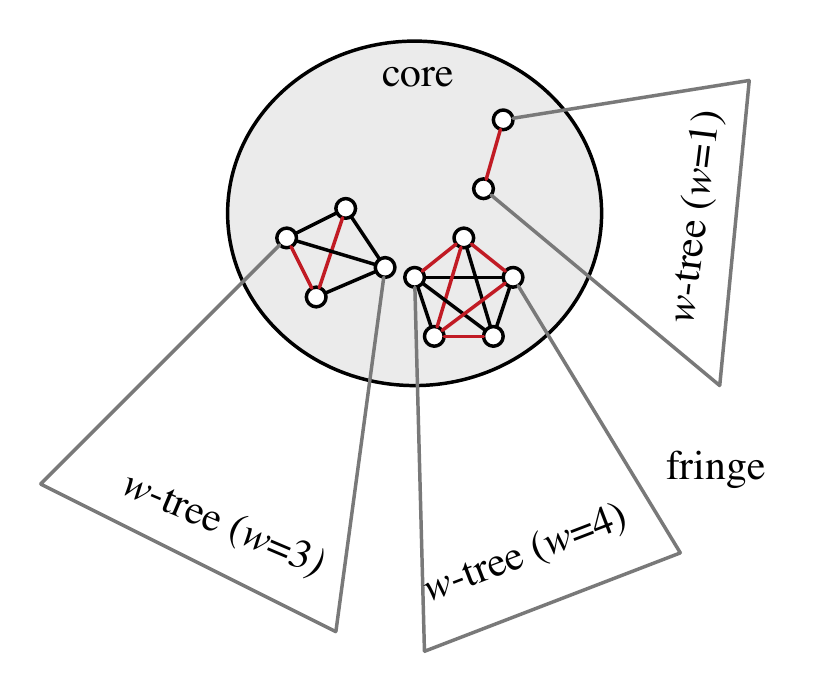}
  \vspace{-1em}
  \caption{An abstract view of partial decompositions. Partial decompositions
  are formed of a \emph{core} graph, which interfaces with $w$-trees through
$w$-cliques (the \emph{fringe}).\label{fig:partial_decomp}}
  \vspace{-1em}
\end{figure}

The manner in which treewidth decomposition can be used starts from a simple
observation made in studies on complex graphs, that is, that they tend to 
exhibit a
\emph{tree-like fringe} and a \emph{densely connected
core}~\cite{newman2001random, newman2002random}. The tree-like fringe
precisely corresponds to bounded-treewidth parts of the network. This
yields an
easy adaptation of the upper bound algorithms based on node ordering: given a
parameter $w$ representing the highest treewidth the fringe can be, we can run
any greedy decomposition algorithm (\textsc{Degree}, \textsc{FillIn},
\textsc{DegreeFillIn}) until we only find nodes of degree $w+1$, at which
point the algorithm stops. At termination, we obtain a data structure
formed of a set of treewidth~$w$
elements ($w$-trees) interfacing through cliques that have size at most
$w+1$ with a \emph{core graph}. The core graph contains all the nodes not
removed in the bag creation process, and has unbounded treewidth. Figure~\ref{fig:partial_decomp} illustrates the notion of partial
decompositions.

The resulting structure can be thought of as a \emph{partial decomposition} (or
\emph{relaxed decomposition}), a concept introduced in~\cite{wei2010tedi,
akiba2012shortest} in the context of answering shortest path queries, and used
in~\cite{maniu2017indexing} for probabilistic distance queries. A partial
decomposition can be extremely useful. The tree-like
fringe can be used to quickly precompute answers to partial queries (e.g.,
precompute distances in the graph).  Once the precomputation is done, these
(partial) answers are added to the core graph, where queries can be answered
directly. 
If the resulting core graph is much smaller than the original graph, the gains
in running time can be considerable, as shown in~\cite{wei2010tedi,
akiba2012shortest, maniu2017indexing}. Hence, the objective of our experiments
in this section is to check how feasible partial decompositions are.

An interesting aspect of greedy upper bound algorithms is that they generate at
any point a partial decomposition, with width equal to the highest degree
encountered in the greedy ordering so far. The algorithm can then be stopped
at any width, and the size of the resulting partial decomposition can be
measured.

To evaluate this, we track here the size of the core graph, in terms of edges. 
We do this because
most algorithms on graphs have a complexity that is directly related to
the number of edges in the graph. As we discussed in the previous section, we
aim that the size of the core graph to be as small as possible. Another aspect to keep in
mind is the number of edges that are added by the fill-in process during the
decomposition: each time a node of degree~$w$ is removed,
at most $\frac{w(w-1)}2$ edges are added to the graph. Finally, for a graph
of treewidth~$k$, the decomposition contains a root of size at most $k^2$ edges. Hence, to ensure
that the core graph is smaller than the original graph we aim for the
treewidth to
be $\sqrt{n}$. As we saw in Section~\ref{sec:results}, this is only
rarely the case, and it is more likely to occur in infrastructure networks. 
%For
%space and legibility reasons, we present here only a selection of results; full results 
%are available in Appendix~\ref{sec:app_partial}.

\begin{figure*}[h]
  \centering
  \begin{subfigure}{0.24\linewidth}
    \includegraphics[width=\textwidth]{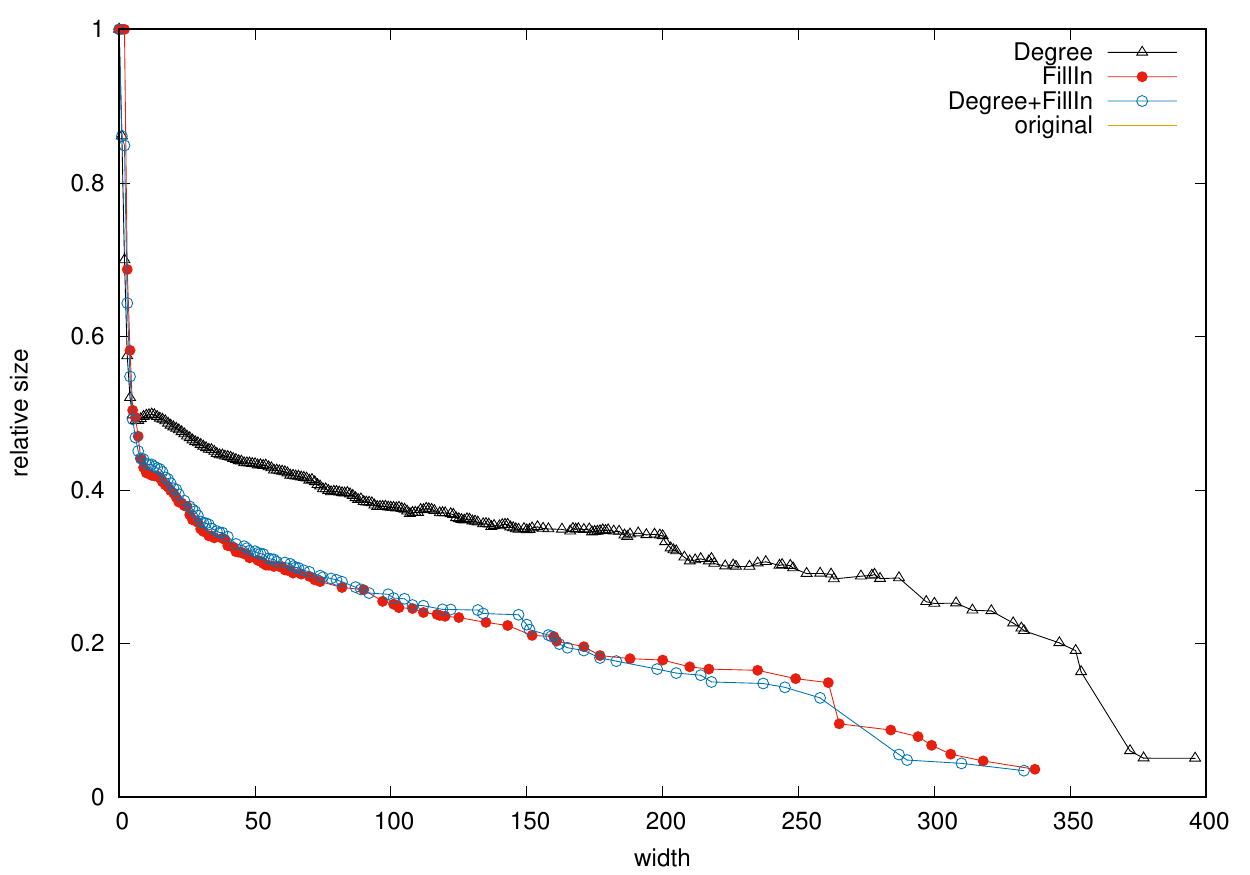}
    \caption{\textsc{Pa}}
  \end{subfigure}
  \begin{subfigure}{0.24\linewidth}
    \includegraphics[width=\textwidth]{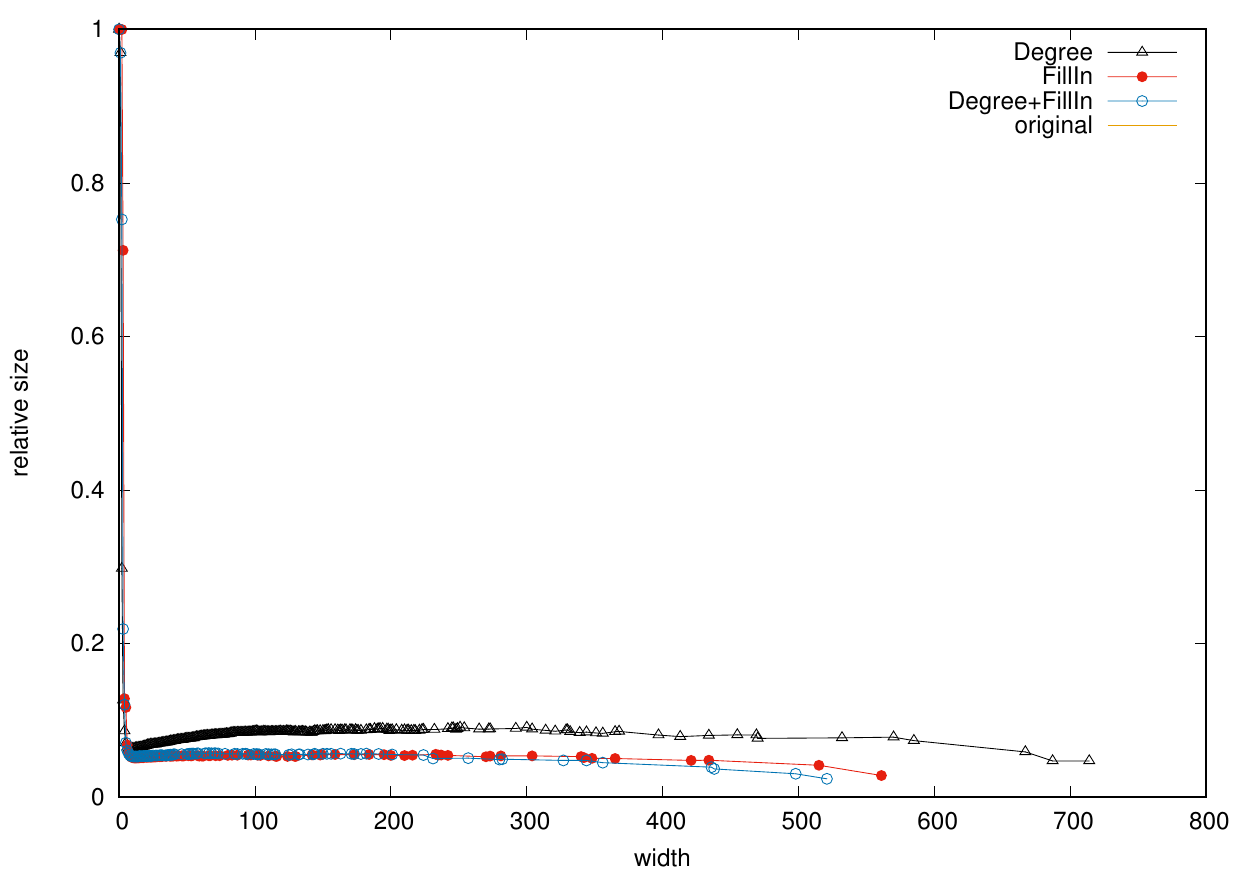}
    \caption{\textsc{Paris}}
  \end{subfigure}
  \begin{subfigure}{0.24\linewidth}
    \includegraphics[width=\textwidth]{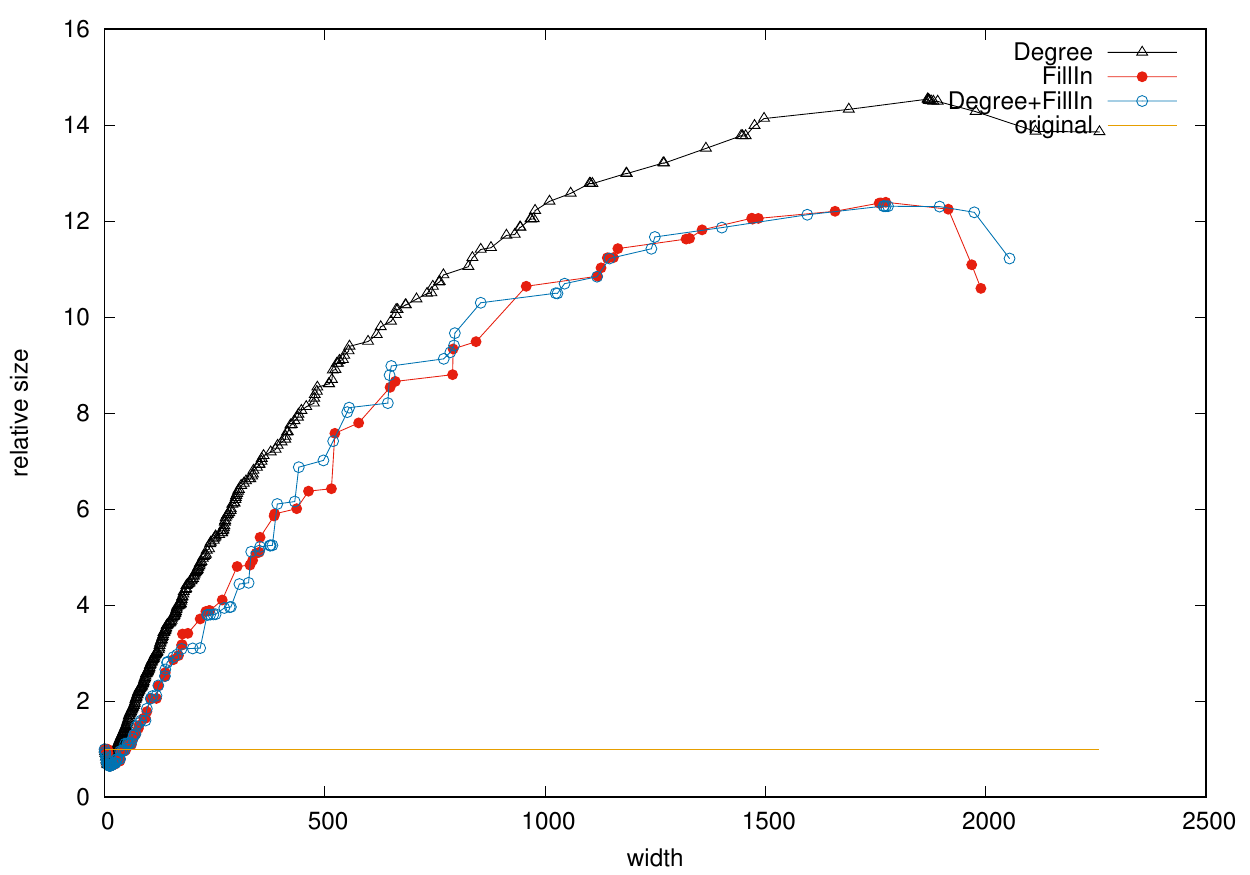}
    \caption{\textsc{Enron}}
  \end{subfigure}
  \begin{subfigure}{0.24\linewidth}
    \includegraphics[width=\textwidth]{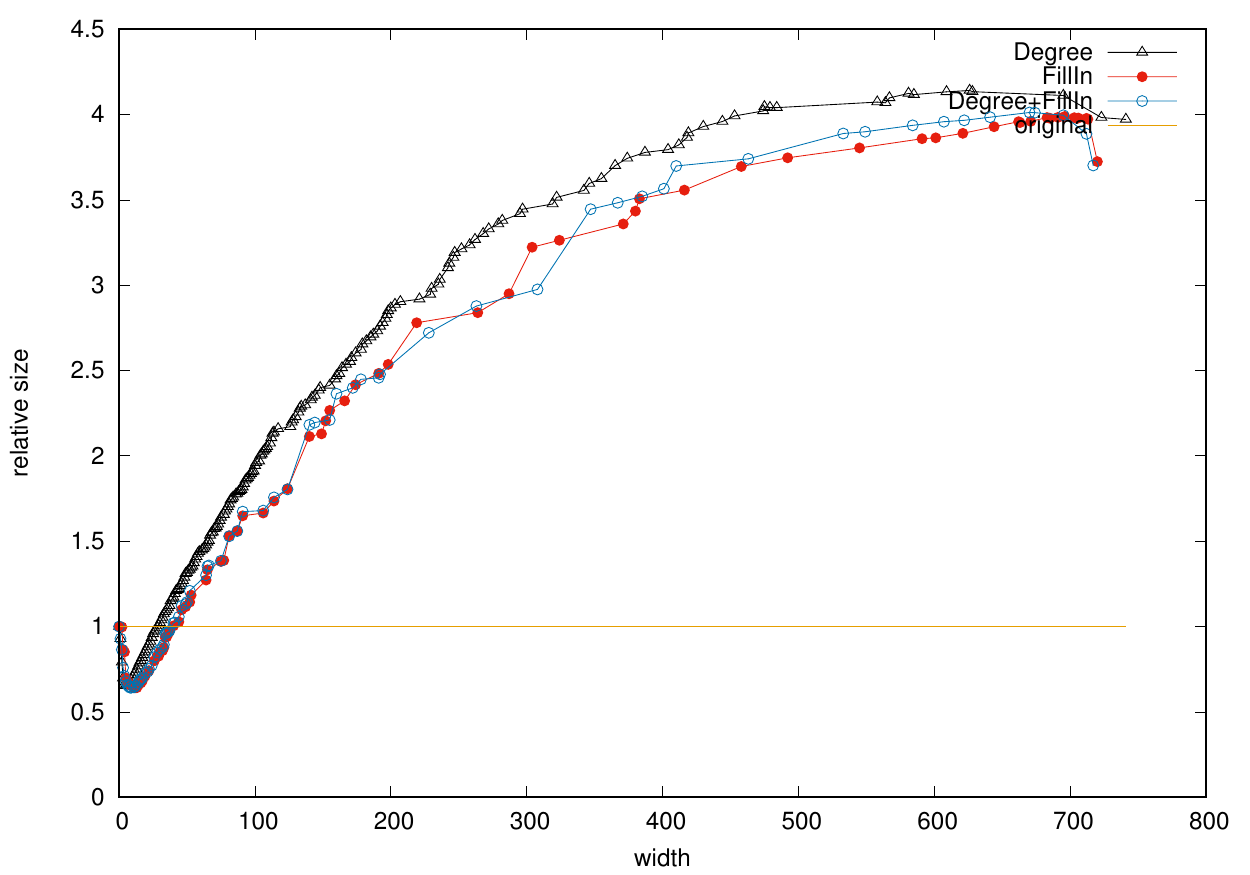}
    \caption{\textsc{Stack-TCS}}
  \end{subfigure}

  \begin{subfigure}{0.24\linewidth}
    \includegraphics[width=\textwidth]{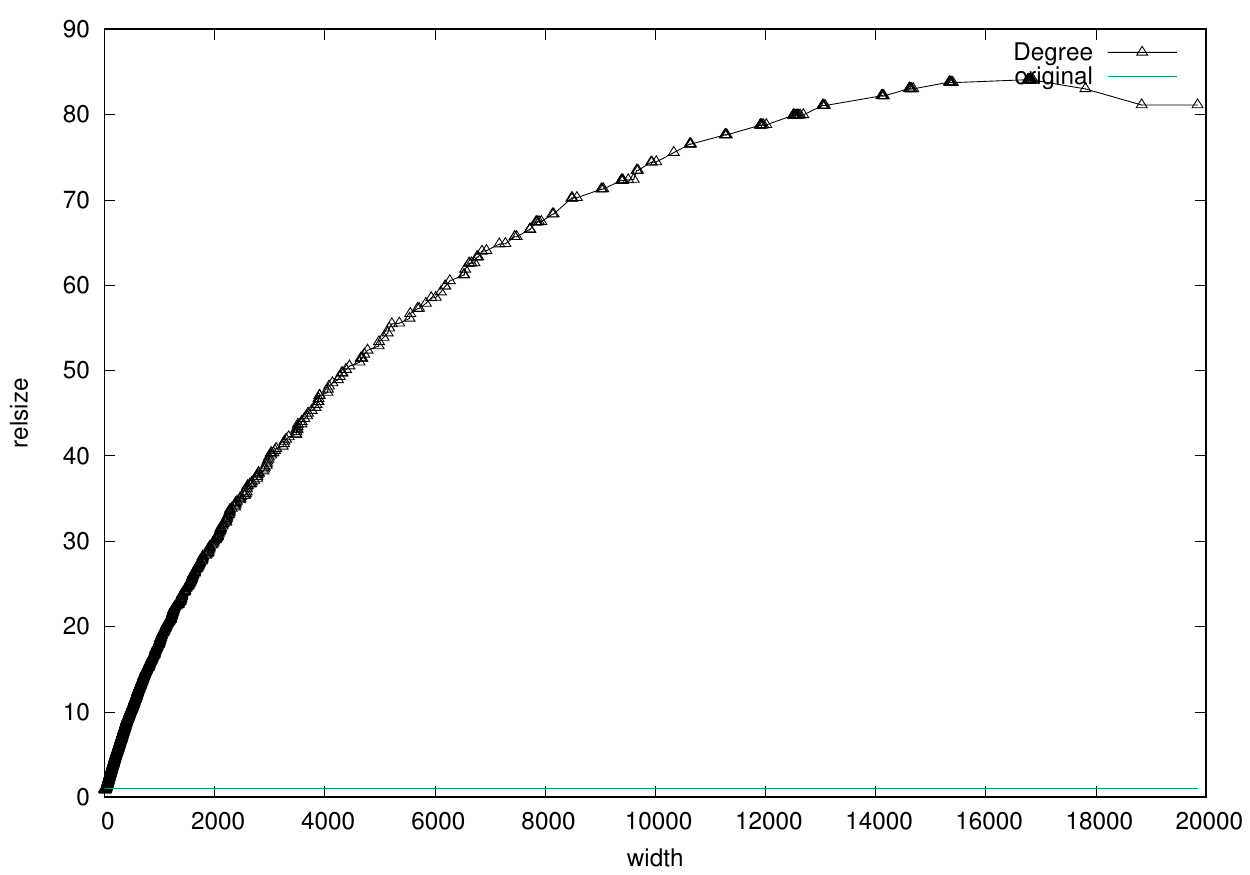}
    \caption{\textsc{Wikipedia}}
  \end{subfigure}
  \begin{subfigure}{0.24\linewidth}
    \includegraphics[width=\textwidth]{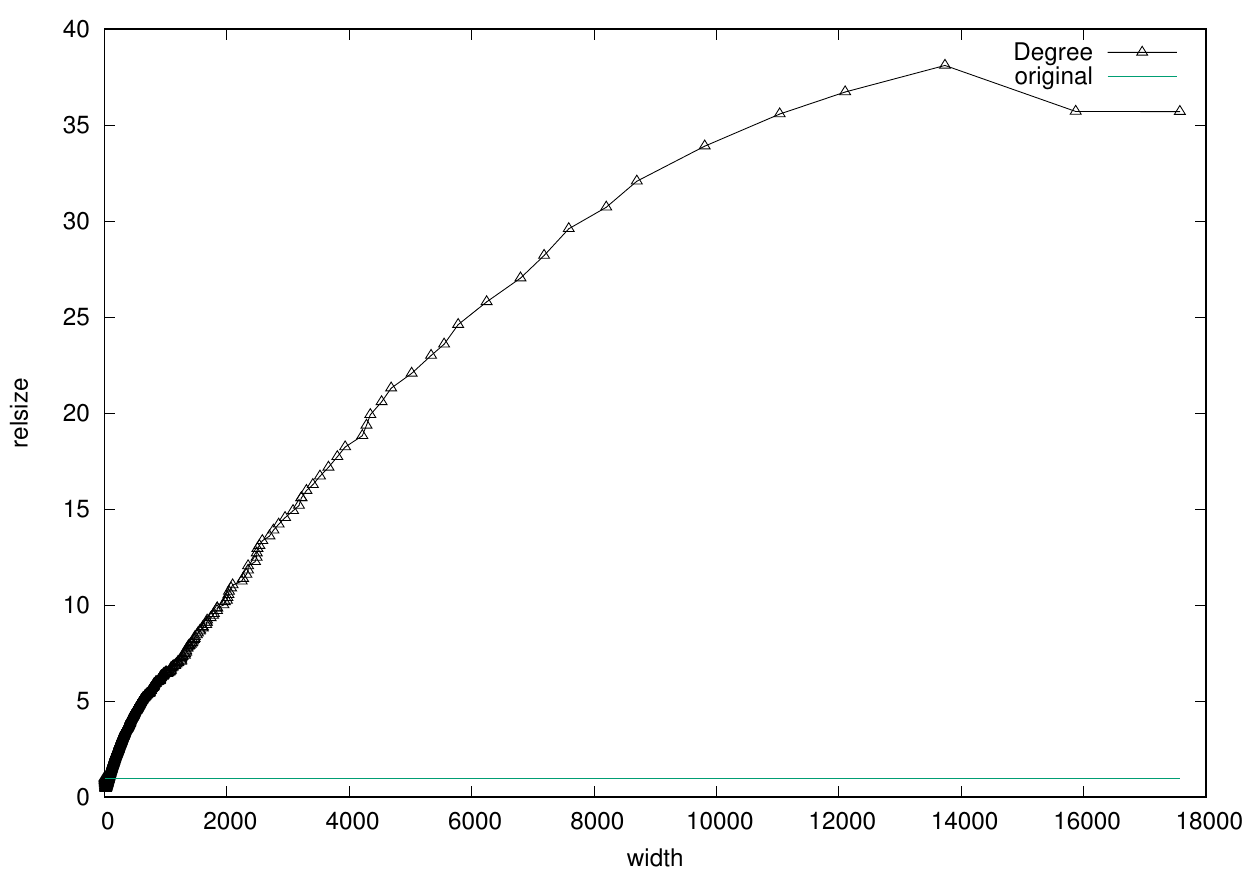}
    \caption{\textsc{Google}}
  \end{subfigure}
  \begin{subfigure}{0.24\linewidth}
    \includegraphics[width=\textwidth]{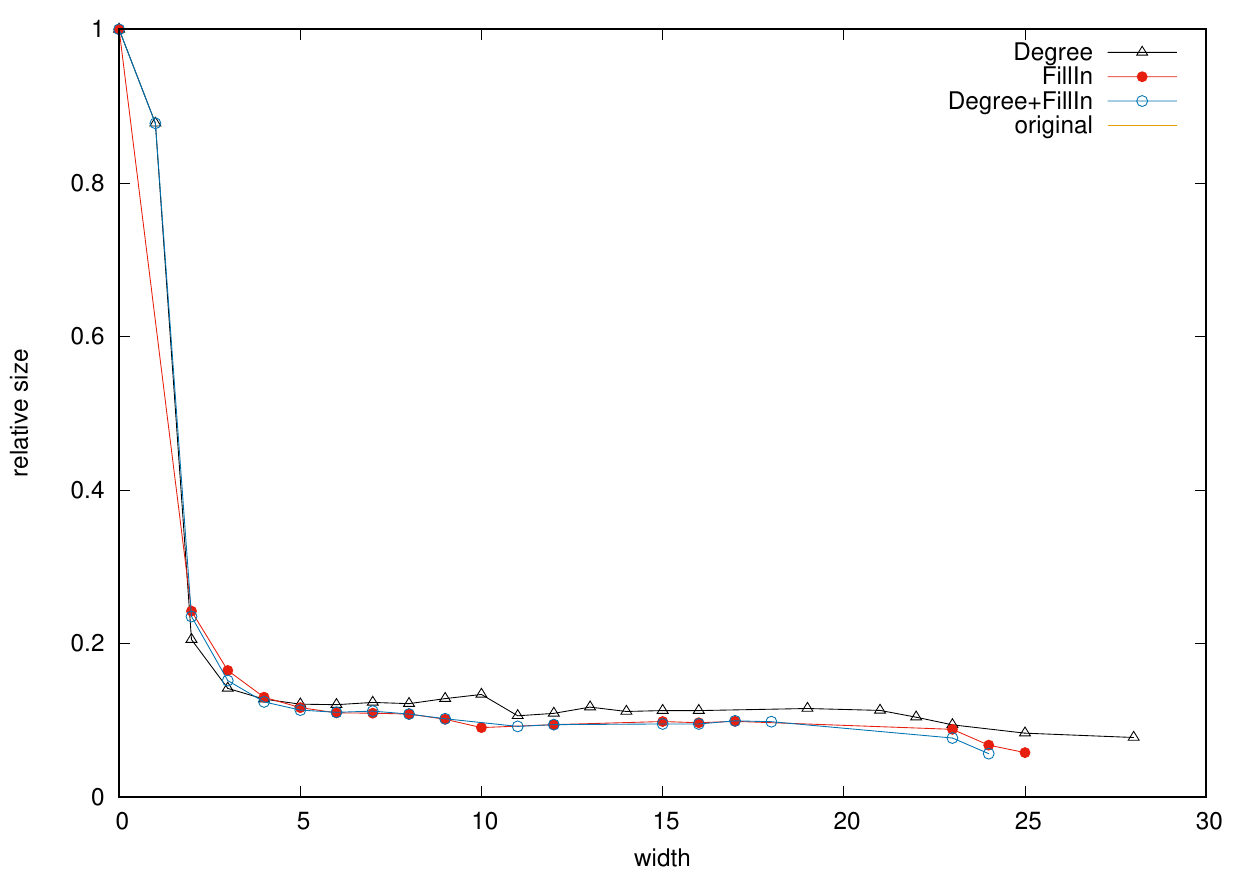}
    \caption{\textsc{Royal}}
  \end{subfigure}
  \begin{subfigure}{0.24\linewidth}
    \includegraphics[width=\textwidth]{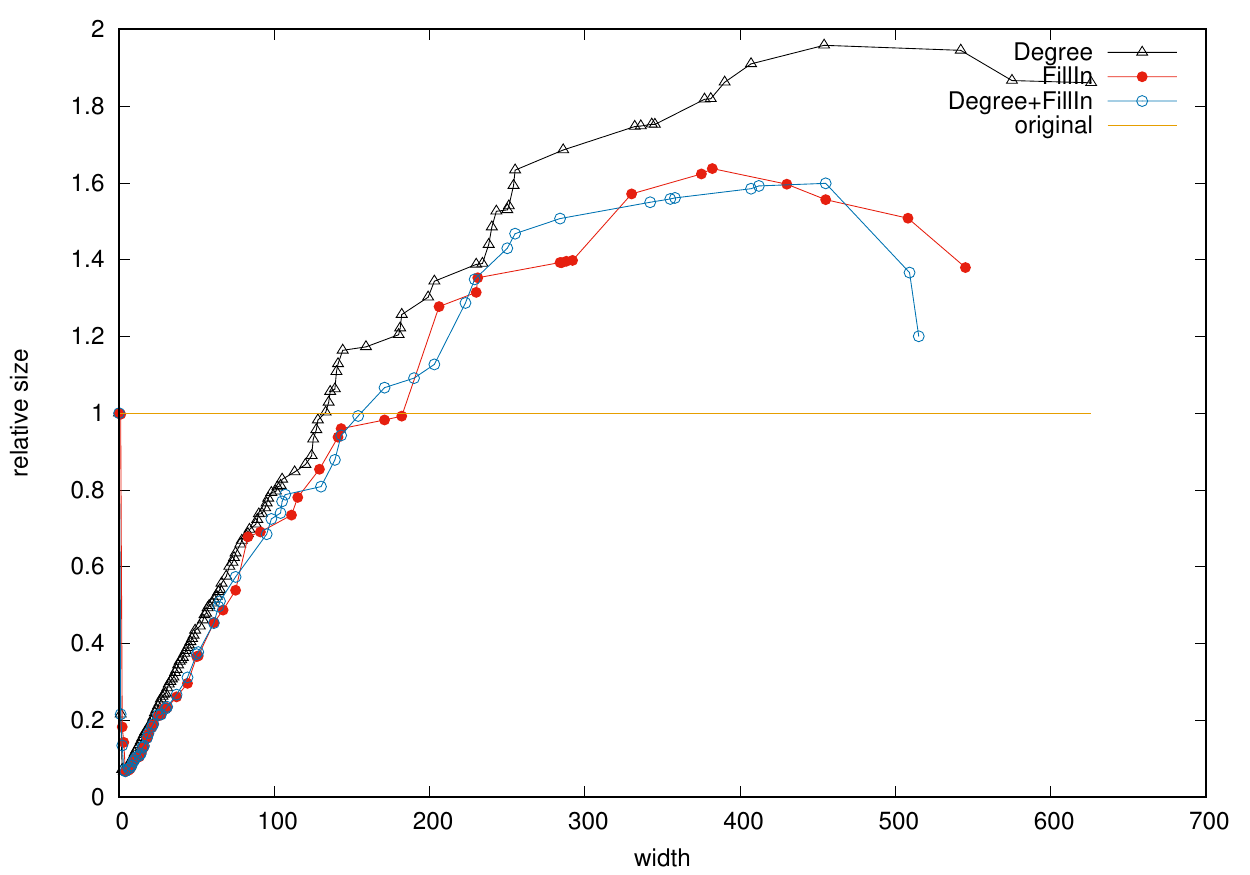}
    \caption{\textsc{Math}}
  \end{subfigure}

    \vspace{-1em}
  \caption{Relative sizes of core graphs in partial
  decompositions, after all bags of a given size
  have been removed in the decomposition.\label{fig:partial_results}}
    \vspace{-1em}
\end{figure*}

Indeed, if we plot the core graph size in the partial decompositions of infrastructure
networks (Figures~\ref{fig:partial_results}a, \ref{fig:partial_results}b), we see that their size is only a
fraction of the original size. We see a large drop in the size for low
widths; this is logical, since the fill-in process does not add edges for
$w=1$ and $w=2$. After this point, the size is relatively stable, and can go as
low as 10\% of the original size. In this sense, infrastructure networks seem the best fit
for indexes based on decompositions, and represents further confirmation of the
good behavior of these networks for hierarchical decompositions.

This desired decomposition behavior no longer occurs for social networks
(Figures~\ref{fig:partial_results}c, \ref{fig:partial_results}d) and the other networks in our dataset
(Figures~\ref{fig:partial_results}e, \ref{fig:partial_results}f). In this case, the decomposition becomes 
too 
large
and even unwieldy -- for some networks, the decomposition started filling the
main memory of our computers (32~GB of RAM); for other networks, they did not finish in
reasonable time (i.e., a few days of computation). For most of these networks, the resulting treewidth is much
larger than our desired bound of at most $\sqrt{n})$; this results in core
graph sizes that can be hundreds of times larger than the original size. The
only exceptions are hierarchical networks, \textsc{Royal} and
\textsc{Math} (Figures~\ref{fig:partial_results}g,
\ref{fig:partial_results}h), which are after all supposed to be close to
a tree -- despite having a relatively high treewidth (remember
Figure~\ref{fig:bounds}b), partial decompositions work well on them. See
Appendix~E
\ifextended\else of \cite{extended} \fi
for full results.

\begin{toappendix}
\section{Full Partial Decomposition Results}\label{sec:app_partial}

The full partial decomposition results are presented in 
Figures~\ref{fig:part_road} and~\ref{fig:part_social}.

\begin{figure*}[p]
  \centering
  \begin{subfigure}{0.24\linewidth}
    \includegraphics[width=\textwidth]{figures/pa.pdf}
    \caption{\textsc{Pa}}
  \end{subfigure}
  \begin{subfigure}{0.24\linewidth}
    \includegraphics[width=\textwidth]{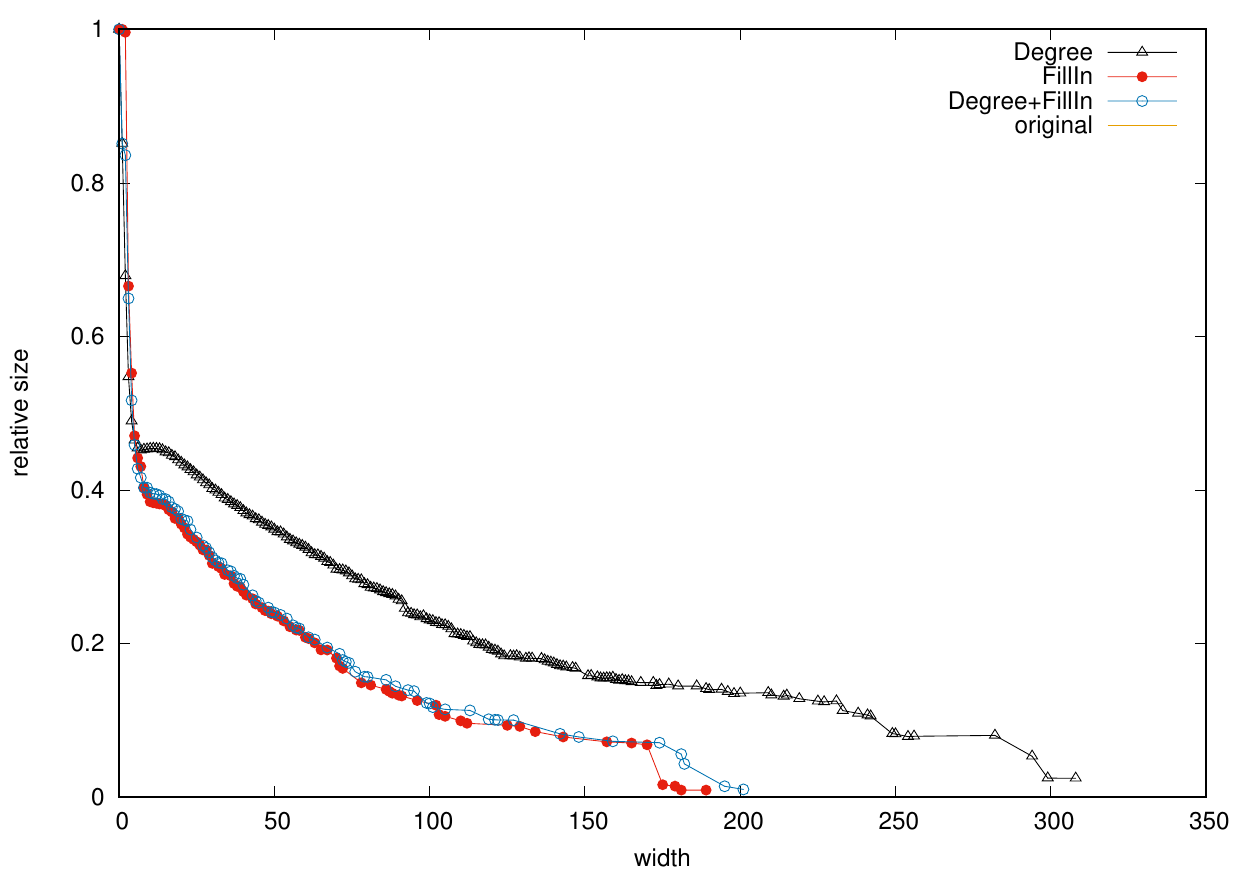}
    \caption{\textsc{Tx}}
  \end{subfigure}
  \begin{subfigure}{0.24\linewidth}
    \includegraphics[width=\textwidth]{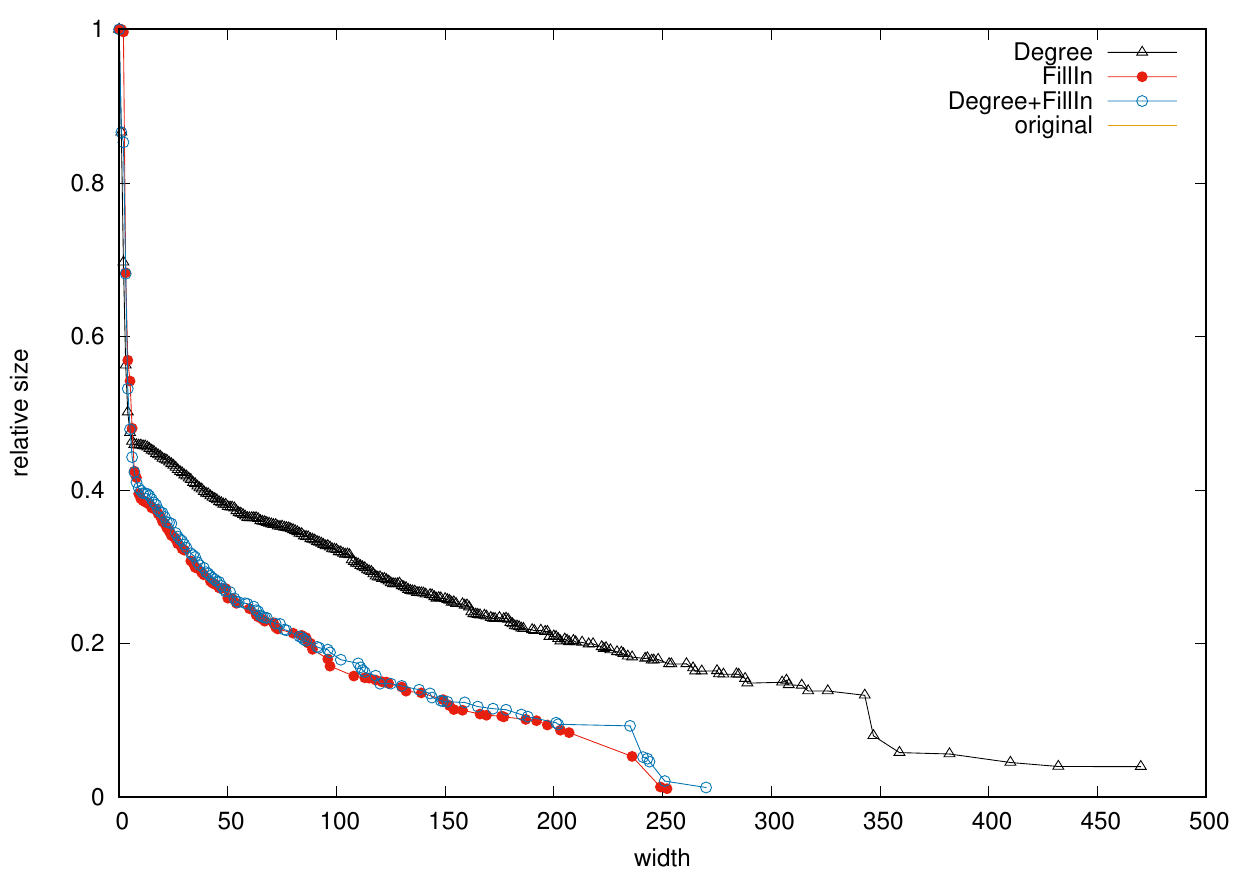}
    \caption{\textsc{Ca}}
  \end{subfigure}
  \begin{subfigure}{0.24\linewidth}
    \includegraphics[width=\textwidth]{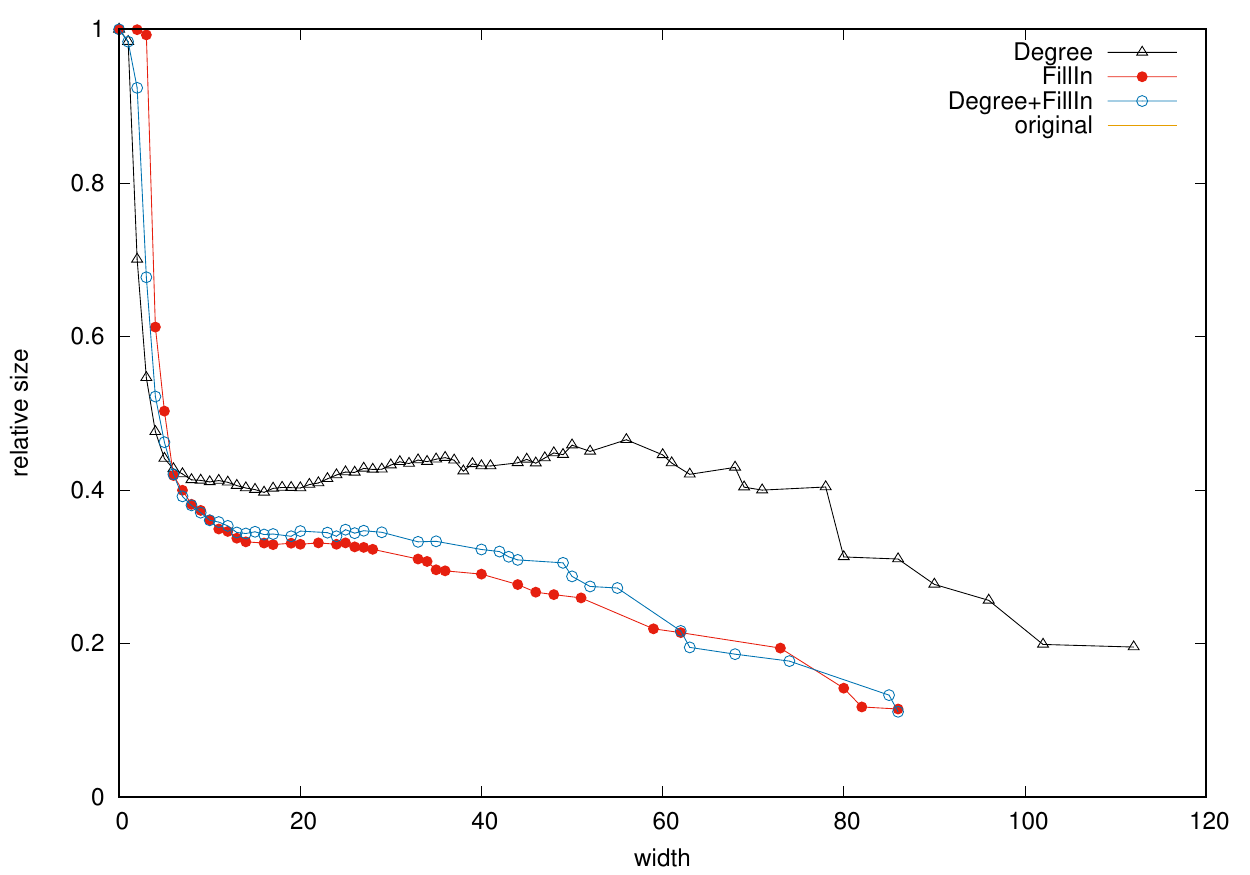}
    \caption{\textsc{Stif}}
  \end{subfigure}
  
  \begin{subfigure}{0.24\linewidth}
    \includegraphics[width=\textwidth]{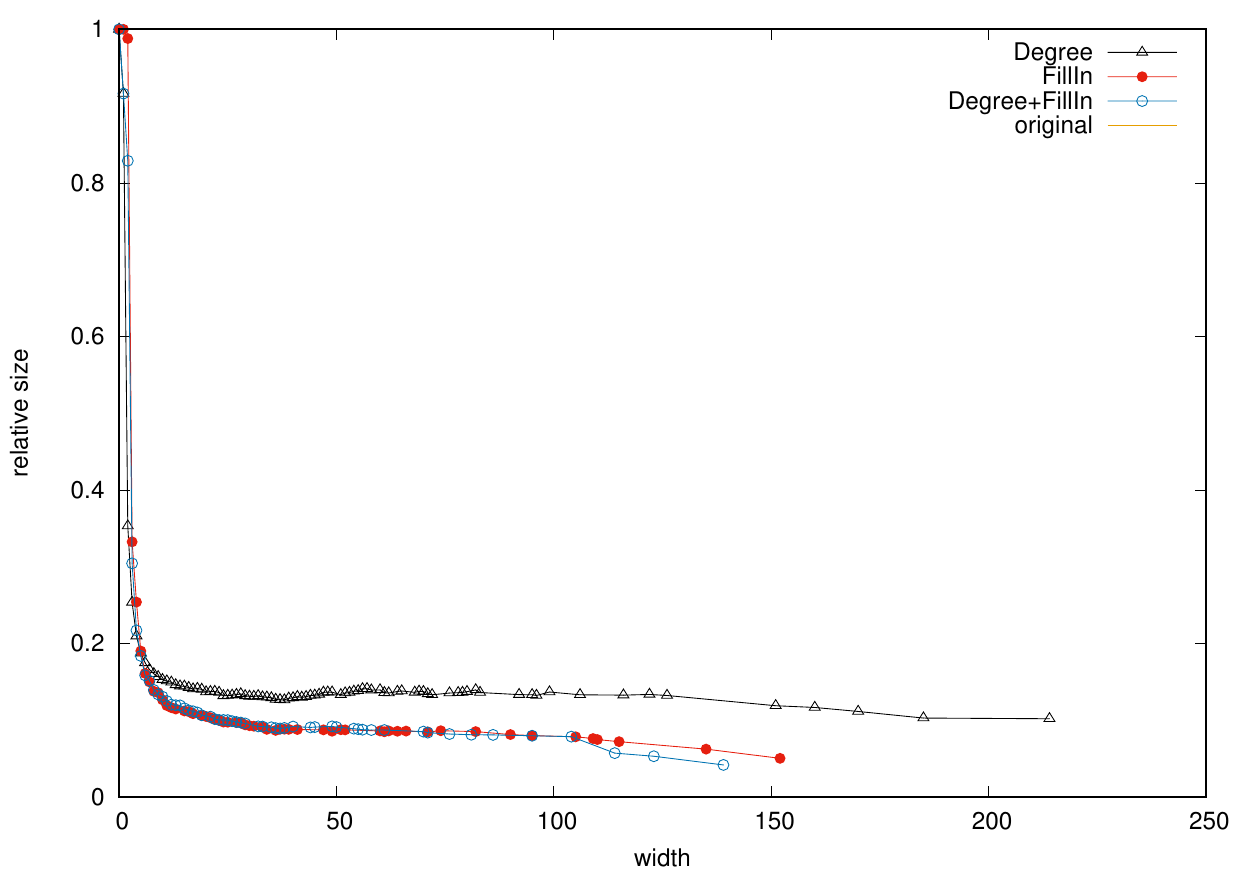}
    \caption{\textsc{Bucharest}}
  \end{subfigure}
  \begin{subfigure}{0.24\linewidth}
    \includegraphics[width=\textwidth]{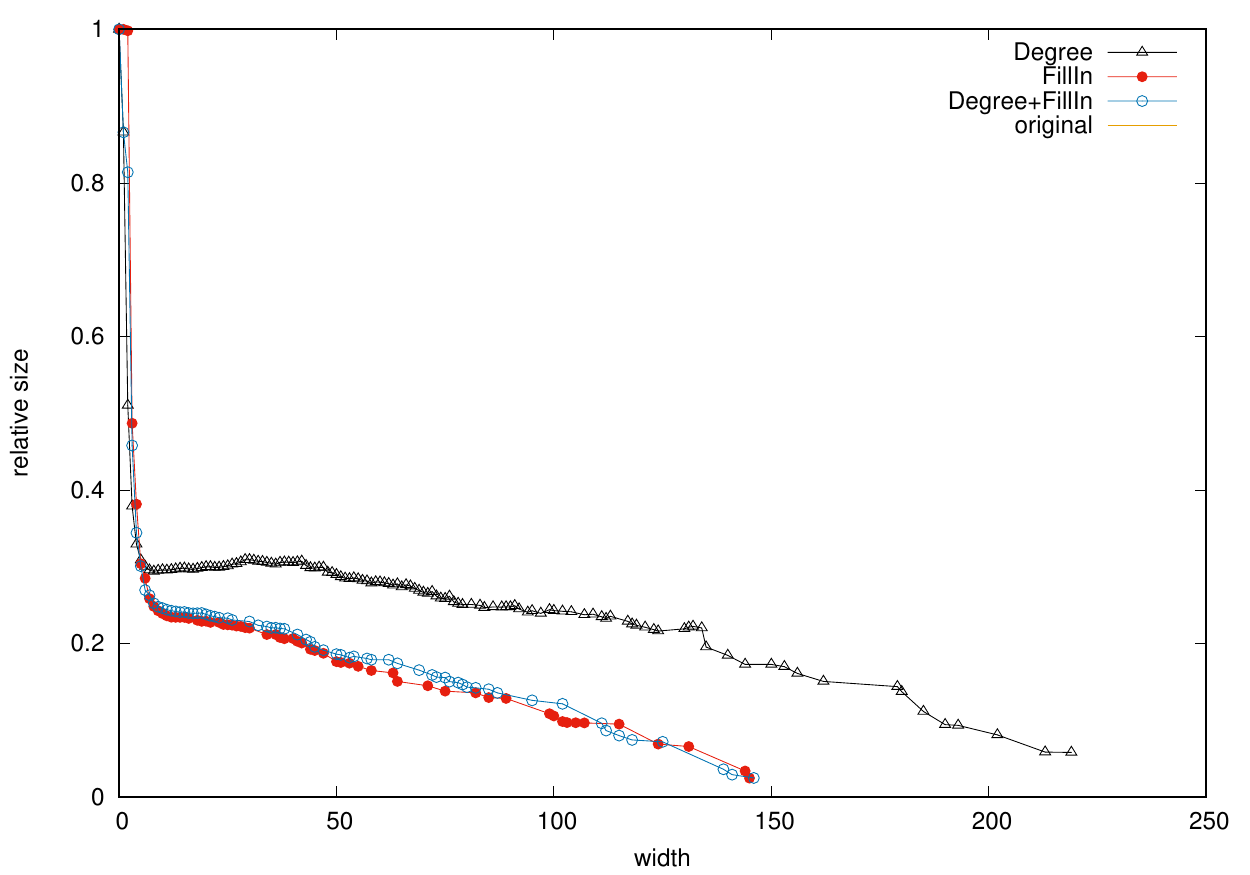}
    \caption{\textsc{HongKong}}
  \end{subfigure}
  \begin{subfigure}{0.24\linewidth}
    \includegraphics[width=\textwidth]{figures/paris.pdf}
    \caption{\textsc{Paris}}
  \end{subfigure}
  \begin{subfigure}{0.24\linewidth}
    \includegraphics[width=\textwidth]{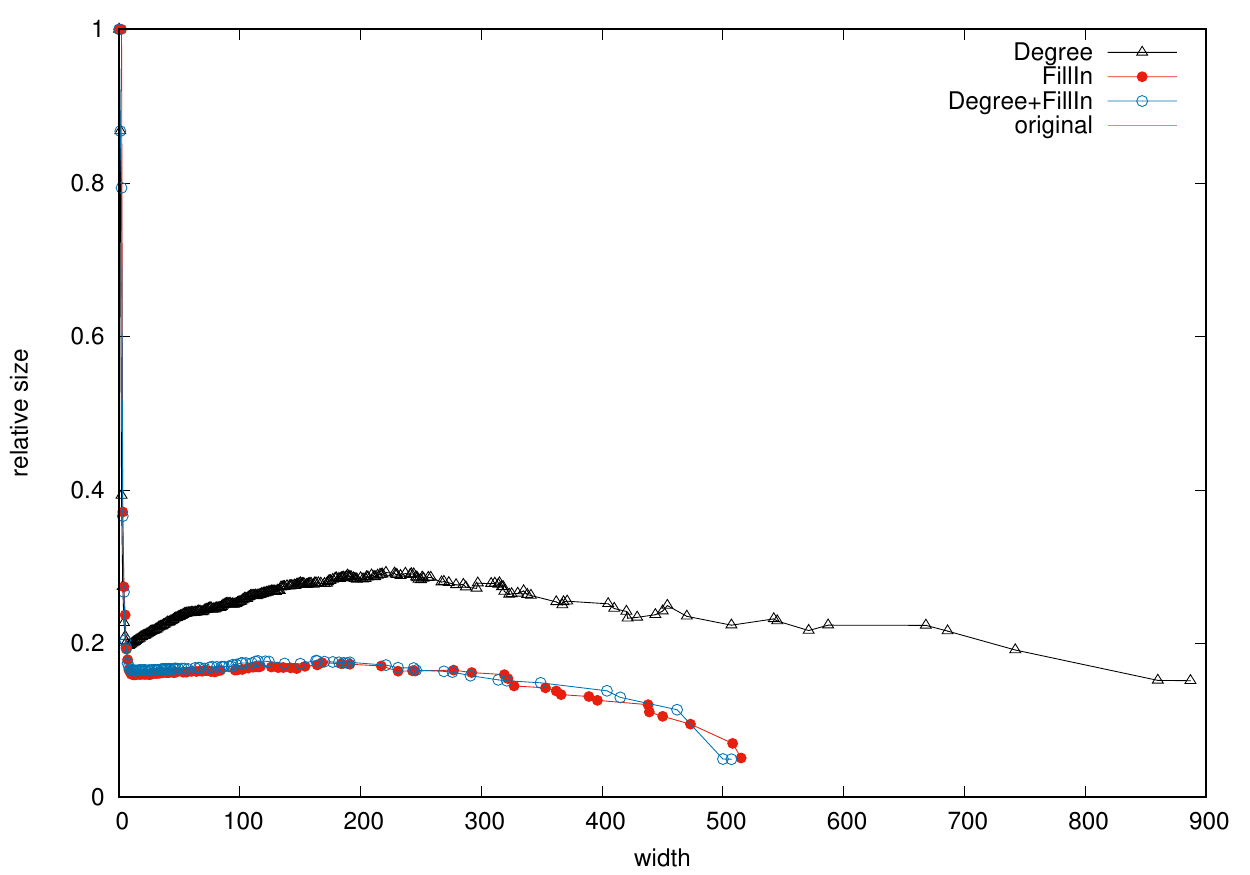}
    \caption{\textsc{London}}
  \end{subfigure}
  
  \caption{{Road/transport networks.} Relative sizes of core graphs 
  in partial
    decompositions, after all bags of a given size
    have been removed in the decomposition.\label{fig:part_road}}
\end{figure*}

\begin{figure*}[p]
  
  \begin{subfigure}{0.24\linewidth}
    \includegraphics[width=\textwidth]{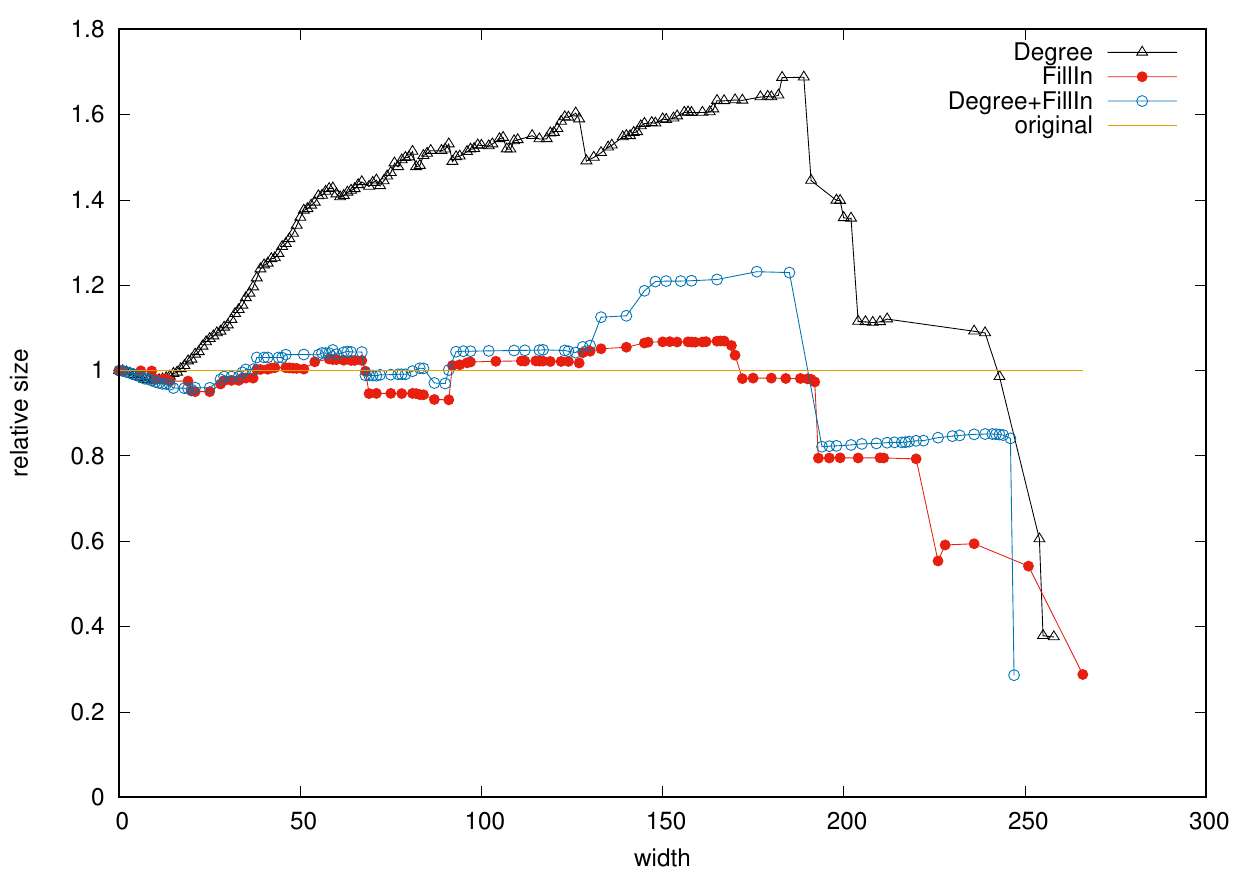}
    \caption{\textsc{Facebook}}
  \end{subfigure}
  \begin{subfigure}{0.24\linewidth}
    \includegraphics[width=\textwidth]{figures/enron.pdf}
    \caption{\textsc{Enron}}
  \end{subfigure}
  \begin{subfigure}{0.24\linewidth}
    \includegraphics[width=\textwidth]{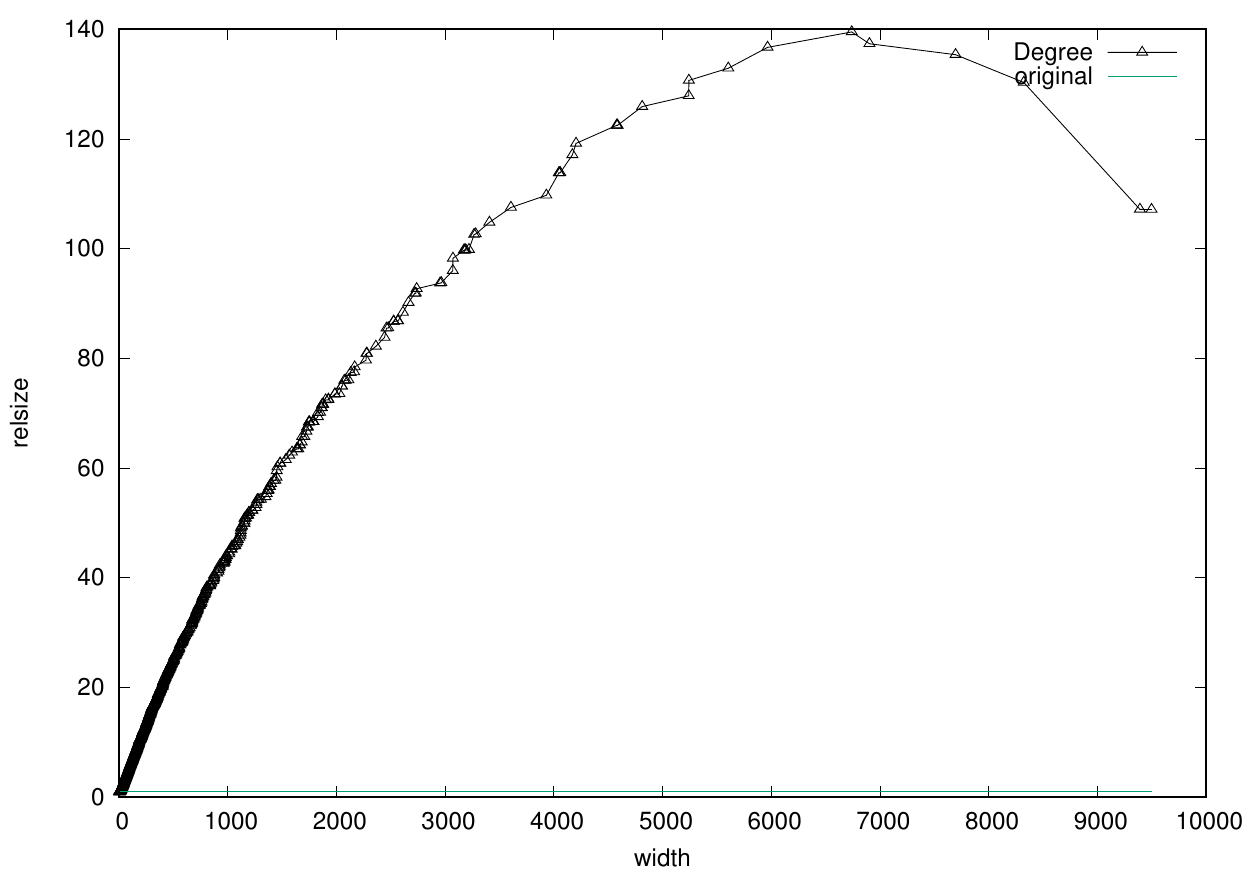}
    \caption{\textsc{CitHeph}}
  \end{subfigure}
  \begin{subfigure}{0.24\linewidth}
    \includegraphics[width=\textwidth]{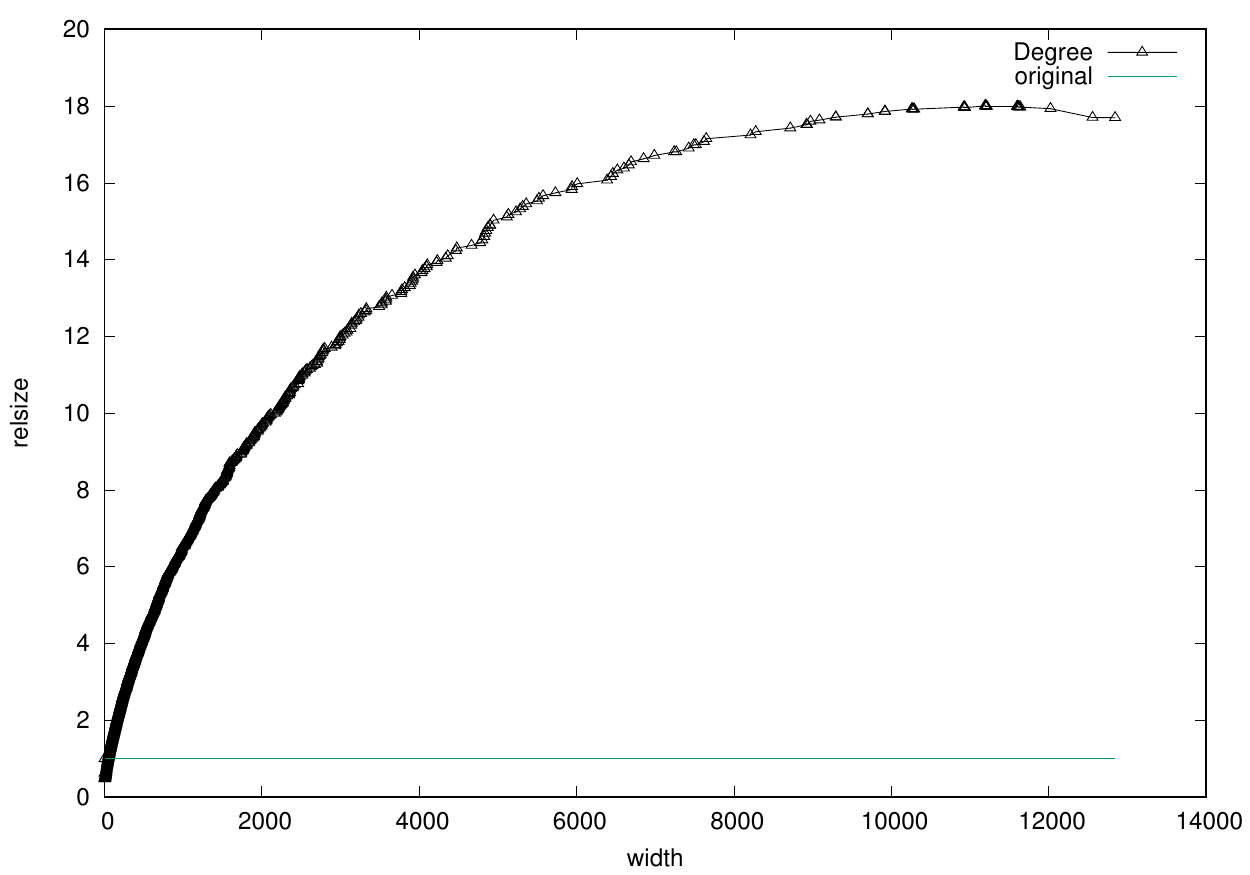}
    \caption{\textsc{WikiTalk}}
  \end{subfigure}
  
  \begin{subfigure}{0.24\linewidth}
    \includegraphics[width=\textwidth]{figures/stacktcs.pdf}
    \caption{\textsc{Stack-TCS}}
  \end{subfigure}
  \begin{subfigure}{0.24\linewidth}
    \includegraphics[width=\textwidth]{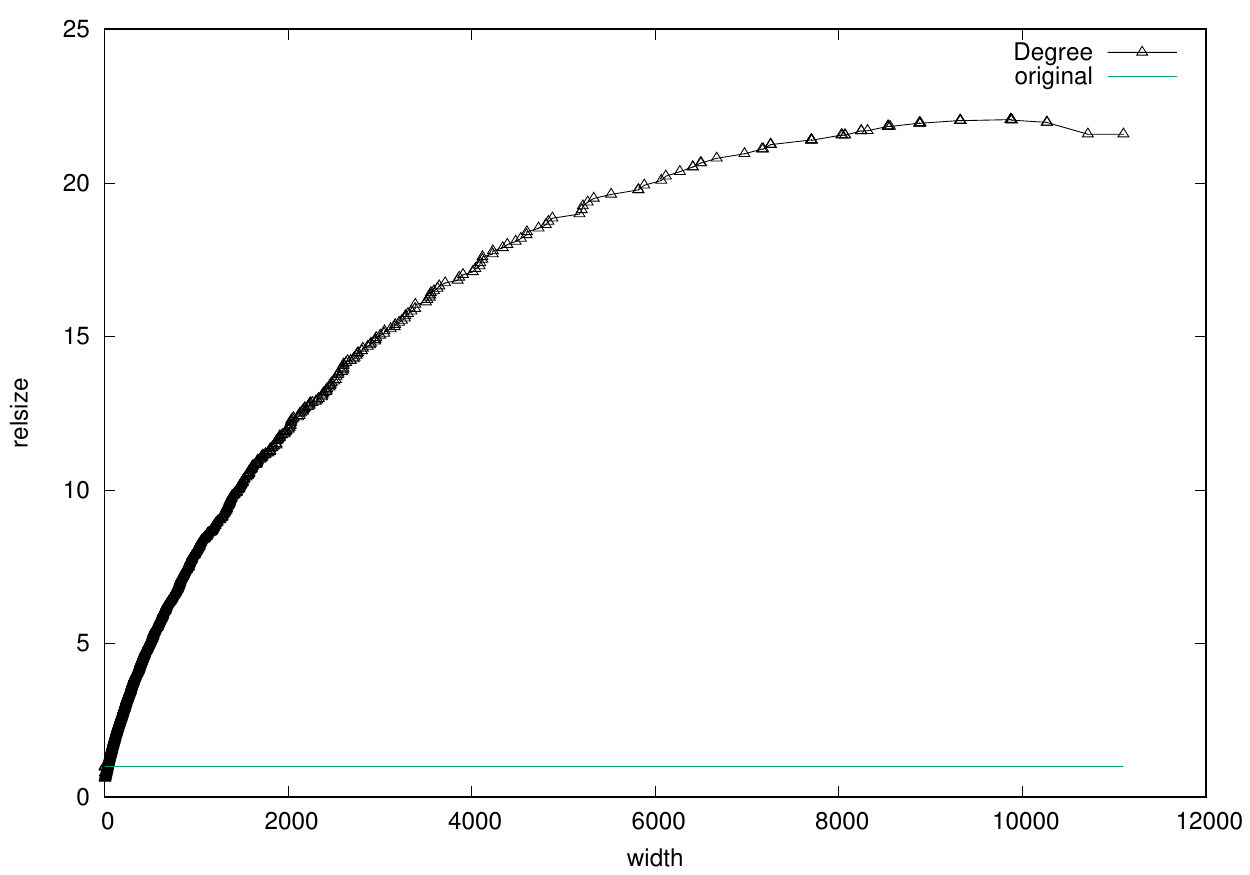}
    \caption{\textsc{StackMath}}
  \end{subfigure}
  \begin{subfigure}{0.24\linewidth}
    \includegraphics[width=\textwidth]{figures/wikipedia.pdf}
    \caption{\textsc{Wikipedia}}
  \end{subfigure}  
  \begin{subfigure}{0.24\linewidth}
    \includegraphics[width=\textwidth]{figures/google.pdf}
    \caption{\textsc{Google}}
  \end{subfigure}

  \begin{subfigure}{0.24\linewidth}
    \includegraphics[width=\textwidth]{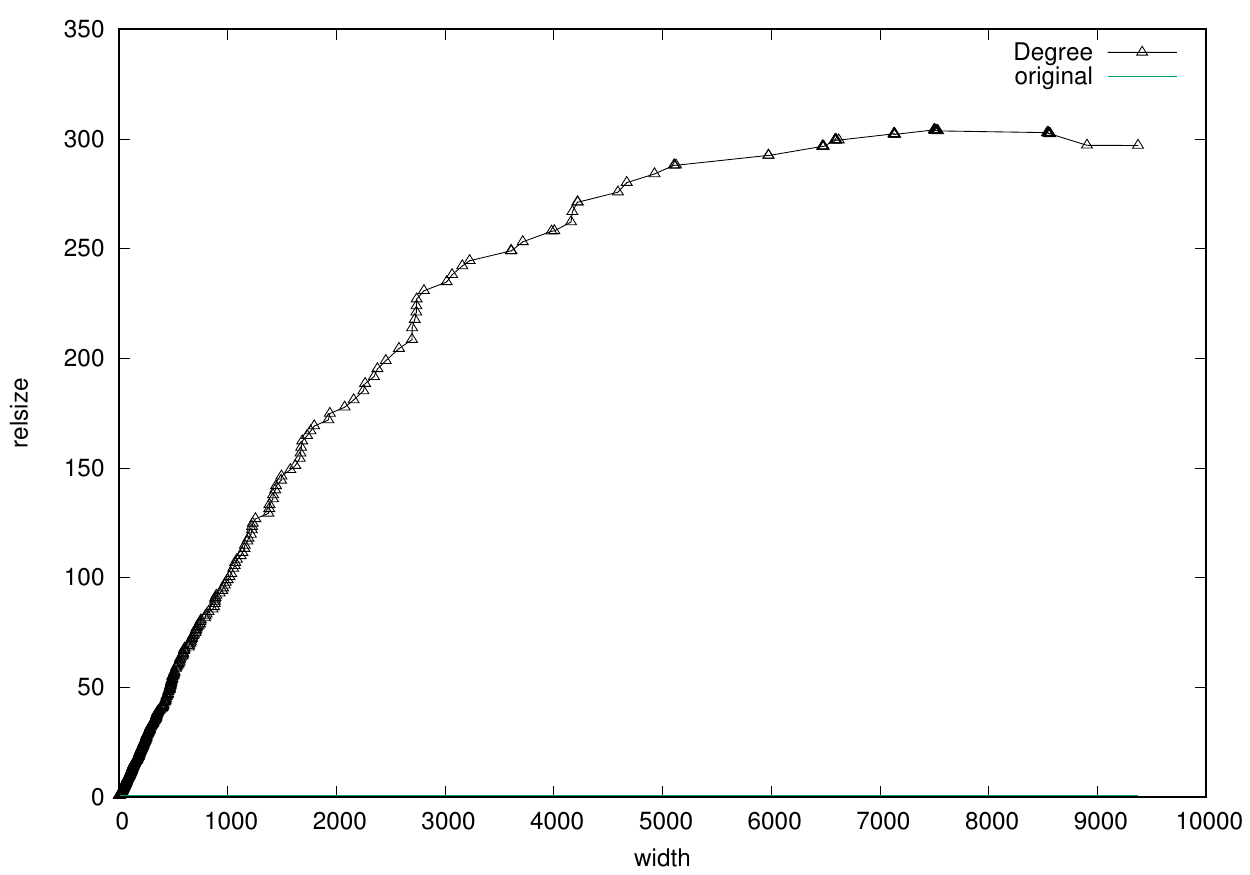}
    \caption{\textsc{Gnutella}}
  \end{subfigure}  
  \begin{subfigure}{0.24\linewidth}
    \includegraphics[width=\textwidth]{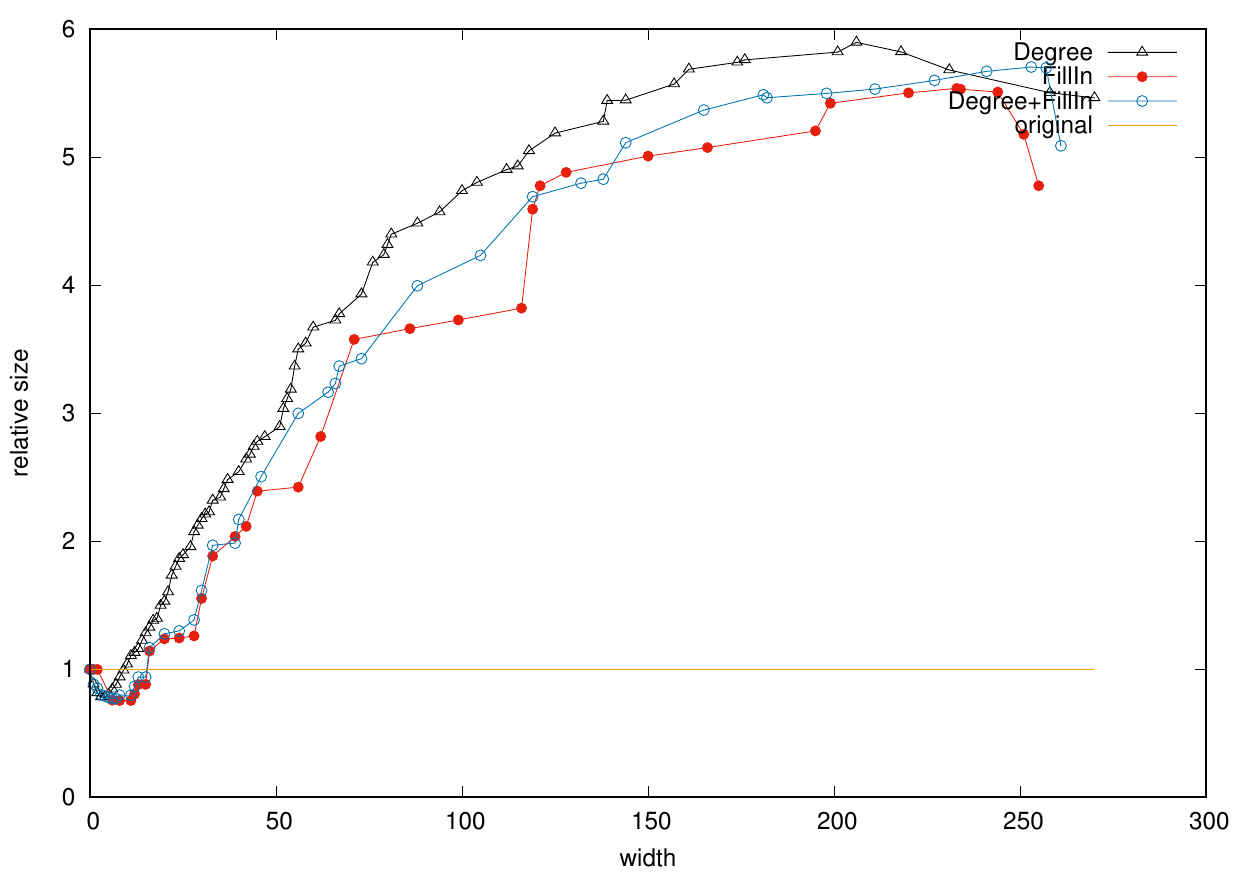}
    \caption{\textsc{Yeast}}
  \end{subfigure}
  \begin{subfigure}{0.24\linewidth}
    \includegraphics[width=\textwidth]{figures/royal.pdf}
    \caption{\textsc{Royal}}
  \end{subfigure}
  \begin{subfigure}{0.24\linewidth}
    \includegraphics[width=\textwidth]{figures/math.pdf}
    \caption{\textsc{Math}}
  \end{subfigure}
  
  \caption{{Social and other networks.} Relative sizes of core graphs 
  in partial
    decompositions, after all bags of a given size
    have been removed in the decomposition.\label{fig:part_social}}
\end{figure*}
\end{toappendix}

\begin{toappendix}
\section{Partial Decompositions of Synthetic Graphs}

\begin{figure*}[h]

  \begin{subfigure}{0.32\linewidth}
    \includegraphics[width=\textwidth]{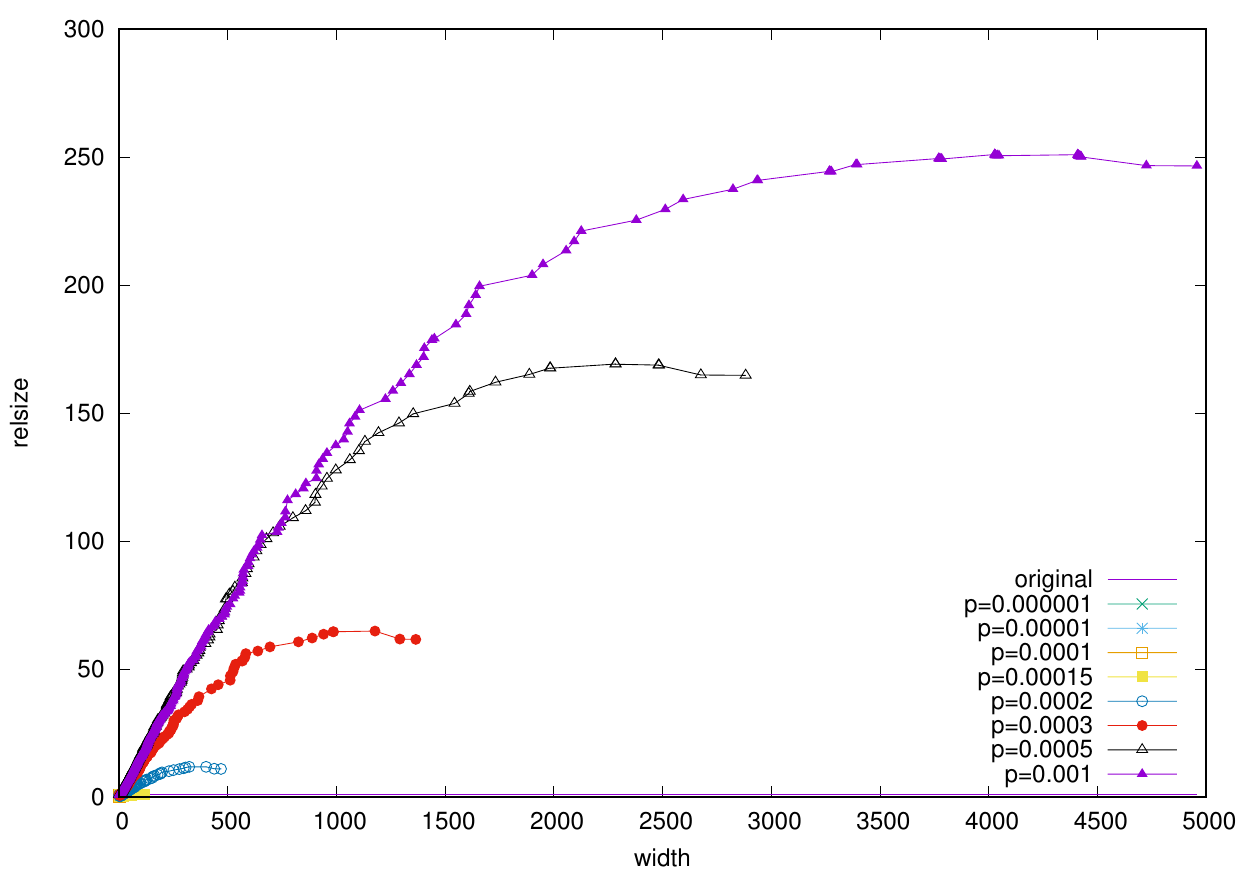}
    \caption{random}
  \end{subfigure}
  \begin{subfigure}{0.32\linewidth}
    \includegraphics[width=\textwidth]{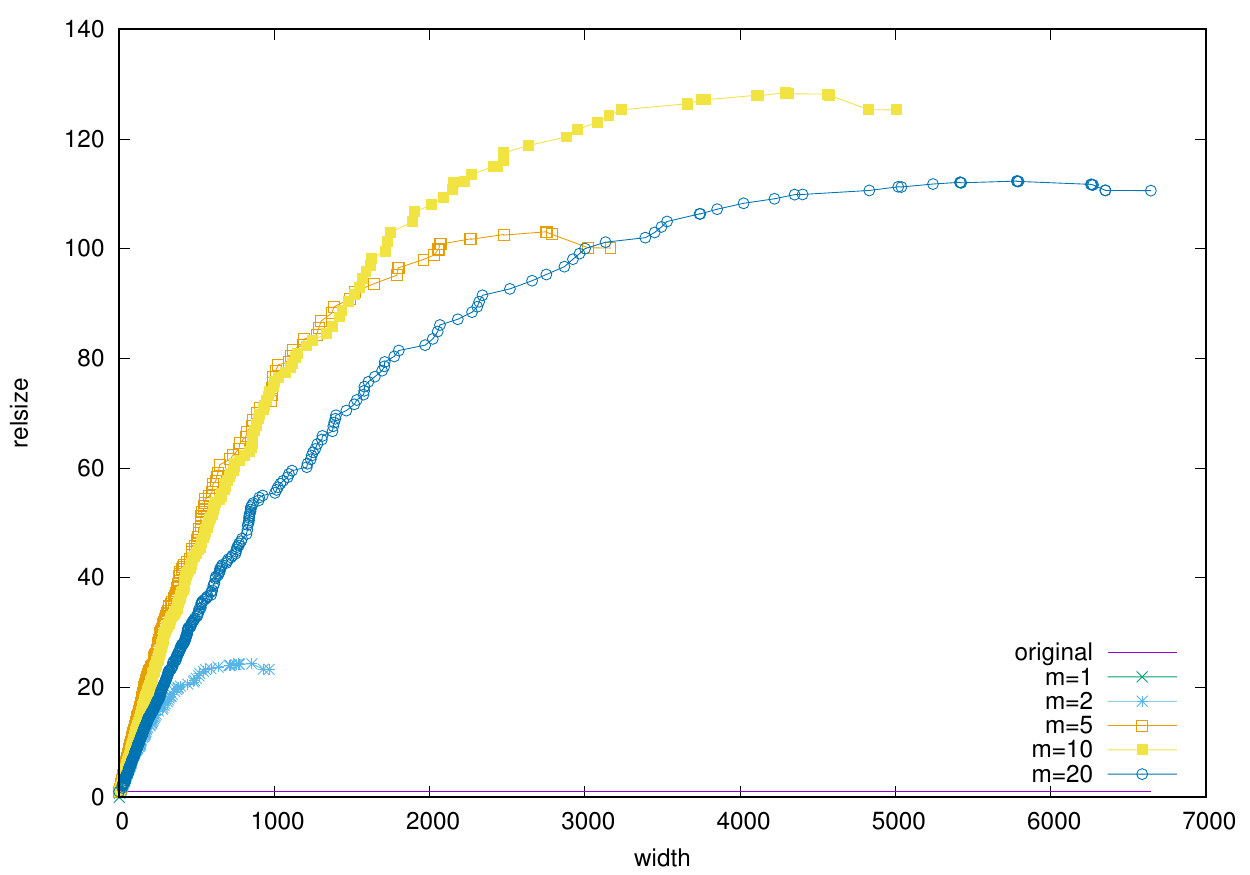}
    \caption{preferential attachment}
  \end{subfigure}
  \begin{subfigure}{0.32\linewidth}
    \includegraphics[width=\textwidth]{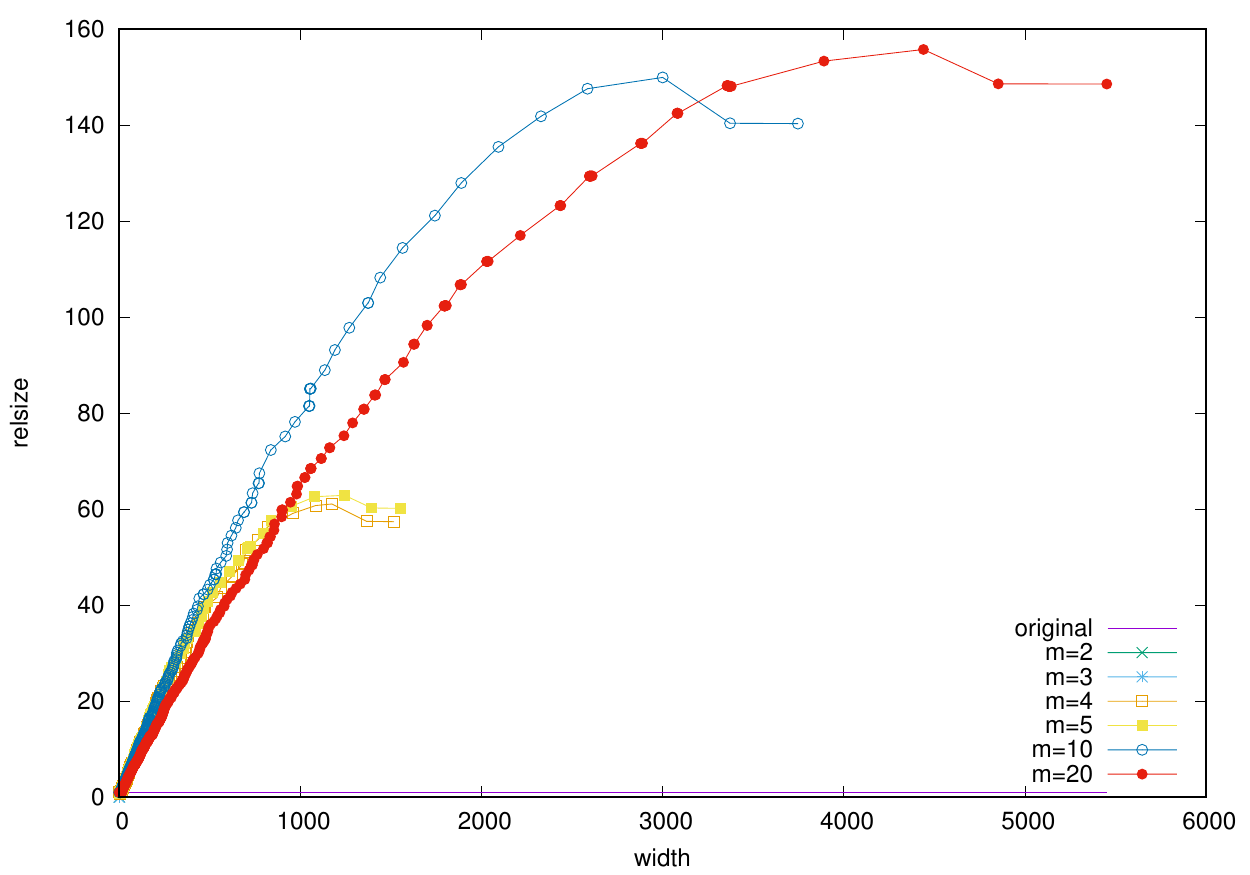}
    \caption{small-world}
  \end{subfigure}

    \vspace{-1em}
  \caption{{\bfseries Synthetic Networks.} Relative sizes of core graphs in partial
  decompositions, after all bags of a given size
  have been removed in the decomposition.\label{fig:part_synthetic}}
    \vspace{-1em}
\end{figure*}
  We show in Figure~\ref{fig:part_synthetic} partial decompositions of
  synthetic graphs, confirming the behavior observed on real-world
  graphs;
the sizes of the resulting core graphs
increase as we increase the parameters of interest.
\end{toappendix}

In practice, however, we do not need to compute full decompositions. For
usable indexes, we may want to stop at low treewidths, which allow exact
computing. This may be useful for graphs whose decompositions have large
core graphs. Indeed (see Appendix~G
\ifextended\else of \cite{extended} \fi
for details), by
stopping at width between 5 and 10, it is often possible to significantly reduce the
original size of the graph, even in cases where core graphs are large,
such as the \textsc{Google} graph.

Such partial decompositions are an extremely promising tool: on databases
where partial decomposition result in a core graph of reduced size, 
efficient algorithms
can be run on the fringe, exploiting its low treewidth, while more costly
algorithms are used on the core, which has reduced size. Combining
results from the fringe and the core graph can be a challenging process,
however, which is worth investigating for separate problems.
As a start, we have shown in~\cite{maniu2017indexing} how to exploit
partial decompositions to significantly improve the running time of
source-to-target queries in probabilistic graphs.

\begin{toappendix}
  \section{Zoomed View of Partial Decompositions} 
  \label{app:zoom}

\begin{figure*}[h]
  \begin{subfigure}{0.24\linewidth}
    \includegraphics[width=\textwidth]{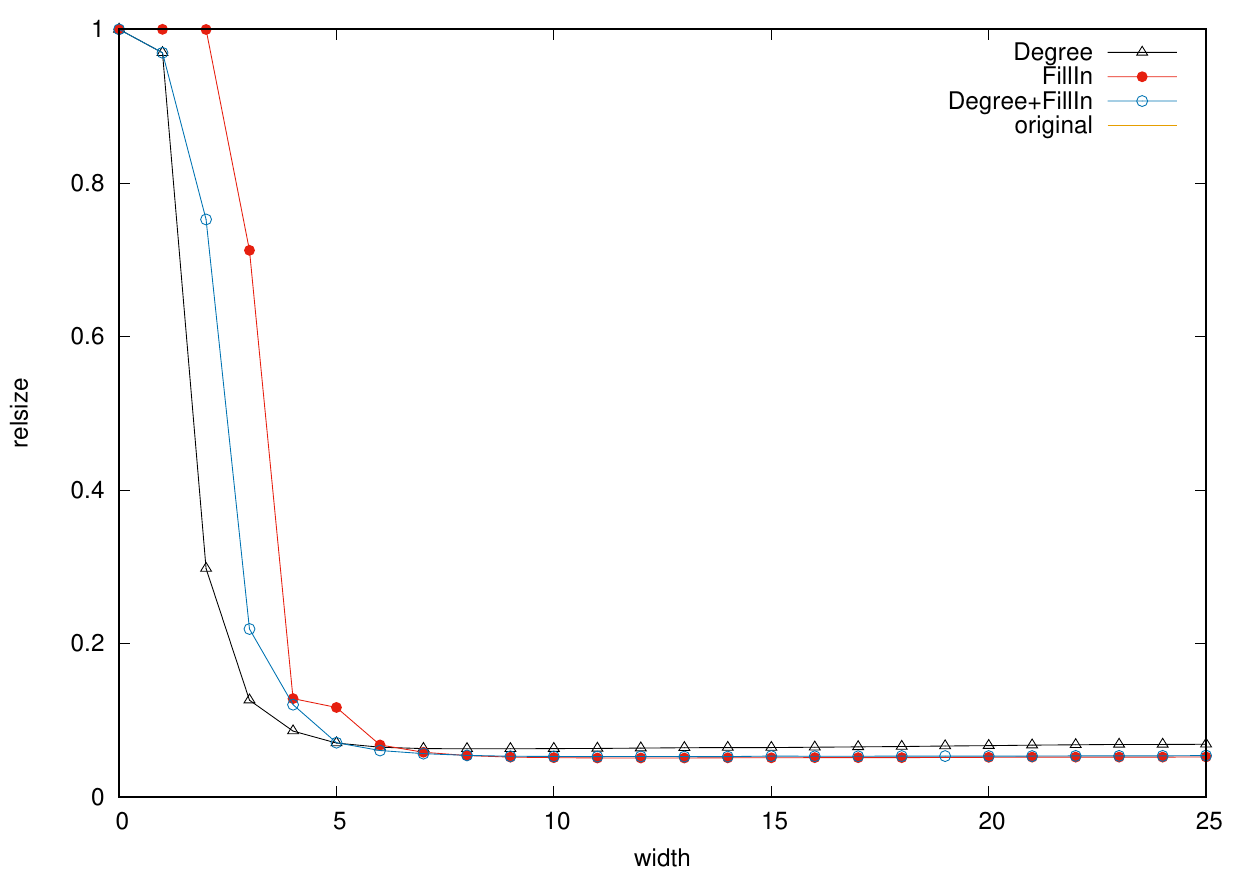}
    \caption{\textsc{Paris}}
  \end{subfigure}
  \begin{subfigure}{0.24\linewidth}
    \includegraphics[width=\textwidth]{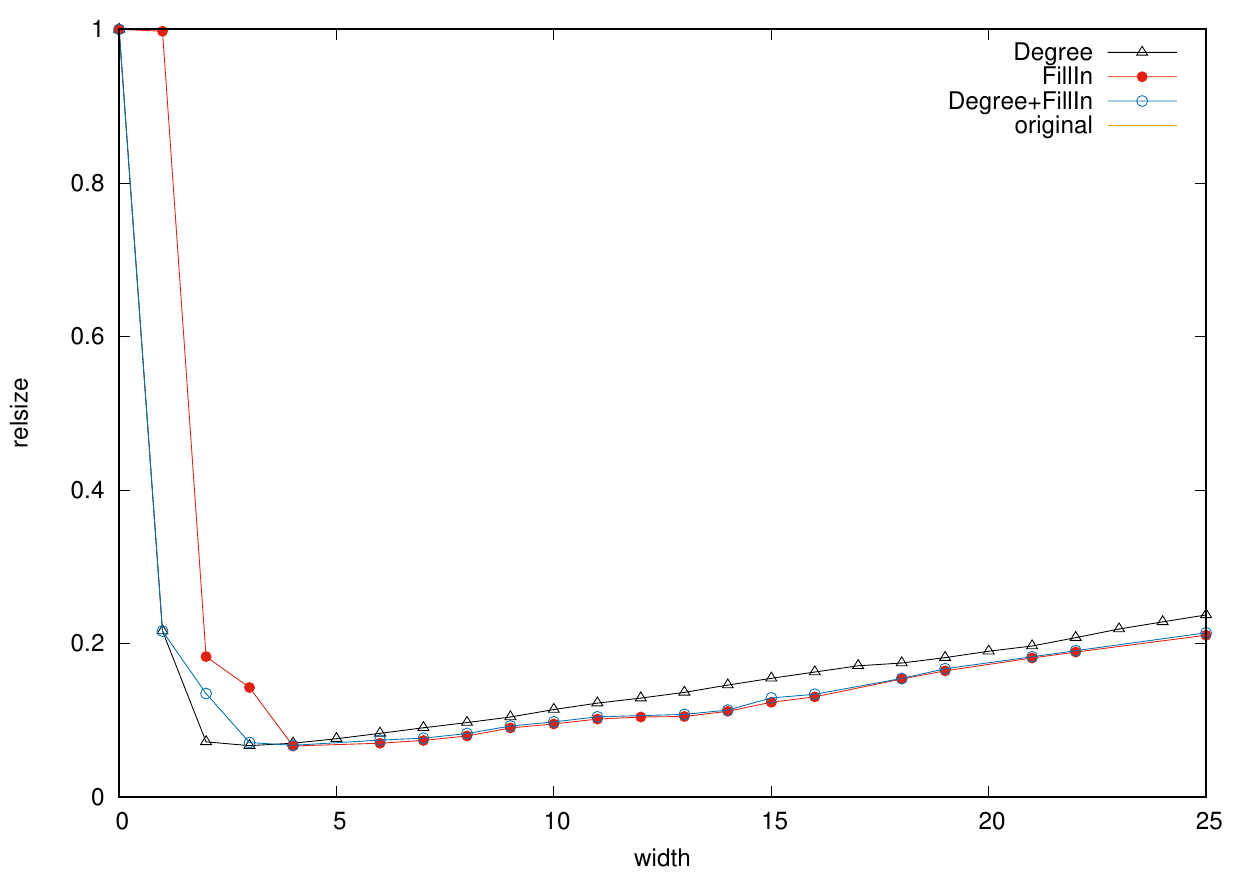}
    \caption{\textsc{Math}}
  \end{subfigure}
  \begin{subfigure}{0.24\linewidth}
    \includegraphics[width=\textwidth]{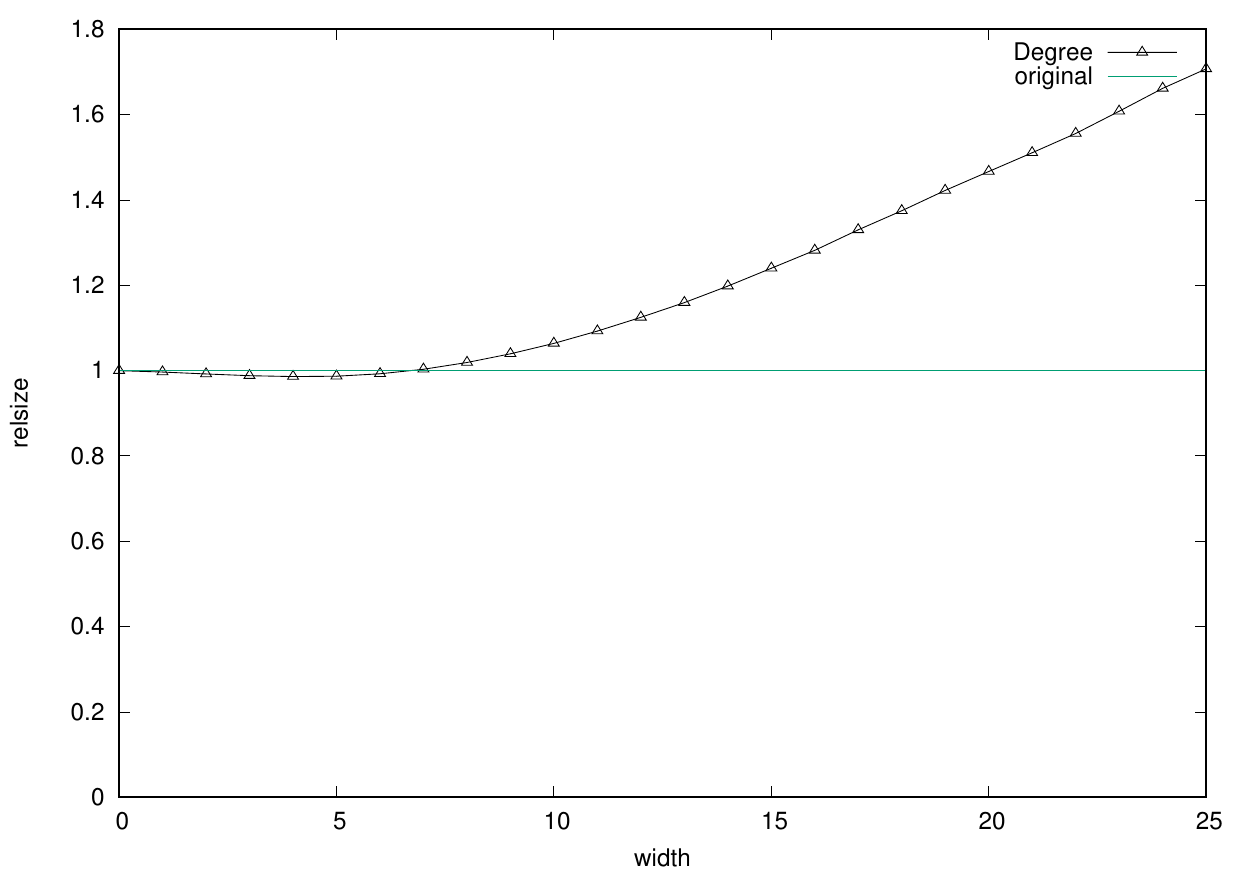}
    \caption{\textsc{CitHeph}}
  \end{subfigure}
  \begin{subfigure}{0.24\linewidth}
    \includegraphics[width=\textwidth]{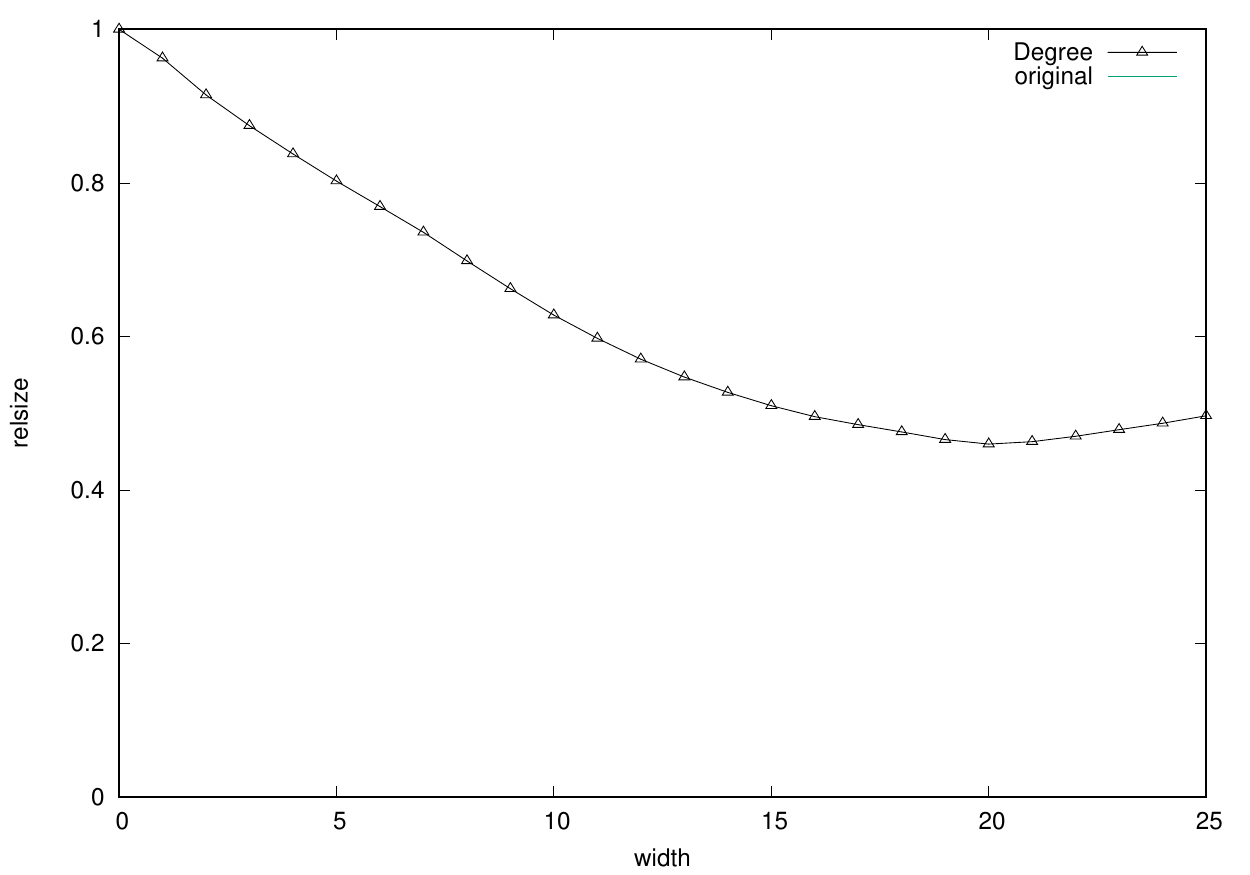}
    \caption{\textsc{Google}}
  \end{subfigure}

  \begin{subfigure}{0.24\linewidth}
    \includegraphics[width=\textwidth]{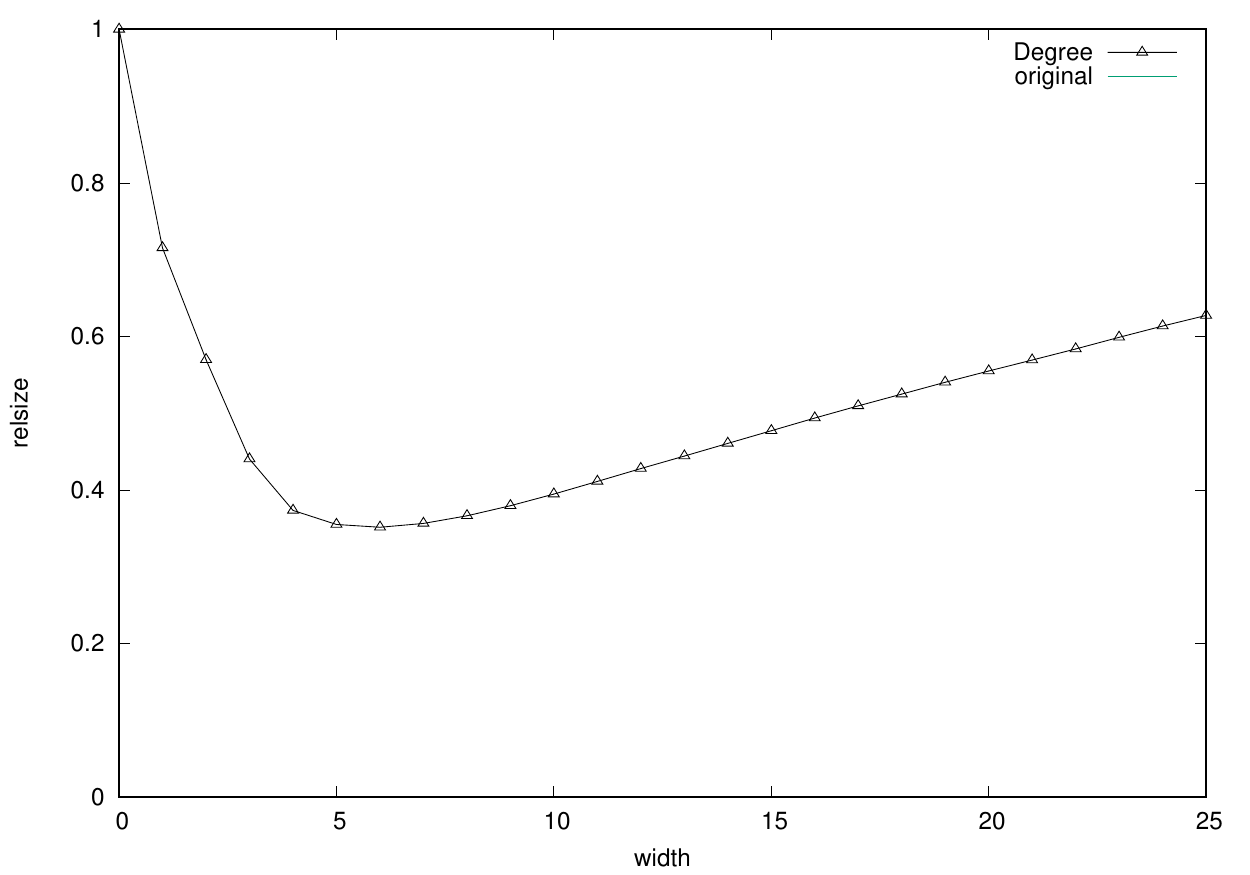}
    \caption{\textsc{Yago}}
  \end{subfigure}
  \begin{subfigure}{0.24\linewidth}
    \includegraphics[width=\textwidth]{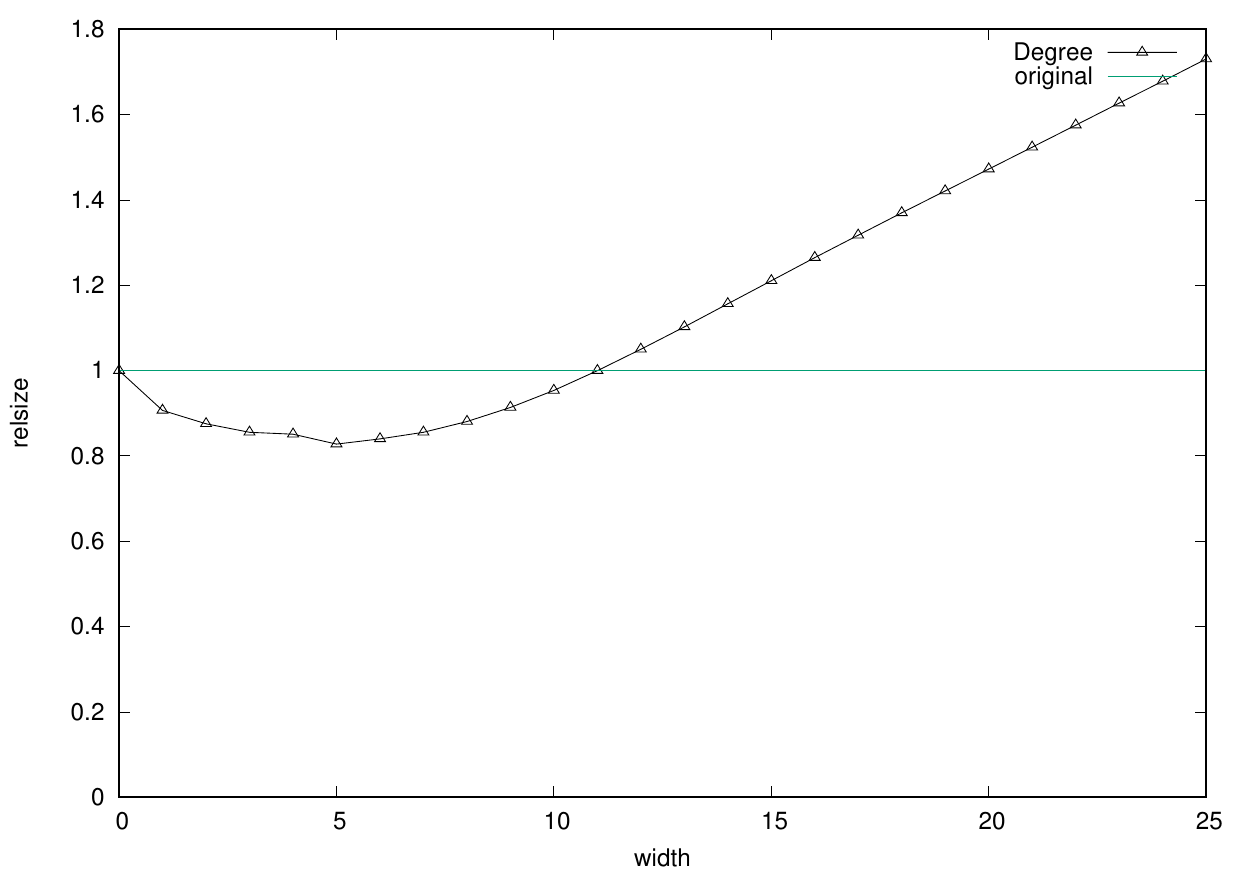}
    \caption{\textsc{DbPedia}}
  \end{subfigure}
  \begin{subfigure}{0.24\linewidth}
    \includegraphics[width=\textwidth]{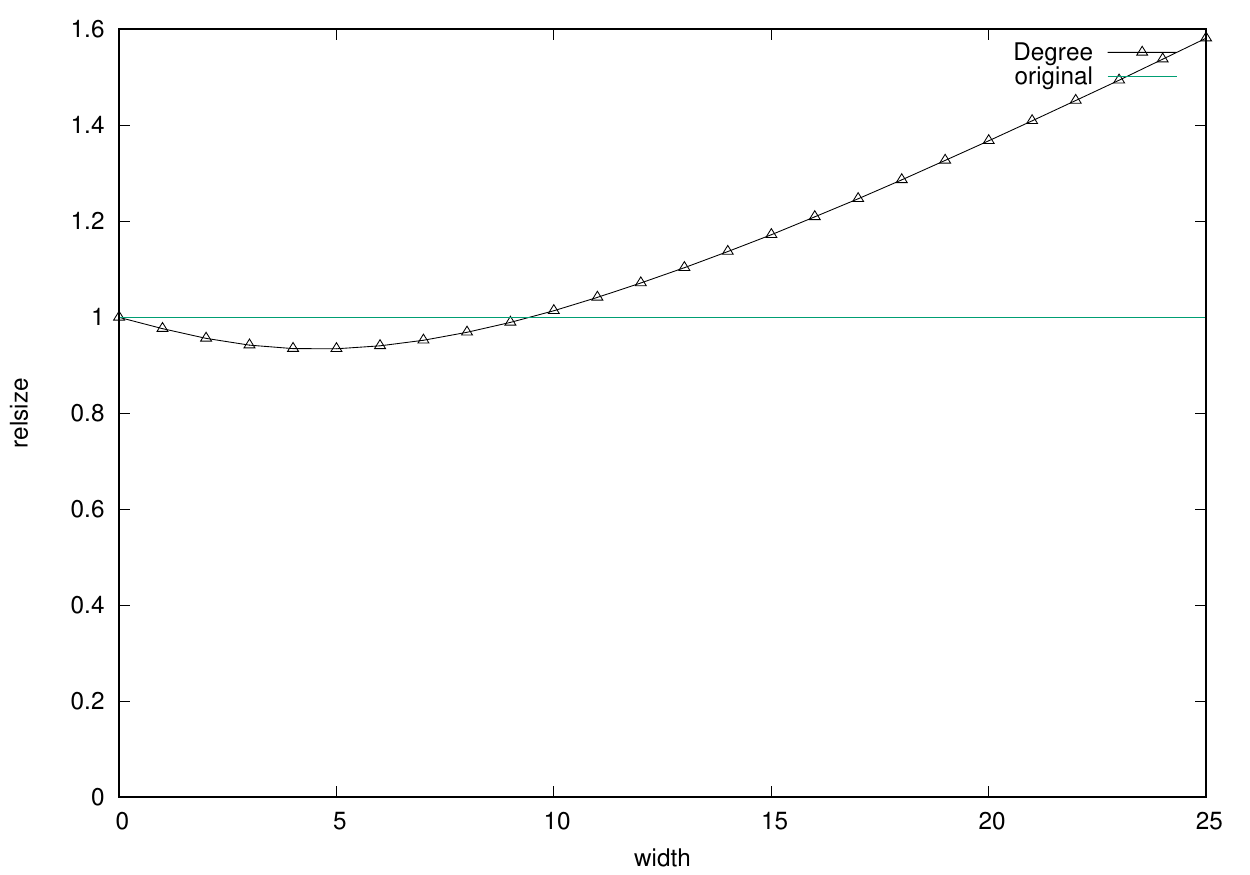}
    \caption{\textsc{LiveJournal}}
  \end{subfigure}
  \begin{subfigure}{0.24\linewidth}
    \includegraphics[width=\textwidth]{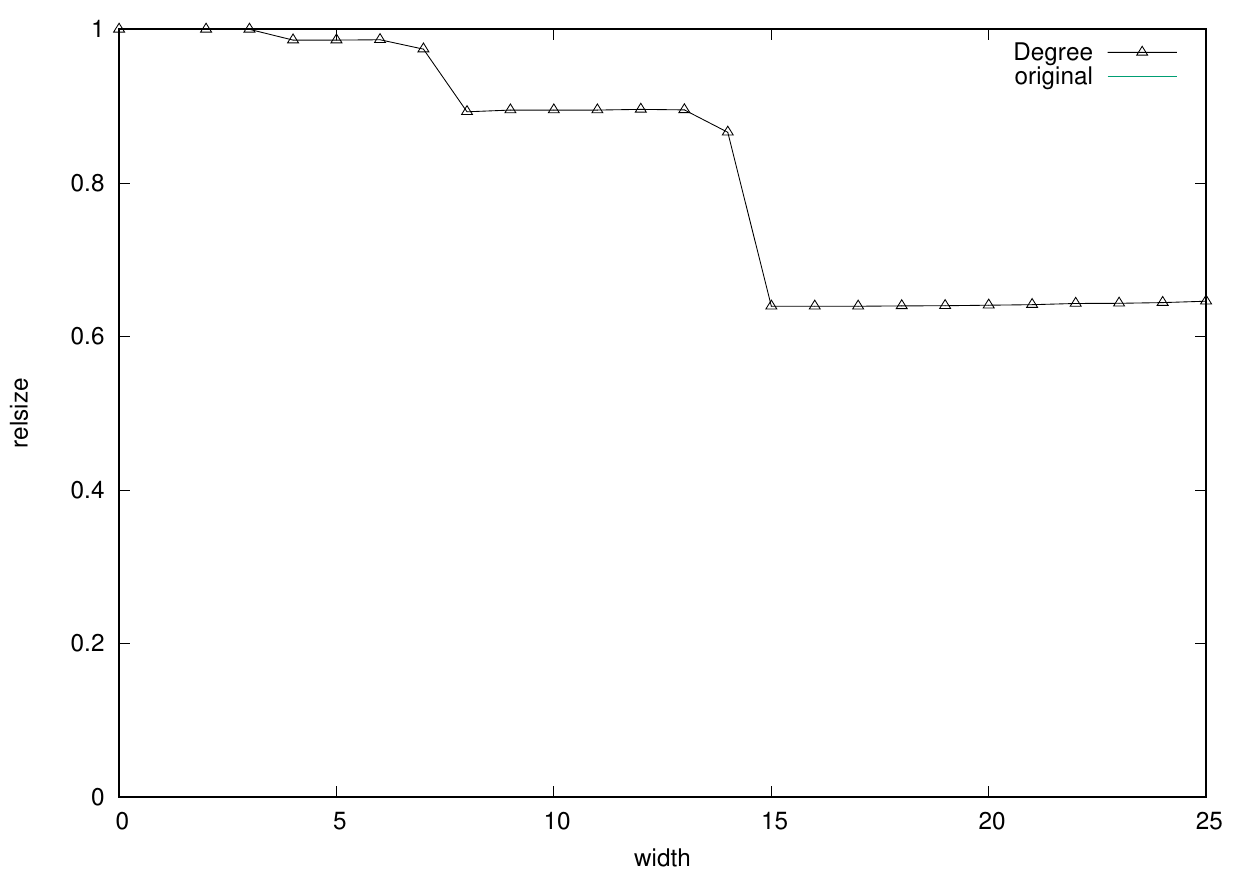}
    \caption{\textsc{Tpch}}
  \end{subfigure}

    \vspace{-1em}
  \caption{Zoomed view of decompositions up to width 25\label{fig:partial_zoom}}
    \vspace{1em}
\end{figure*}
We plot in
Figure~\ref{fig:partial_zoom} decompositions up to a width of 25, for a
selection of graphs, including the graphs for which no upper bound algorithm has
finished (\textsc{Yago}, \textsc{DbPedia}, \textsc{LiveJournal}, \textsc{Tpch}).
Immediately apparent is the fact that the minimal size of the core graph occurs
at very low widths: in almost all cases, it occurs around a width between 5
and 10. This is low enough that running algorithms even doubly exponential
in the width on the resulting fringe graphs
can be performed. In terms of actual size, it can vary greatly. In road
networks, it even reaches~10\%, compared to around 50\% for other graphs. The
exception to this are denser networks (\textsc{CitHeph} and
\textsc{LiveJournal}) where almost no benefit is visible for partial
decompositions of small width. An interesting behavior occurs in \textsc{Tpch}, where
the decomposition size changes in steps; we conjecture this is due to the fact
that database instances contain many cliques -- specifically, one clique for
each tuple in each relation.
\end{toappendix}

\begin{toappendix}
  \clearpage
\end{toappendix}

\section{Discussion}\label{sec:discussion}
In this article, we have estimated the treewidth of a variety
of graph datasets from different domains, and we have studied the running time
and effectiveness of estimation algorithms. This study was motivated by the central role
treewidth plays in many theoretical work attacking query
evaluation tasks that are otherwise intractable, where 
low
treewidths are an indicator for low complexity.

Our study on the \emph{algorithms} for estimation leads to results that may seem
surprising. For upper bounds, we discovered that greedy treewidth estimators,
based on elimination orderings, provide the best cost--estimation trade-off; they
also have the advantage of outputting readily-usable tree decompositions. In
the case of lower bounds, we discovered that degeneracy-based bounds display the
best behavior; moreover, the algorithms which aim to improve the bounds (LBN
and LBN+) only very rarely do so.

In terms of \emph{treewidth estimations}, we have discovered that, generally,
the treewidth of real-world graphs is quite large. With few exceptions, most of
the graphs we have tested in this study are \emph{scale-free}. This may partly
explain our findings -- scale-free networks exhibit a number of high-degree, or
\emph{hub}, nodes that force high values for the treewidth.  The few exceptions to this rule
are \emph{infrastructure networks}, where treewidths are comparatively
lower. Indeed,
we were able to reproduce a $\mathcal{O}(\sqrt[3]{n})$ bound on the treewidth of
road networks. We conjecture these relatively low bounds are explained by
characteristics of infrastructure networks: specifically, they are similar to very sparse
random networks.

Even in the case of infrastructure networks, the absolute value of
treewidth still renders algorithms that are exponential in
the treewidth impractical. However, one of the main lesson of this work is that it is
still possible to exploit the structure of such datasets, by computing
partial tree decompositions: following the approach of
Section~\ref{sec:partial}, it is often possible to decompose a dataset into
a fringe of low treewidth and a smaller core of high treewidth. This has
been used in~\cite{maniu2017indexing}, but for a very specific
(and limited) application: connectivity queries in probabilistic graphs.

In brief, though low-treewidth data guarantees a wealth of theoretical
results on the tractability of various data management tasks, this is
unexploitable in most real-world datasets, which do not have low
enough treewidth. One direction is of course to find relaxations of the treewidth
notion, such as 
in~\cite{DBLP:journals/jacm/FrickG01,DBLP:conf/pods/KazanaS13}. But since
low treewidth is sometimes the only notion that ensures tractability
\cite{kreutzer2010lower,ganian2014lower,amarilli2016tractable}, other
approaches are needed; a promising one lies in the
notion of partial tree decompositions. We believe future work in database
theory should
study in more detail the nature of partial tree decompositions, how to
obtain optimal or near-optimal partial decompositions, and how to exploit
them for improving the efficiency of a wide range of data management
problems, from  query evaluation, to enumeration, probability estimation,
or knowledge compilation.

\bibliographystyle{plainurl}
\bibliography{bibliography}

\end{document}